\documentclass[12pt]{article}

\usepackage[margin=1in]{geometry}
\usepackage{setspace}
\usepackage{amsmath,amsthm,amssymb}
\usepackage{nicefrac}

\usepackage{booktabs,multirow}
\usepackage{array}
\usepackage{float}
\usepackage{arydshln}
\usepackage[table]{xcolor}
\usepackage{siunitx}
\usepackage{dcolumn}

\usepackage{graphicx}
\usepackage[skip=0.5ex]{subcaption}
\usepackage{caption}

\usepackage{soul}
\usepackage{comment}
\usepackage{placeins}
\usepackage{enumitem}
\usepackage{url}
\usepackage[framemethod=tikz]{mdframed}
\newmdenv[splitbottomskip=2\baselineskip, skipabove=10pt, skipbelow=10pt]{myframe}

  
\usepackage{natbib}
\bibpunct[, ]{(}{)}{,}{a}{}{,}%
\theoremstyle{plain}

\theoremstyle{definition}

\theoremstyle{remark}

\definecolor{Gray}{gray}{0.8}
\definecolor{Gray2}{gray}{0.5}
\definecolor{olive}{cmyk}{0.64,0,0.95,0.4}

\newcolumntype{L}[1]{>{\raggedright\let\newline\\\arraybackslash\hspace{0pt}}m{#1}}
\newcolumntype{R}[1]{>{\raggedleft\let\newline\\\arraybackslash\hspace{0pt}}m{#1}}
\newcolumntype{C}[1]{>{\centering\let\newline\\\arraybackslash\hspace{0pt}}m{#1}}
\newcolumntype{J}[1]{>{\let\newline\\\arraybackslash\hspace{0pt}}m{#1}}

\newcommand{\sym}[1]{^{#1}}
\newcolumntype{d}[1]{D{.}{.}{#1}}
\newcommand{\dor}[1]{\multicolumn{1}{c}{#1}}

\newcounter{marginparcounter}

\usepackage{fancyhdr}
\usepackage{hyperref}
\hypersetup{colorlinks=true,linkcolor=blue,citecolor=blue,urlcolor=blue}

\begin{document}

\title{\textbf{Chasing Tails: How Do People Respond to Wait Time Distributions?}}

\author{
Evgeny Kagan\thanks{Carey Business School, Johns Hopkins University. Email: \href{mailto:ekagan@jhu.edu}{ekagan@jhu.edu}}
\and
Kyle Hyndman\thanks{Naveen Jindal School of Management, University of Texas at Dallas. Email: \href{mailto:KyleB.Hyndman@utdallas.edu}{KyleB.Hyndman@utdallas.edu}}
\and
Andrew M.\ Davis\thanks{Samuel Curtis Johnson Graduate School of Management, SC Johnson College of Business, Cornell University. Email: \href{mailto:adavis@cornell.edu}{adavis@cornell.edu}}
}

\date{\today}

\maketitle

\begin{abstract}
We use a series of pre-registered, incentive-compatible online experiments to investigate how people evaluate and choose among different waiting time distributions. Our main findings are threefold. First, consistent with prior literature, people show an aversion to both longer expected waits and higher variance. Second, and more surprisingly, moment-based utility models fail to capture preferences when distributions have thick-right tails: indeed, decision-makers strongly prefer distributions with long-right tails (where probability mass is more evenly distributed over a larger support set) relative to tails that exhibit a spike near the maximum possible value, even when controlling for mean, variance, and higher moments. Conditional Value at Risk (CVaR) utility models commonly used in portfolio theory predict these choices well. Third, when given a choice, decision-makers overwhelmingly seek information about right-tail outcomes. These results have practical implications for service operations: (1) service designs that create a spike in long waiting times (such as priority or dedicated queue designs) may be particularly aversive; (2) when informativeness is the goal, providers should prioritize sharing right-tail probabilities or percentiles; and (3) to increase service uptake, providers can strategically disclose (or withhold) distributional information depending on right-tail shape.
\end{abstract}

\medskip

 
\section{Introduction \label{sec:intro}}
Waiting times in service systems are rarely deterministic. Instead, they vary considerably, with some customers served quickly while others experience extensive delays. In customer service, call routing can expedite assistance for some requests, while others may need to wait longer for a specialist or a callback. A pharmacy might fill one prescription in minutes, while the next customer may need to wait longer for  insurance verification or stock replenishment. Uber and Lyft use a batched matching mechanism that can delay assignment as the system widens its search for available drivers  \citep{UberMatching2025,LyftDispatch2025}, while DoorDash often batches multiple deliveries into a single multi-stop trip  \citep{DoorDashBatched2024}. Realized waits on these digital platforms tend to be short when idle capacity is available, but can be quite long when the system expands its search or aggregates jobs. The same customer requesting the same service could experience very different waiting times depending solely on chance.

Queuing theory provides some insight into how different waiting time distributions arise. For example, dedicated (as opposed to pooled) queues can improve welfare in some settings \citep{sunar2021}, but they may do so at the cost of creating a waiting time distribution with a spike at zero and a thicker right tail \citep{cao2021}. Similarly, priority queue models show how class-specific waiting time distributions combine to form unconditional distributions \citep{stanford2014,wang2015}. Such mixture distributions can be complex and can exhibit multiple modes corresponding to the different priority classes \citep{horvath2020}. Equal balancing of server loads can negatively impact some of the queueing nodes, and that can increase variance \citep{do2020}. A common feature throughout this literature is that policies designed to maximize system efficiency typically also reshape the entire waiting time distribution. A system optimized for average performance may look different from a system optimized based on a mean-variance criterion or based on a maximum wait guarantee \citep{do2020,cao2021}.

While the distributional effects of different service designs on waiting times have been examined in the literature, little is known about the decision-making process of customers joining a service system. This motivates our first research question: \textit{How do decision-makers evaluate waiting time distributions?} A second, more practical question, particularly for service providers, is how to communicate this distributional information: \textit{What distributional information do people value the most, and how do they respond to the different types of information provided?} In practice, information may be displayed as a point estimate (usually mean), a range (minimum/maximum time), a percentile (e.g., 80\% of customers are served within $X$ minutes), or a fuller representation of the distribution. In this paper, we aim to address both these questions using a series of incentivized online experiments.

\subsection{Experimental Approach and Preview of Results}
Our experiments include, in total, 1,114 participants.  The sample sizes, pre-registration links, research questions, and key findings of each of our studies are summarized in Table \ref{tab:design}.\footnote{In addition to the studies reported in the main text, we conducted two further experiments: a pilot study and a follow-up study, with a total of 510 additional participants. See Appendix B for details on the pilot study and Appendix D for the follow-up study.}

Study 1 (\textsection 2) constitutes our baseline set of experimental treatments, in which we examine the response to three classic probability distributions (Bernoulli, Uniform and Exponential). The initial results align with the conventional wisdom that the cost of waiting increases with the expected duration and uncertainty of a wait. In particular, a 1-minute increase in mean waiting time is valued at approximately \$0.10, and a 1-minute increase in standard deviation of a wait is valued at approximately \$0.05. These results are consistent with hourly wages earned on online recruitment platforms \citep{palan2018}, and extend prior results, obtained mainly with hypothetical elicitation \citep{kroll2008,leclerc1995,festjens2015risk} and normally distributed waits \citep{flicker2022}, to a wider range of distributions using an incentive-compatible protocol. Interestingly, we also find differences between distributions. While Uniform and Bernoulli waits become significantly less desirable with increased variance, the evaluation of Exponential waits does not change with variance.  Additionally, we find significant differences between distributions, even when mean and variance are held constant. This provides some suggestive evidence that higher moments and/or tail shape may play a role in determining customer preferences.
\begin{table}[b!]
\centering
\scriptsize
\caption{Pre-registration Links, Research Questions, and Key Results}\label{tab:design}
    \begin{tabular}{L{1.7cm}L{4cm}L{4.6cm}L{5.2cm}}
    \toprule
    \textbf{Study} & \textbf{Pre-registration} & \multicolumn{1}{c}{\textbf{Research Questions}} & \multicolumn{1}{c}{\textbf{Key Results}} \\
    \midrule
    \textbf{Study 1} (\textsection 2) \newline $N = 316$ & 
       \url{https://aspredicted.org/C32_J22} & 
    Are uncertain waits valued less than certain? 
    Does distributional form matter beyond mean/variance? &
    Greater variance makes the wait less attractive. Significant differences between distribution shapes. \\
    \midrule
    \textbf{Study 2} (\textsection 3) \newline $N = 479$ & 
    \textbf{Study 2A} \newline \url{https://aspredicted.org/sgxr-zzgm.pdf} \newline $~$ \newline
    \textbf{Study 2B}   \newline \url{https://aspredicted.org/gdw5-dbkv.pdf} \newline $~$ \newline
    \textbf{Study 2C}   \newline \url{https://aspredicted.org/4g2p-y2dy.pdf} & 
    Do higher moments predict decisions? \newline $~$ \newline $~$ \newline
    What types of right tails are most aversive? \newline $~$ \newline $~$ \newline
    Does thick-tail aversion persist under incomplete information? & 
    Moment-based models fail to consistently predict choices. \newline $~$ \newline $~$ \newline
    Long-right tails preferred over thick-right tails. \newline $~$ \newline $~$ \newline
    Yes. \newline $~$ \\
    \midrule
    \textbf{Study 3} (\textsection 4) \newline $N = 319$ & 
    \url{https://aspredicted.org/Y27_X27} & 
    What distributional information do people value most? & 
    Most interested in right-tail information. \\
    \bottomrule
    \end{tabular}
\end{table}

To resolve the questions raised in Study 1, Study 2  (\textsection 3) introduces experiments that examine higher moments and different right-tail shapes, while controlling for the various features of the distributions. In particular, Study 2A isolates specific drivers of the departures from the mean-variance benchmark by varying one moment (variance, skewness, kurtosis) at a time and holding the remaining moments constant.  Study 2B evaluates whether the shape of the right tail (``thick-right'' when the probability of a long wait is lumped at a singular outcome versus ``long-right'' when this probability is spread out across multiple high wait outcomes) affects decisions (See Fig. 2c on p. 13 for an illustration).  Study 2C then examines whether the preferences and behaviors observed in Studies 2A and 2B are robust to a more realistic setting where distributional information may be incomplete. 

The results of Study 2 show that statistical moments alone do not provide an adequate framework for customer preferences.  Study 2A shows that higher variance and higher kurtosis distributions need not be aversive, particularly when choices are made between distributions with different shapes. Instead, Study 2B reveals that tail shape is a key driver of customer waiting preferences: participants show a robust preference for long-right tails over thick-right tails, conditional on mean and variance. This suggests that people distinguish between different types of tail risk: they prefer distributions where extreme delays are possible but improbable (long-right tail) over distributions where moderately long delays are relatively likely (thick-right tail). Study 2C confirms that the observed ``thick-right-tail aversion'' phenomenon persists in an environment with incomplete information showing that preferences can be ranked as: known long-right tail $\succ$ unknown $\succ$ known thick-right tail. 

Study 3 (\textsection 4) complements the findings of Study 2 by investigating what specific distributional information decision-makers find most valuable when facing uncertain waits.  This is especially relevant to service providers deciding whether or not to provide distributional information to customers and, if so, what specific type of information to share.  Study 3 addresses this through eliciting what distributional information (i.e., probability mass in the left tail, in the midrange, or in the right tail) a decision-maker is most interested in learning.  

The results of Study 3 suggest a robust preference for learning right-tail (i.e., worst-case) information about a waiting time distribution. This reinforces the main result from Study 2 that people respond mainly to the shape of the right tail when making decisions. A follow-up study further shows that the effects of revealing wait-time information can have a positive or negative effect on how a wait is evaluated, depending on the shape of the right tail. 

We conclude our analysis in \textsection 5 by pooling decision data from multiple studies and by showing that utility models that incorporate right-tail characteristics explain the data well. We then embed the estimated utility model into a discrete-event simulation of a queueing system and show that tail-based utility can lead to different optimal service designs (pooled vs. dedicated queues) than the classic mean-only utility model.
 
\subsection{Literature\label{sec:literature}}
The relevant literature includes: (i) human behavior in the time domain, (ii) disclosure of waiting times in service systems (iii) preferences over higher moments in the money domain, and (iv) transportation research on travel-time reliability. Together, these works show that averages are frequently insufficient to characterize preferences and behavior, but they leave open the following question: which \emph{features} of a waiting-time distribution explain people's preferences when evaluating different distributions?

\subsubsection*{Human Behavior in the Time Domain} 
The literature on preferences for different waiting formats is quite limited. Most studies focus either on deterministic  \citep{lin2015customer} or Bernoulli-distributed waits \citep{leclerc1995,abdellaoui2014,kroll2008,festjens2015risk} and rely largely on hypothetical elicitation. The closest studies that use incentive-compatible elicitation are \cite{flicker2022} who examine normally distributed waits, and \cite{luo2022} who compare experienced and prospective waits in queues. \cite{kremer2016} do not present participants with an explicit time distribution but rather let them infer duration from queue length or configuration. \cite{buell2021} finds that people exhibit a last-place aversion in queues. Our finding that people avoid distributions with thick right tails echoes this result and suggests that a salient spike in ``worst-case'' outcomes is aversive even in the absence of a social comparison. Also related is the literature on the response to delay information \citep{althenayyan2022,aksin2022,ansari2022}. Different from us, these studies focus mainly on events and behaviors that occur \textit{while} waiting, rather than on ex-ante choices.

\subsubsection*{Waiting Time Information Sharing} 
Recent service operations literature has examined strategic decisions about when and how to communicate delay information, both prior to a customer (or patient) joining a service system \citep{debo2023} and during the wait \citep{kim2025}, using analytical models. Empirical studies have further examined how delay announcements affect customer behavior, abandonment decisions, and satisfaction \citep{yu2017, yu2022}. \cite{liu2025} provides a comprehensive survey of patient-provider communication in healthcare systems. While this literature generates insights about disclosure policies and their effects, it generally takes customer (or patient) preferences as given. Our work complements this literature by examining, at the micro-level, what these preferences may look like. 

\subsubsection*{Higher Moments in the Money Domain}
Several studies have examined responses to higher moments (beyond means and variances) in the money domain. Some studies find that people may prefer right-skewed distributions, even when means and variances are held constant \citep{brunner2011,wu2011}. This has been attributed to factors such as over-optimism or overweighting of small probabilities in the right tail \citep{kahneman1979,astebro2015}. Unlike money, for prospects involving time, a longer right tail (e.g., in an Exponential distribution) would correspond to a longer wait, potentially leading to aversive preferences. It is therefore not clear how people would respond to a right skew in our setting. \cite{Frechette-et-al-2017} show that decision-makers are interested in learning more about the left tail (as opposed to midrange or right tail) of a probability distribution for money. Similar to their paper, we use a rank-order-based information elicitation protocol (In Study 3) and also find that decision-makers are most interested in the worst case (in our case, the right tail), when outcomes are measured in time units. 

\subsubsection*{Transportation}
The transportation literature provides additional insights into how people experience time uncertainty. Empirical studies document that travel time distributions exhibit complex shapes: sometimes unimodal, sometimes bimodal, but frequently feature thick or long right tails \citep{vanLintvanZuylen2005,FosgerauFukuda2012,SusilawatiEtAl2013}. These distributional properties have motivated measures of ``travel time reliability'' that quantify predictability through right-tail behavior \citep[e.g.][]{KimMahmassani2015,Taylor2017,ZangEtAl2024}. While most transportation research characterizes travel time distributions through empirical datasets and simulation models, experimental studies typically rely on non-incentivized stated preferences, asking drivers to choose among hypothetical routes \citep{BogersEtAl2006}. Different from this literature, we examine waiting time preferences using incentivized human-subject experiments where participants face real consequences from their choices. For comprehensive reviews of travel time uncertainty research, see \cite{CarrionLevinson2012} and \cite{ZangEtAl2022}.

\subsection{Contributions}
Our contributions are threefold. First, we present evidence for a novel preference. We show that the shape of the right tail is a key predictor of choices: decision-makers exhibit an aversion to waits with a ``thick-right'' tail. This result is distinct from prior results that  ``variance matters'' within Bernoulli \citep{leclerc1995,festjens2015risk} or Normal \citep{flicker2022} distributions, which, by design, do not vary tail shapes. Second, our results inform models of service operations. Service designs with the same mean wait can have very different distributional shapes \citep{cao2021,do2020}. To develop more realistic predictions for joining behavior, revenues, and welfare, analytical models of service operations may benefit from incorporating utility specifications that are empirically grounded. Our estimation results (in \textsection 5.1) offer a plausible alternative for what these utility models may look like. Third, our results contribute to the practice of service operations. To maximize informativeness, service providers should disclose right-tail information to customers in addition to sharing mean waiting times. However, service providers should be aware that disclosing distributional information when the right tail is thick may reduce demand.

\section{Study 1: Mean and Variance Matter but Fail to Capture All Behavior
\label{sec:study1}}
Study 1 focuses on the first two moments of a waiting time distribution: mean and variance. We examine three different probability distributions: Binary (Bernoulli), Exponential and Uniform. We chose these distributions because they span a range of different real-life service settings, from a binary short/long wait, to an uncertain wait that involves a \textit{range} of potential waiting times with a flat prior on the likelihood of each time, to a distribution that is right-skewed and has a long-right tail as is typical for call center wait times \citep{brown2005}. Within each distribution, we first verify that longer average waits and higher variance make the wait less attractive. We then compare choices across distributions while holding mean and variance constant, to test whether distributional form affects decisions beyond the first two moments.\footnote{Prior to conducting Study 1, we ran a pilot study, in which we tested several alternative designs. Please see Appendix B for details.} 

\subsection{Experiment Design}
The treatments and waiting scenarios are in Table \ref{tab:study:1:scenarios}. We carefully parameterized the distributions to hold the mean and variance constant within each scenario across all three treatments. To elicit the value of each distribution (i.e., the amount of money someone needs to be paid to experience that distribution), we used multiple price lists \citep{Holt02}. In particular, participants saw six scenarios (in random order). In each scenario, Option A was: ``\textit{Receive \$1.00 and wait 1 minute}", while Option B was ``\textit{Receive \$$X$ and wait $Y$ minutes}", where $X$ and $Y$ depended on the scenario and the row of the price list. In particular, $X$ was varied within a scenario, while $Y$ was varied across scenarios. By fixing the distribution of $Y$ and examining the monetary value $X$ at which participants switched from Option A to Option B, we can find the dollar amount they require to experience that distribution (see Figure \ref{fig:study1_screenshot} for a screenshot of the setting).\footnote{The multiple price list 
elicitation is similar to the Becker-DeGroot-Marschak method used in \cite{luo2022} in that it elicits a monetary value of a given wait.}

A total of 372 Prolific workers were recruited (average age: 38.5, 58\% female). Only US-based workers with an approval rating of at least 99\% were eligible. A total of 56 participants were excluded based on pre-registered comprehension and attention checks, resulting in a sample size of 316. Recruitment was stopped once the target sample of 100 participants per treatment was reached. Participants were incentivized to report their preferences truthfully by experiencing one of their choices (both in terms of time and money) in real-time, chosen at random at the end of the experiment. All participants received a \$3 show-up fee in addition to their earnings from the experiment. Please see Table \ref{tab:demographics} for demographic details of the study participants.

\begin{table}[tb]
    \centering \scriptsize
    \caption{Study 1: Treatments and Waiting Scenarios}
    \label{tab:study:1:scenarios}
    \begin{tabular}{C{2.6cm}C{3.2cm}C{2.6cm}C{4cm}C{2.75cm}} \toprule
    &  \multicolumn{3}{c}{Treatment (varied between-subject): Distribution used for Option B} & \\ \cmidrule{2-4}
    Scenario (varied within-subject)   &  Binary & Uniform & Exponential & Mean and Standard deviation   \\  \midrule
     1      &  deterministic   &  deterministic &  deterministic & $\mu=5,\sigma=0$  \\
     2    &  deterministic   &  deterministic &  deterministic & $\mu=10,\sigma=0$  \\
     3  &   4 or 6, with prob. 0.5  & $U[3.27, 6.73]$ &  exp($\lambda=1, \textrm{mean shift} = 4$) &  $\mu=5,\sigma=1$  \\
     4 &   3 or 7, with prob. 0.5  & $U[1.54, 8.56]$ &  exp($\lambda=1/2, \textrm{mean shift} = 3$) &$\mu=5,\sigma=2$     \\
     5   &  9 or 11, with prob. 0.5  & $U[8.27, 11.73]$& exp($\lambda=1, \textrm{mean shift} = 9$)&$\mu=10,\sigma=1$  \\
     6    &  5 or 15, with prob. 0.5   & $U[1.34, 18.66]$ & exp($\lambda=1/5, \textrm{mean shift} = 5$)&$\mu=10,\sigma=5$ \\  \bottomrule
    \end{tabular} 
    \begin{minipage}{\textwidth} \scriptsize \vspace{0.1cm}
    Note: The sequence of scenarios within each block (deterministic/stochastic) was randomized. All parameters are in minutes. 
    \end{minipage}
\end{table}

\subsection{Hypotheses}
As noted in \textsection\ref{sec:literature}, research on choices in the time domain is quite limited. Because time spent (in a service setting) may be seen as a loss, prospect theory would predict risk seeking in the loss domain \citep{kahneman1979,tversky1992}. However, prior work on waiting-time choices does not support risk seeking; instead, people tend to prefer options with lower uncertainty  \citep{leclerc1995,festjens2015risk}. Evidence on sensitivity to distributional form is equally sparse: most existing studies use binary waits \citep{leclerc1995,abdellaoui2014}, bypassing other distributional shapes. Given the lack of strong theory in support of a directional hypothesis, we therefore hypothesize that the first two moments fully account for the choices.

\smallskip
\noindent \textit{H1A: The attractiveness of a wait decreases in its variance.}

\noindent \textit{H1B: Holding constant the mean and variance of waiting times, the attractiveness of a wait is the same for Binary, Uniform, and Exponential waiting time distributions.}

\subsection{Results}
To conserve space, we focus on hypothesis tests in the main text and present detailed summary statistics and other results in Appendix C.2. To test the hypotheses, we estimated logistic regressions of switching choices (i.e., the minimum dollar amount that prompts a participant to accept an uncertain wait) on the mean and standard deviation of the waiting-time distribution. 
The results in Table \ref{tab:Study:1:hypothesis:test} strongly support H1A but reject H1B. Doubling the mean from 5 to 10 minutes produces the largest effect ($\beta = 0.491$, $p < 0.001$), indicating that participants demanded nearly \$0.49 for a mean increase of 5 minutes. Increases in variance also significantly reduced attractiveness: moving from $\sigma = 0$ to $\sigma = 1$ ($\beta = 0.118$, $p = 0.002$); from $\sigma = 1$ to $\sigma = 2$ at $\mu = 5$ ($\beta = 0.046$, $p = 0.026$); and from $\sigma = 1$ to $\sigma = 5$ at $\mu = 10$ ($\beta = 0.193$, $p < 0.001$).\footnote{We do, however, find that the variance effect is quite weak, and not statistically significant, if we focus solely on the Exponential Distribution. Please see Appendix C.2 for detailed analysis. In \textsection 3, we will show that this is due to the long-right tail being evaluated quite favorably relative to thick tails.}

To test H1B, we compared the coefficients across distributional forms (Binary, Uniform, Exponential) using Wald tests. Equality between Uniform and Exponential could not be rejected ($p = 0.411$), nor between Binary and Exponential ($p = 0.187$). However, Binary and Uniform differed significantly ($p = 0.015$), with the Binary distribution having a lower price. That is, Binary waits are less aversive. Further, the joint test of equality across all three shapes was rejected ($p = 0.045$). These results provide some initial evidence that while mean and variance are important determinants of choice, distributional form also affects behavior. 

\begin{table}[bt]\centering \scriptsize    \renewcommand{\arraystretch}{1.2}
\caption{Study 1: Hypothesis Tests}
\label{tab:Study:1:hypothesis:test}
\begin{tabular}{L{9.2cm}ccc} \toprule
  Comparison     & \multicolumn{3}{c}{Test Results}   \\ \cmidrule{1-4}
$\mu = 5 \to \mu = 10$  & \multicolumn{3}{c}{$\beta = 0.491$, $p < 0.001$}            \\
$\sigma = 0 \to \sigma = 1$   & \multicolumn{3}{c}{$\beta = 0.118$, $p = 0.002$}      \\
$\sigma = 1 \to \sigma = 2$ (at $\mu = 5$) & \multicolumn{3}{c}{$\beta = 0.046$, $p = 0.026$}     \\
$\sigma = 1 \to \sigma = 5$ (at $\mu = 10$) & \multicolumn{3}{c}{$\beta = 0.193$, $p < 0.001$}     \\
\midrule
\texttt{isUniform = isExponential}      & \multicolumn{3}{c}{$p = 0.411$} \\
\texttt{isBinary = isUniform}      & \multicolumn{3}{c}{$p = 0.015$}\\
\texttt{isBinary = isExponential}      & \multicolumn{3}{c}{$p = 0.187$} \\
\texttt{isBinary = isUniform = isExponential}      & \multicolumn{3}{c}{$p = 0.045$} \\
\bottomrule
\end{tabular}
\begin{minipage}{0.76\textwidth}
        \scriptsize
        \vspace{0.1cm}
        Note: $\beta$ coefficients are marginal effects, measured in US Dollars. Higher $\beta$ indicates a higher cost of waiting. For example, the average increase in compensation when going from $\mu = 5$ to $\mu = 10$ is \$0.49. $p-$values are based on logistic regression coefficients and parameter tests. See Table \ref{tab:regression:study:2} for full regression results.
    \end{minipage}
\end{table}

\subsection{Discussion}
The results presented so far echo the findings from prior hypothetical-choice studies \citep{leclerc1995,kroll2008,festjens2015risk}. Consistent with these studies, participants demanded more money for longer expected waits and for higher variance. The results also align with \cite{flicker2022}, who show that a mean--variance utility model captures choices well for normally distributed waits. What Study 1 results add is that two waits with identical means and variances can still be valued differently when they follow different distributions: binary waits are most attractive, while uniform waits require higher compensation. That is, some property beyond the first two moments appears to drive decisions. 

Because we have relied on off-the-shelf distributions, we cannot tell with the available data which property this is. Uniform and exponential waits differ from the binary wait (and from each other) in several ways: they have wider bounds, a larger worst-case outcome, and in the exponential case a right skew. Discreteness, range, and tail shape therefore move together, which means that we cannot tell whether participants are reacting to continuity, to extreme realizations, or to the shape of the right tail. For example, it is possible that participants are unable to correctly interpret continuous distributions and therefore find them more aversive relative to discrete ones. The next section introduces more controlled manipulations that vary one feature at a time in order to isolate the specific drivers of the departures from the mean--variance benchmark.

\section{Study 2: People Are Averse to Thick-Right Tails}
To better characterize behavior, in Study 2 we extend our approach as follows. First, we make all distributions bounded and discrete, such that the nature of the distribution (and the comprehension difficulties potentially introduced by unboundedness and continuity) cannot affect behavior. Second, we carefully control for the range of the distributions. Third, we fix three of the four moments (mean, variance, kurtosis, skewness) to isolate the response to each moment (Study 2A). Fourth, to better understand how tail shape drives behavior, we examine the role of tail length, thickness and mode location (Study 2B). Fifth, we evaluate whether the preference patterns observed with complete distributional information persist in more realistic settings with incomplete information (Study 2C).

An overview of the experiments and decisions is in Table \ref{tab:study:2ABC:scenarios}. In all three experiments (2A, 2B and 2C), participants choose between two distributions: A and B. Both options earn the decision-maker the same monetary reward of \$6 for successfully completing the wait but differ in the waiting time distribution. That is, rather than using multiple price lists to determine the cost of a given wait (as in Study 1), we now use a binary choice format where participants directly compare two distributions that differ in only one key characteristic. This has the advantage of simplifying choices, producing more robust within-subject comparisons and reducing noise from individual differences in the valuation of time versus money.

\subsection{Study 2A}
Study 2A was designed to more precisely characterize the effects of distributional shape on waiting preferences. To do so, we fixed three of the four moments of the waiting time distribution and varied the fourth. The decisions are summarized in the top panel of Table \ref{tab:study:2ABC:scenarios}. 

\begin{table}[b!]
    \centering \scriptsize
    \caption{Study 2 Experiment Design}
    \label{tab:study:2ABC:scenarios}
    \renewcommand{\arraystretch}{1.0}
\begin{tabular}{lC{12.8cm}} \toprule
    & \textbf{Study 2A: Moment-Based Comparisons} \\
  Thirteen Decisions in Total  & Binary Choice (A vs B) \\ \midrule
     Type 1 (Three decisions)  & $\mu_A=\mu_B$, \colorbox{gray!20}{$\sigma_A \neq \sigma_B$}, $\mathbb{S}_A=\mathbb{S}_B$, $\mathbb{K}_A=\mathbb{K}_B$ \\
     Type 2 (Three decisions)  & $\mu_A=\mu_B$, $\sigma_A=\sigma_B$, \colorbox{gray!20}{$\mathbb{S}_A \neq \mathbb{S}_B$}, $\mathbb{K}_A=\mathbb{K}_B$ \\ 
     Type 3 (Three decisions) & $\mu_A=\mu_B$, $\sigma_A=\sigma_B$, $\mathbb{S}_A=\mathbb{S}_B$, \colorbox{gray!20}{$\mathbb{K}_A \neq \mathbb{K}_B$} \\
     Four additional decisions & $A$ and $B$ randomly selected from Type 1-3 decisions\\
\midrule
    & \textbf{Study 2B: Long vs. Thick Right Tails} \\
  Thirteen Decisions in Total  & Binary Choice (A vs B) \\ \midrule
     Type 1 (Three decisions)  & $\mu_A=\mu_B$, $\sigma_A=\sigma_B$, \colorbox{gray!20}{$|\text{supp}(A)| \neq |\text{supp}(B)|$} \\
     Type 2 (Three decisions)  & $\mu_A=\mu_B$, $\sigma_A=\sigma_B$, \colorbox{gray!20}{$\text{mode}_A \neq \text{mode}_B$} \\ 
     Type 3 (Three decisions) & $\mu_A=\mu_B$, $\sigma_A=\sigma_B$, \colorbox{gray!20}{$\text{max}_A \neq\text{max}_B$} \\
     Four additional decisions & $A$ and $B$ randomly selected from Type 1-3 decisions \\
\midrule
    & \textbf{Study 2C: Full/Incomplete Info vs. Unknown} \\
   Eight Decisions in Total & Binary Choice (A vs B)\\ \midrule
     Type 1 (Three decisions) & $\mu_A=\mu_B,~\text{max}_A = \text{max}_B$, Long-tailed distribution with full info vs Unknown distribution \\
     Type 2 (Three decisions) & $\mu_A=\mu_B,~\text{max}_A = \text{max}_B$, Thick-tailed distribution with full info vs Unknown distribution \\
     Type 3 (One decision) & $\mu_A=\mu_B,~\text{max}_A = \text{max}_B$, Long-tailed distribution with incomplete info vs Unknown distribution \\
     Type 4 (One decision) & $\mu_A=\mu_B,~\text{max}_A = \text{max}_B$, Thick-tailed distribution with incomplete info vs Unknown distribution \\
\bottomrule
\end{tabular}
\begin{minipage}{\textwidth} \scriptsize \vspace{0.1cm}
Note: Decision sequence was randomized within-subject. Notation: $\mu$ = mean, $\sigma$ = standard deviation, $\mathbb{S}$ = skewness, $\mathbb{K}$ = kurtosis, $\text{supp}(X)$ = support set of distribution $X$, $|\cdot|$ = cardinality (number of elements), $\text{max}_X$ = maximum value of distribution $X$.
\end{minipage}
\end{table}
\subsubsection{Experiment Design}
To isolate the specific effects of distributional moments, all distributions were discrete with a mean of 9 minutes and the same support set of $\{1,3,5,...,17\}$ minutes. See Table \ref{tab:study:2A:distributions} for a detailed description of all choices. A total of 156 participants were recruited from Prolific with the same eligibility criteria as in Study 1. 

\begin{figure}[tbh!]
    \centering     
    \captionsetup[subfigure]{font=scriptsize}
    \caption{Study 2A: Sample Screenshots}
    \label{fig:study2a:screenshots}
        \vspace{0.1cm}
    \begin{subfigure}[c]{0.49\textwidth}
        \centering
        \caption{Type 1 Decision: $\mu_A=\mu_B$, $\sigma_A > \sigma_B$, $\mathbb{S}_A=\mathbb{S}_B$, $\mathbb{K}_A=\mathbb{K}_B$}
        \includegraphics[width=\textwidth]
        {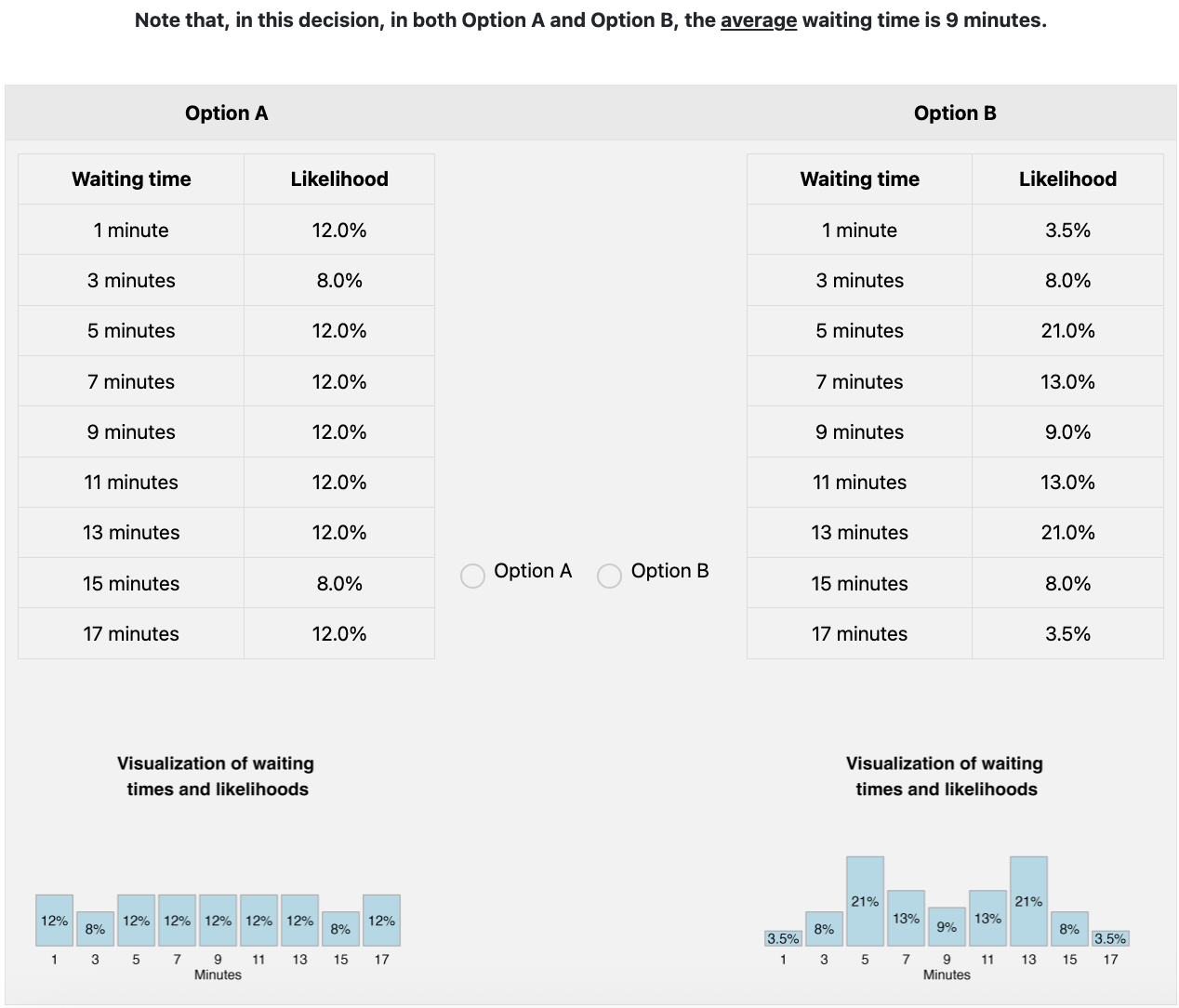}
        \label{fig:study2a:block1}
    \end{subfigure}
    \hfill    
    \begin{subfigure}[c]{0.49\textwidth}
        \centering
        \caption{Type 2 Decision: $\mu_A=\mu_B$, $\sigma_A=\sigma_B$, $|\mathbb{S}_A| < |\mathbb{S}_B|$, $\mathbb{K}_A=\mathbb{K}_B$}
        \includegraphics[width=\textwidth]{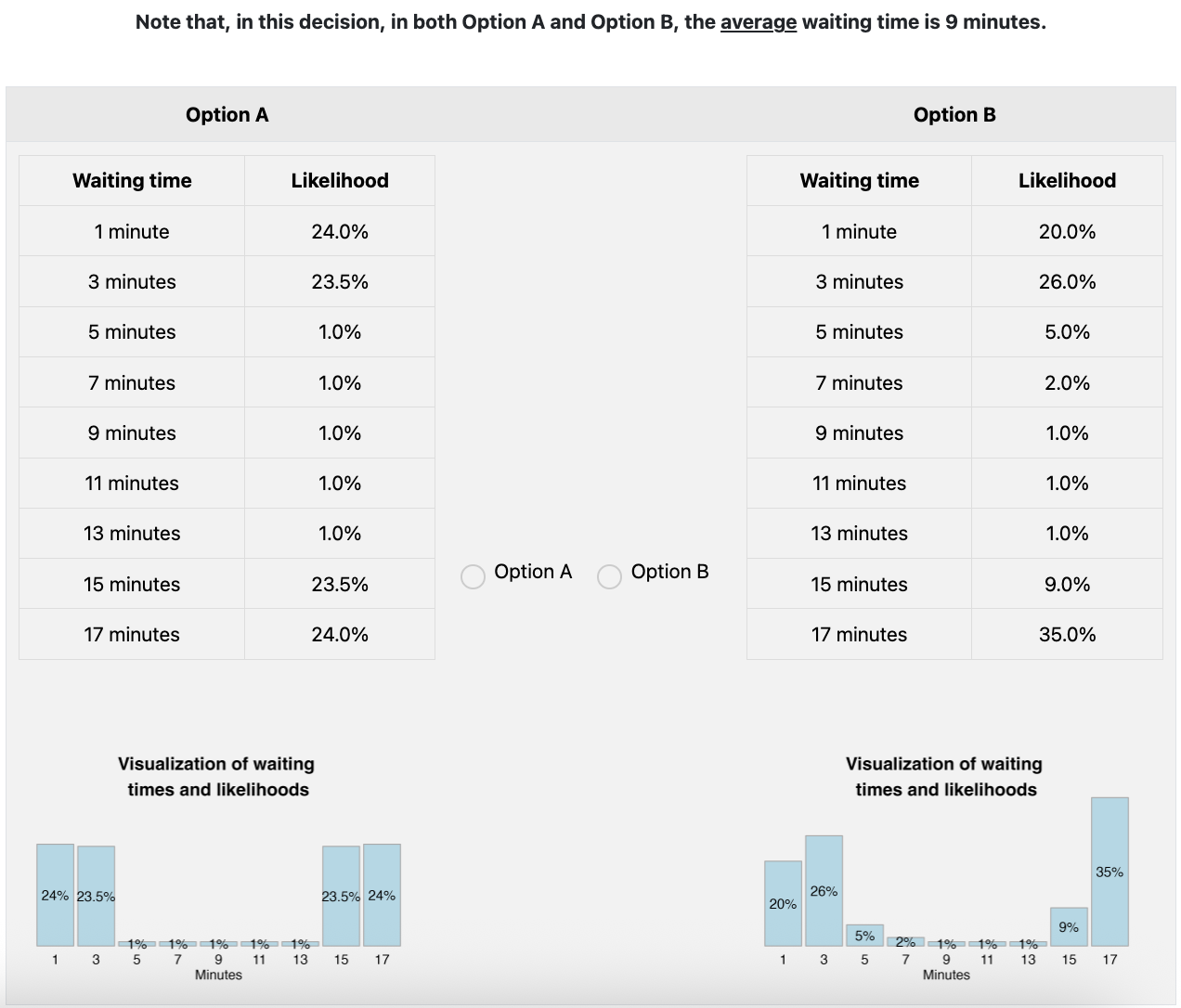}
        \label{fig:study2a:block2}
    \end{subfigure}
    
\vspace{0.1cm}
\begin{center}
\begin{subfigure}[c]{0.49\textwidth}
    \centering
    \caption{Type 3 Decision: $\mu_A=\mu_B$, $\sigma_A=\sigma_B$, $\mathbb{S}_A=\mathbb{S}_B$, $\mathbb{K}_A > \mathbb{K}_B$}
        \includegraphics[width=\textwidth]{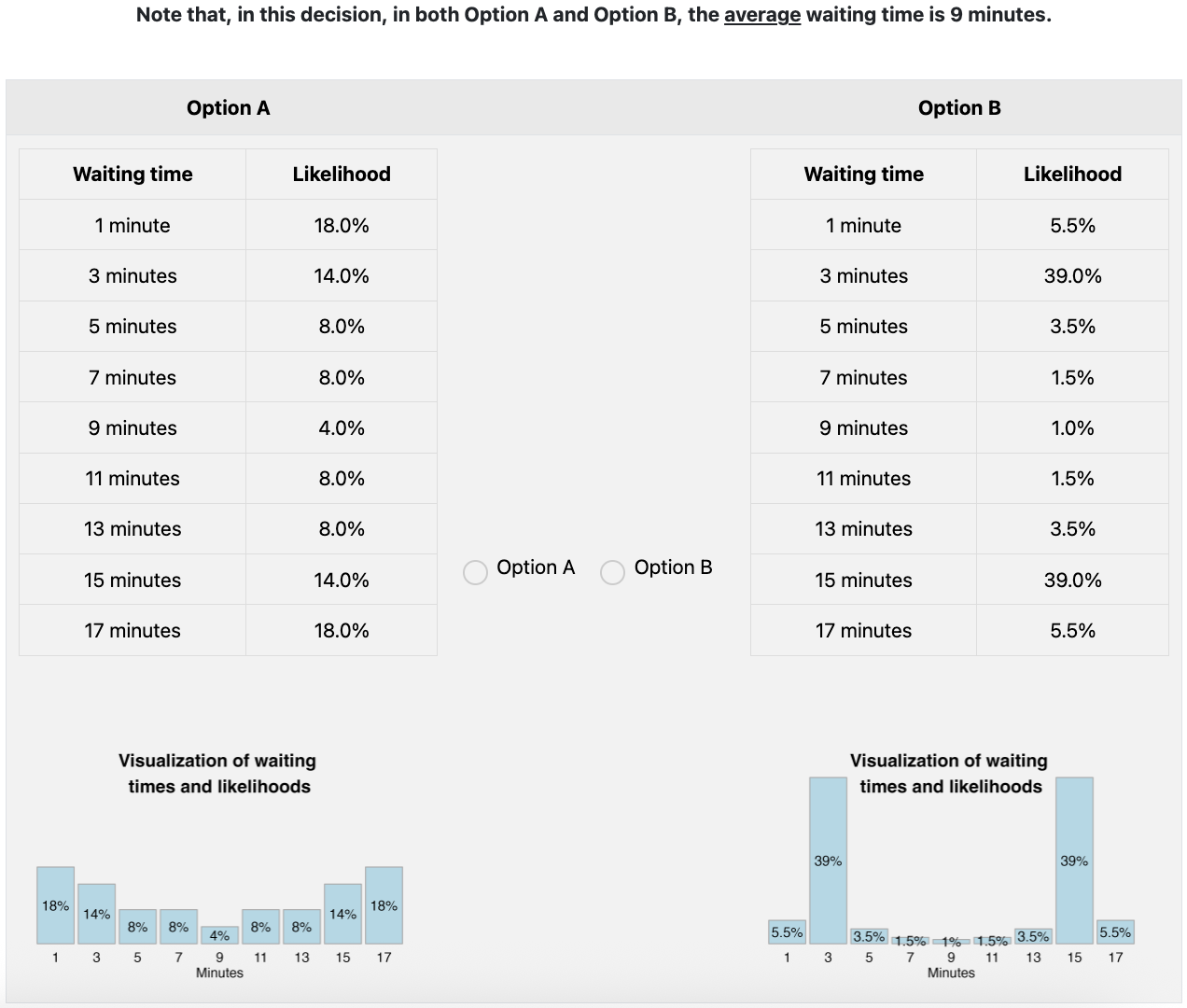}
        \label{fig:study2a:block3}
    \end{subfigure}
\end{center}
\end{figure}

The experiment consisted of 13 binary choices presented in random order.\footnote{In addition to the 13 decisions, there were two decisions that were used as attention and consistency checks: one that presented subjects with a (stochastically) dominated option and a second one that presented subjects with a deterministic vs. a stochastic wait.} There were three types of decisions.  For Type 1 decisions, variance was varied between options A and B, while mean, skewness, and kurtosis were held constant. For Type 2  decisions, skewness was varied, while mean, variance, and kurtosis were held constant. For Type 3  decisions, kurtosis was varied, while mean, variance, and skewness were held constant. The remaining decisions used a random sampling design from the distributions used in the first nine decisions to allow for more precise estimation of the utility function that best describes choices (see \textsection 5.1 for utility estimation). As in Study 1, one randomly selected choice was implemented for a real-time wait at the end of the experiment. Figure \ref{fig:study2a:screenshots} presents sample screenshots of each decision type. Participants were presented with tables showing each support value and corresponding probability of that support value being realized as well as with histograms of each distribution. A and B positions (left or right) were randomized, as was the sequence of decisions.

\subsubsection{Hypotheses}
We are not aware of existing studies examining the response to higher moments when evaluating time durations; however, some testable predictions can be derived from our prior results in Study 1. Specifically, Study 1 suggests that people respond negatively to the first two moments of a distribution. Further, in Study 1 we saw that people preferred simple, two-point waits to more complex, continuous waits, suggesting a preference for simplicity \citep[also see][for related work on simplicity]{sonsino2002complexity,moffatt2015heterogeneity,puri2025simplicity}. Because symmetric distributions are arguably easier to evaluate (one side mirrors the other), we anticipate a negative response to (the absolute value of) skewness. Finally, of the three distributions examined in Study 1, binary distributions, which were the most preferred ones, also had the lowest kurtosis. We therefore hypothesize that, all else equal, people will prefer symmetric over asymmetric distributions and distributions with a lower kurtosis over those with a higher kurtosis. 

\smallskip
\noindent \textit{H2A: The attractiveness of a wait decreases in its variance, (the absolute value of) skewness, and kurtosis.}

\subsubsection{Results}
The results of Study 2A are shown in Table \ref{tab:Study:2A:moment:test}. For variance comparisons (Type 1), 41-42\% of participants chose the lower variance option across the three decisions, with only 36.5\% making at least two of three choices consistent with variance aversion and only 14.7\% making all three choices consistent with this preference. Given that all distributions had identical mean (9 minutes), skewness, and kurtosis, these results indicate that most participants preferred \textit{higher} variance when other moments were controlled. For skewness comparisons (Type 2) results were mixed. In decision 1, only 34\% preferred the symmetric distribution, whereas in decisions 2 and 3, 64-67\% preferred symmetry (both $p < 0.001$). Overall, 58.3\% of participants showed a preference for symmetric distributions ($p = 0.019$), but only 17.9\% made all three choices consistent with preferring symmetry, and the substantial variation across decisions suggests that skewness preferences depend heavily on the specific distributions being compared (the asymmetric distribution that was viewed more favorably exhibited a long-right tail -- a key driver we will discuss in Study 2B). For kurtosis comparisons (Type 3), participants showed a preference for higher kurtosis distributions. While decision 1 showed no preference (49.4\%), decisions 2 and 3 revealed that only about 35\% chose the lower kurtosis option. Overall, only 36.5\% of participants made choices consistent with preferring lower kurtosis, and just 14.7\% made all three choices consistent with this preference.

In sum, the data do not support hypothesis H2A. On the contrary, participants showed preferences for higher variance and higher kurtosis distributions, which suggests that moment-based models may not be reflective of how people make decisions in this context. Instead, the results indicate that tail shape might be driving decisions. For example, in Figure \ref{fig:study2a:block2},  distribution B has a thick-right tail with an accumulation of probability mass at the highest support value (17 minutes). This was the least preferred distribution across all comparisons in Study 2A. Conversely, the probability mass in both distributions in Figure \ref{fig:study2a:block1} is well spread out over the support values. Both these distributions were among the most favored ones among the nine distributions used in  Study 2A. These comparisons suggest that participants may be responding to how probability mass is distributed across the tails in ways that are not well described by statistical moments. 

\begin{table}[t]\centering \scriptsize    \renewcommand{\arraystretch}{1.2}
\caption{Study 2A: Hypothesis Tests}
\label{tab:Study:2A:moment:test}
\begin{tabular}{L{3.2cm}cccccccc} \toprule
       & \multicolumn{2}{C{3.9cm}}{Type 1 Decisions: \newline Low Var.\ $\succ$ High Var.} && \multicolumn{2}{C{3.9cm}}{Type 2 Decisions:\newline Symmetric $\succ$ Asymmetric} && \multicolumn{2}{C{3.9cm}}{Type 3 Decisions:\newline Low Kurt.\ $\succ$ High Kurt.} \\ \cmidrule{2-3}\cmidrule{5-6}\cmidrule{8-9}
  & Frequency        & Test $=50\%$     && Frequency        & Test $=50\%$     && Frequency        & Test $=50\%$     \\ \cmidrule{1-3}\cmidrule{5-6}\cmidrule{8-9}
Decision 1  & 41.0\%      & 0.988      && 34.0\%      & 1.000            && 49.4\%      & 0.564            \\
Decision 2   & 41.0\%      & 0.988      && 64.1\%      & $<0.001$            && 35.9\%      & 1.000      \\
Decision 3 & 41.7\%      & 0.981      && 66.7\%      & $<0.001$            && 34.0\%      & 1.000     \\
\midrule
Majority of Decisions & 36.5\%      & 1.000      && 58.3\%      & 0.019            && 36.5\%      & 1.000     \\
All Three Decisions & 14.7\%      & 1.000      && 17.9\%      & 1.000            && 14.7\%      & 1.000     
 \\
\bottomrule
\end{tabular}
\begin{minipage}{\textwidth}
        \scriptsize
        \vspace{0.1cm}
        \textit{Note: }``Majority of Decisions'' row tests whether a subject made at least two of the three decisions consistent with the hypothesis. ``All Three Decisions'' row tests whether a subject made all three decisions consistent with the hypothesis. $p-$values are based on one-sided proportion tests.
    \end{minipage}
\end{table}

\subsection{Study 2B}
Study 2A has shown that distributional moments do not consistently predict responses, suggesting that features of distributions other than moments may be driving decisions. Study 2B examines what these features might be. While many potential designs are possible, we focused on three distributional features that prior research has identified as psychologically relevant: support set size (cardinality), modal value of the distribution, and tail shape. Support set size was chosen because experimental economics research has shown systematic ``range effects,'' where elicited preferences are affected by the number of possible outcomes \citep{blavatskyy2009range,johnson2013understanding,fudenberg2022}. The modal value manipulation was chosen based on research showing that visually prominent features (such as a modal value in a histogram) may receive disproportionately high weight in choices \citep{orquin2018visual, lu2022histogram}. Tail shape was chosen because it captures worst-case outcomes. Building on prior work that worst-case outcomes are especially salient \citep{cohn2015evidence,ackerman2017worst,buell2021}, we focus on right-tail properties and test which tail profiles (e.g., thinner/shorter vs. thicker/longer) are preferred or avoided.

\subsubsection{Experiment Design}
\begin{figure}[tb!]
\centering
\caption{Study 2B: Sample Screenshots}
\label{fig:study2b:screenshots}
\vspace{0.1cm}

\begin{subfigure}[c]{0.49\textwidth}
    \centering
    \caption{Type 1 Decision (Spread Out vs. Concentrated):}
    \includegraphics[width=\textwidth]{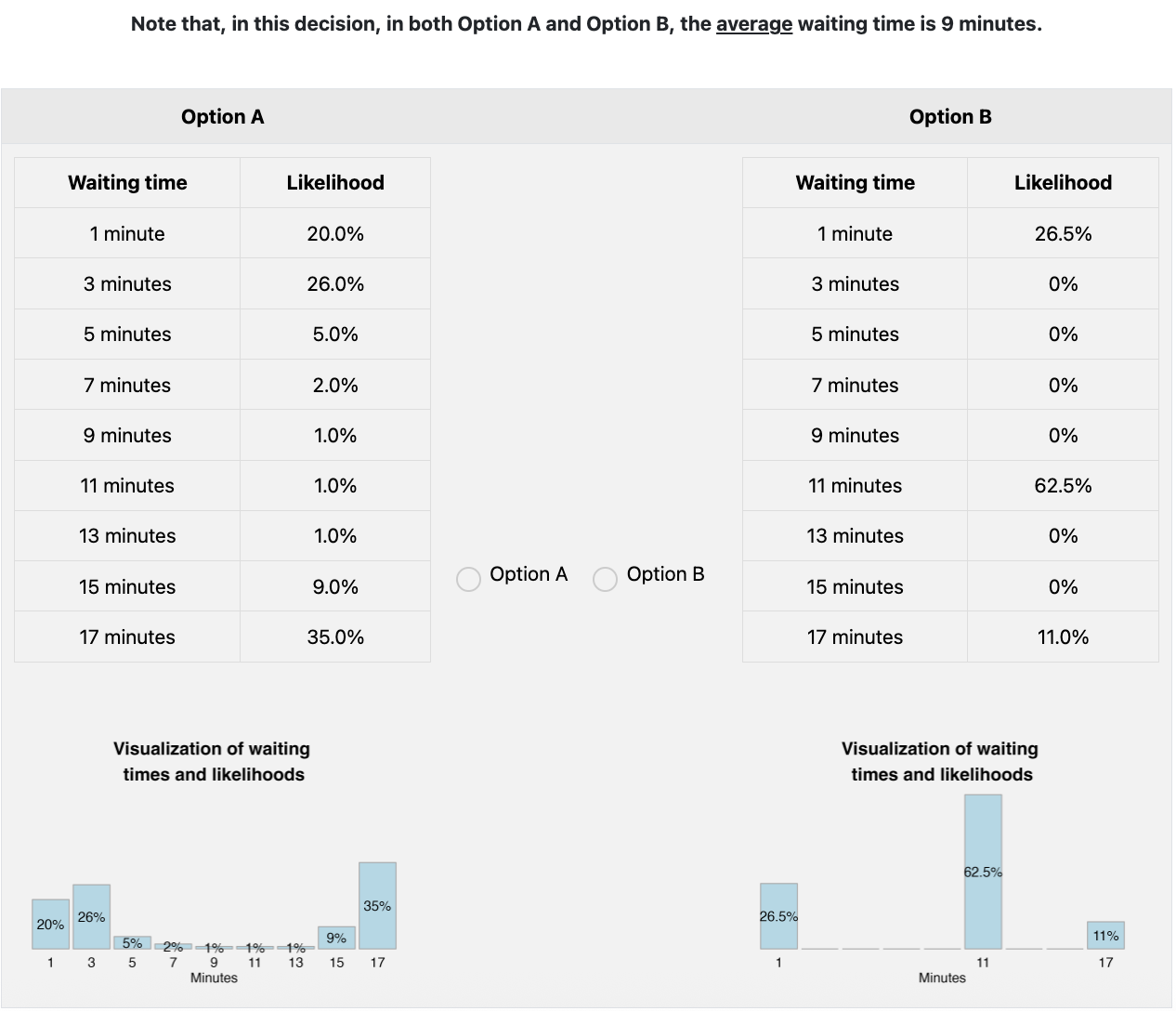}    
    \label{fig:study2b:block1}
\end{subfigure}
\hfill
\begin{subfigure}[c]{0.49\textwidth}
    \centering
    \caption{Type 2 Decision (High Mode vs. Low Mode):}
    \includegraphics[width=\textwidth]{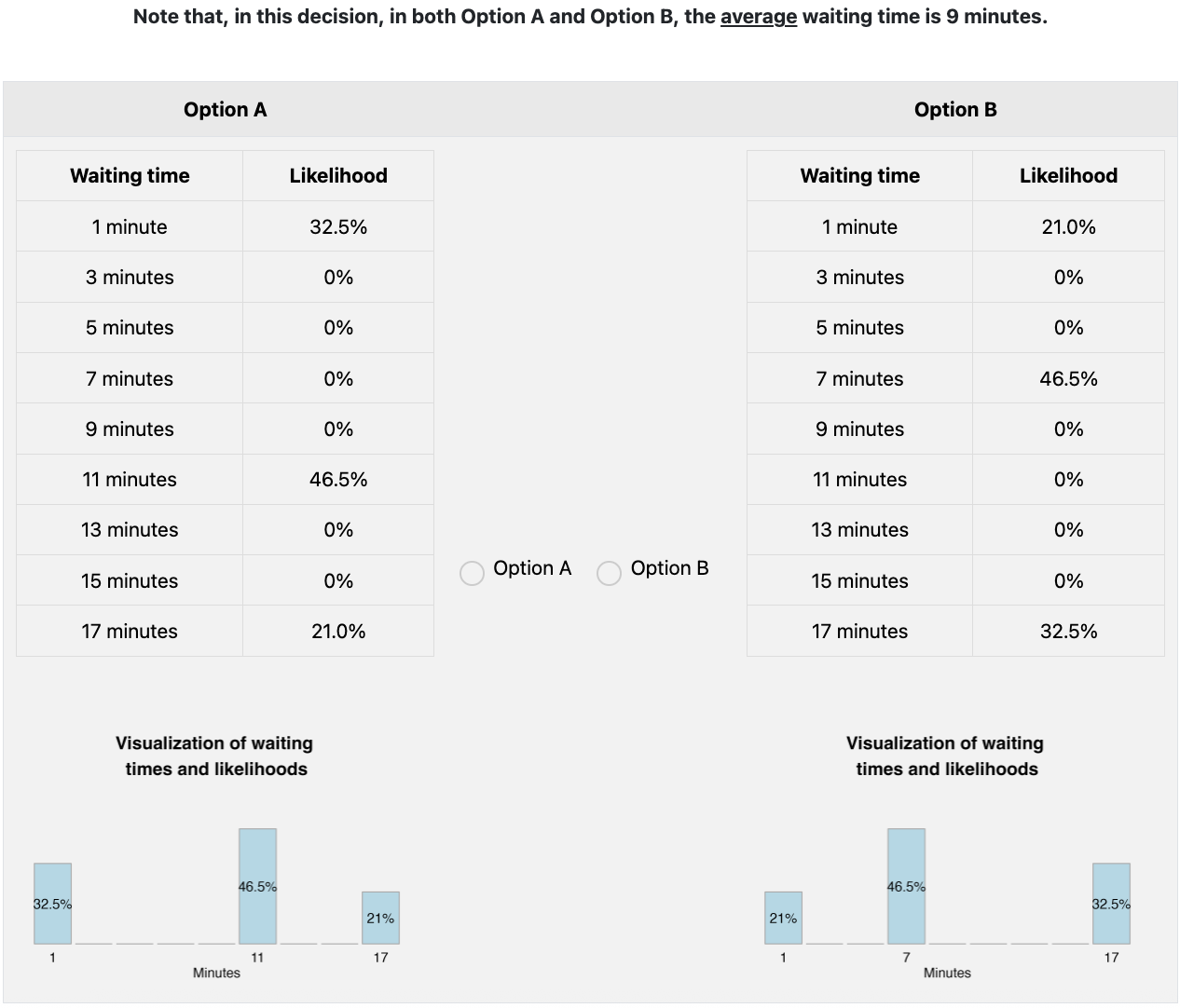}
    \label{fig:study2b:block2}
\end{subfigure}

\vspace{0.1cm}

\begin{subfigure}[c]{0.49\textwidth}
    \centering
    \caption{Type 3 Decision (Long Tail vs. Thick Tail):}
    \includegraphics[width=\textwidth]{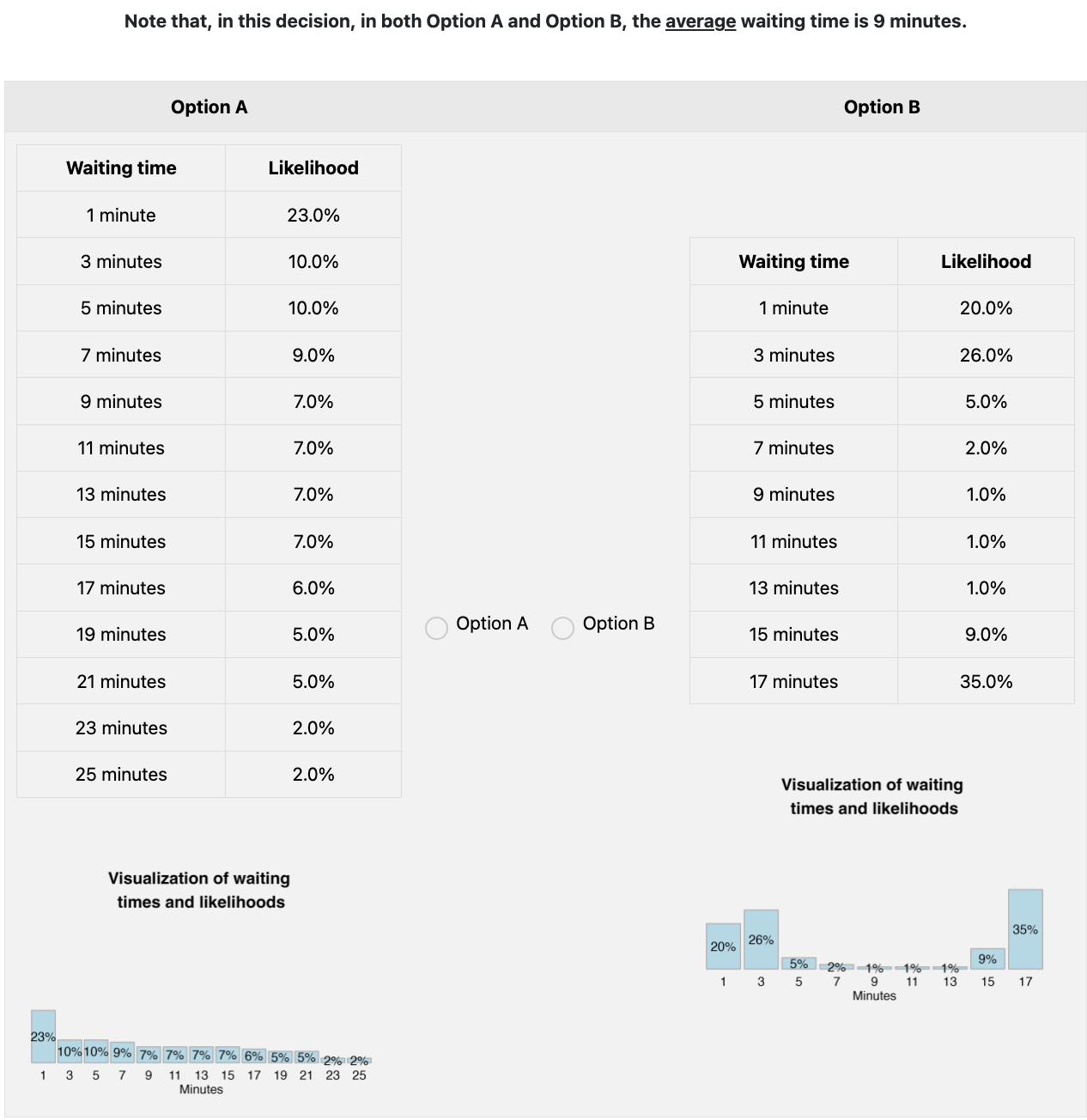}
    \label{fig:study2b:block3}
\end{subfigure}
\end{figure}

The decisions faced by participants are shown in the middle panel of Table \ref{tab:study:2ABC:scenarios}. Like Study 2A, all distributions were discrete with a mean of 9 minutes. A total of 165 participants were recruited from Prolific with the same eligibility criteria as in Studies 1 and 2A. The experiment consisted of 13 incentivized binary choices, further subdivided in several decision types. Each decision type focused on one aspect of tail shape. Type 1 compared distributions with different support set sizes ($|\text{supp}(A)| \neq |\text{supp}(B)|$) while holding mean and variance constant. Type 2 varied the location of the mode. This was achieved by using two distributions that had their skewness sign flipped ($\mathbb{S}_A = -\mathbb{S}_B$) while holding mean, variance and kurtosis constant. Finally, Type 3 varied the maximum value ($\text{max}_A \neq\text{max}_B$) while holding mean and variance constant, testing preferences for long vs. thick tails. As in Study 2A, the remaining decisions used a random sampling design, drawing from the distributions used in decisions 1-9. Importantly, mean and variance were the same for Option A and Option B in all decisions. Figure \ref{fig:study2b:screenshots} presents sample screenshots of each decision type.

\subsubsection{Hypotheses}
A key observation from Study 2A was that thick-tailed distributions (like Option B in Fig. 1c) were less preferred \textit{even when they had lower variance}, suggesting that tail characteristics may override traditional risk preferences. This points toward more fundamental preferences regarding how probability mass is distributed over the available support values. In particular, we hypothesize that preferences are driven primarily by tail shape, such that people are averse to thick tails where probability mass accumulates at longer wait times. 

The aversion to thick tails may be driven by several mechanisms that we seek to identify and test. People may respond negatively to spikes in probability mass in the right tail, particularly when these spikes are prominent relative to the rest of the distribution. Therefore, a distribution with fewer support points, each of which has a high concentration of probability mass, may be perceived more favorably than a distribution where the right tail is ``fatter'' than the surrounding values.  This is also supported by the preference for discrete, binary distributions identified in Study 1. 
Further, people may also be sensitive to modal value location, showing greater aversion when the most likely outcome occurs at longer wait times due to the psychological salience of modal realizations \cite[see][and references therein]{lu2022histogram}. Finally, people may respond differently to long versus thick tails even when variance is held constant, preferring scenarios where extreme outcomes are possible but unlikely to scenarios with moderately extreme outcomes that are more probable. Therefore, we hypothesize:

\smallskip
\noindent \textit{H2B: Holding mean and variance constant, waiting preferences are well-explained by psychologically salient tail characteristics, with preferences favoring: (1) distributions with fewer non-zero mass points, (2) a mode located at a lower value, and (3) long over thick right tails.}

\subsubsection{Results}
The results of Study 2B are in Table \ref{tab:Study:2B:tail:test}. For Type 1 decisions comparing concentrated versus spread-out distributions and Type 2 decisions examining mode location preferences, there are no discernible patterns. Specifically, only 41.2\% of participants made choices consistent with preferring concentrated distributions, with the proportion of participants acting consistent with the hypothesis ranging from 33.3\% to 58.2\% across individual decisions. Similarly, for Type 2, results were mixed with 56.4\% showing an overall preference for low modes but substantial variation across decisions (50.9\% to 61.2\%).

\begin{table}[h!]\centering \scriptsize \renewcommand{\arraystretch}{1.15}
\caption{Study 2B: Hypothesis Tests}
\label{tab:Study:2B:tail:test} 
\begin{tabular}{L{3.2cm}cccccccc} \toprule
       & \multicolumn{2}{C{4cm}}{Type 1 Decisions: \newline Concentrated $\succ$ Spread Out} && \multicolumn{2}{C{3.6cm}}{Type 2 Decisions:\newline Low Mode $\succ$ High Mode} && \multicolumn{2}{C{4cm}}{Type 3 Decisions:\newline Long Tail $\succ$ Thick Tail} \\ \cmidrule{2-3}\cmidrule{5-6}\cmidrule{8-9}
  & Frequency        & Test $=50\%$     && Frequency        & Test $=50\%$     && Frequency        & Test $=50\%$     \\ \cmidrule{1-3}\cmidrule{5-6}\cmidrule{8-9}
Decision 1  & 42.4\%      & 0.974      && 51.5\%      & 0.349            && 83.6\%      & $<0.001$     \\
Decision 2   & 33.3\%      & 1.000      && 61.2\%      & 0.002            && 82.4\%      & $<0.001$     \\
Decision 3 & 58.2\%      & 0.018      && 50.9\%      & 0.408            && 75.8\%      & $<0.001$     \\
\midrule
Majority of Decisions & 41.2\%      & 0.988      && 56.4\%      & 0.051            && 86.7\%      & $<0.001$     \\
All Three Decisions & 12.7\%      & 1.000      && 15.8\%      & 1.000            && 59.4\%      & 0.008     
 \\
\bottomrule
\end{tabular}
\begin{minipage}{\textwidth}
        \scriptsize
        \vspace{0.1cm}
        \textit{Note: }``Majority of Decisions'' row tests whether a subject made at least two of the three decisions consistent with the hypothesis. ``All Three Decisions'' row tests whether a subject made all three decisions consistent with the hypothesis. $p-$values are based on one-sided proportion tests.
    \end{minipage}
\end{table}

In contrast, for Type 3 decisions comparing long versus thick right tails, participants showed a strong and consistent preference for long-tailed distributions, with 75.8\% to 83.6\% choosing the long-tail option across all three decisions (all $p < 0.001$), and 86.7\% making the majority of their choices consistent with this preference ($p < 0.001$). These results provide partial support for hypothesis H2B and suggest that a key feature of a waiting time distribution is the shape of the right tail, with people preferring scenarios where extreme outcomes are possible but unlikely (long-right tails) to scenarios with moderately extreme outcomes that are more probable (thick-right tails). That is, decision-makers dislike having a large chance of experiencing the worst-case, where ``worst-case'' is defined relative to the available support values.

\subsection{Study 2C}
The results of Studies 2A-B suggest that people are averse to distributions with a thick-right tail. These results were shown under full distributional information. However, practical constraints typically prevent providers from sharing such detailed information with decision-makers. In Study 2C we examine whether thick-right-tail aversion persists under more realistic conditions where participants receive only partial information about wait time distributions.

\subsubsection{Experiment Design}
\begin{figure}[tb!]
    \centering
    \caption{Study 2C: Sample Screenshots}
    \label{fig:study2c_screenshots}
        \vspace{0.2cm}

    \begin{subfigure}[c]{0.48\textwidth}
        \centering
        \caption{Decision Type 1: Unknown vs. Long Tail Full Info}
        \includegraphics[width=\textwidth]{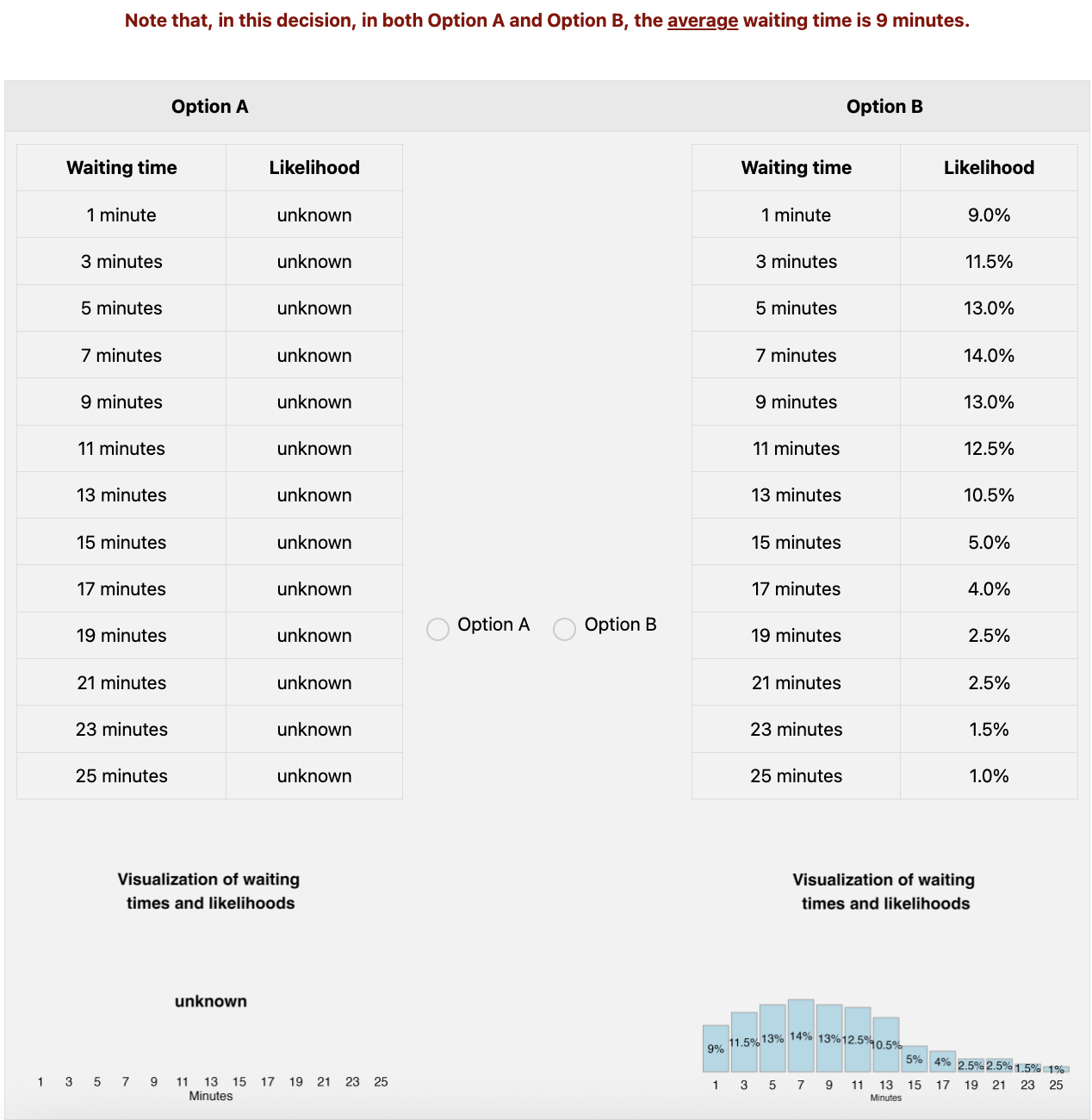}
        \label{fig:decision_type_1}
    \end{subfigure}
    \hfill
    \begin{subfigure}[c]{0.48\textwidth}
        \centering
        \caption{Decision Type 2: Unknown vs. Thick Tail Full Info}
        \includegraphics[width=\textwidth]{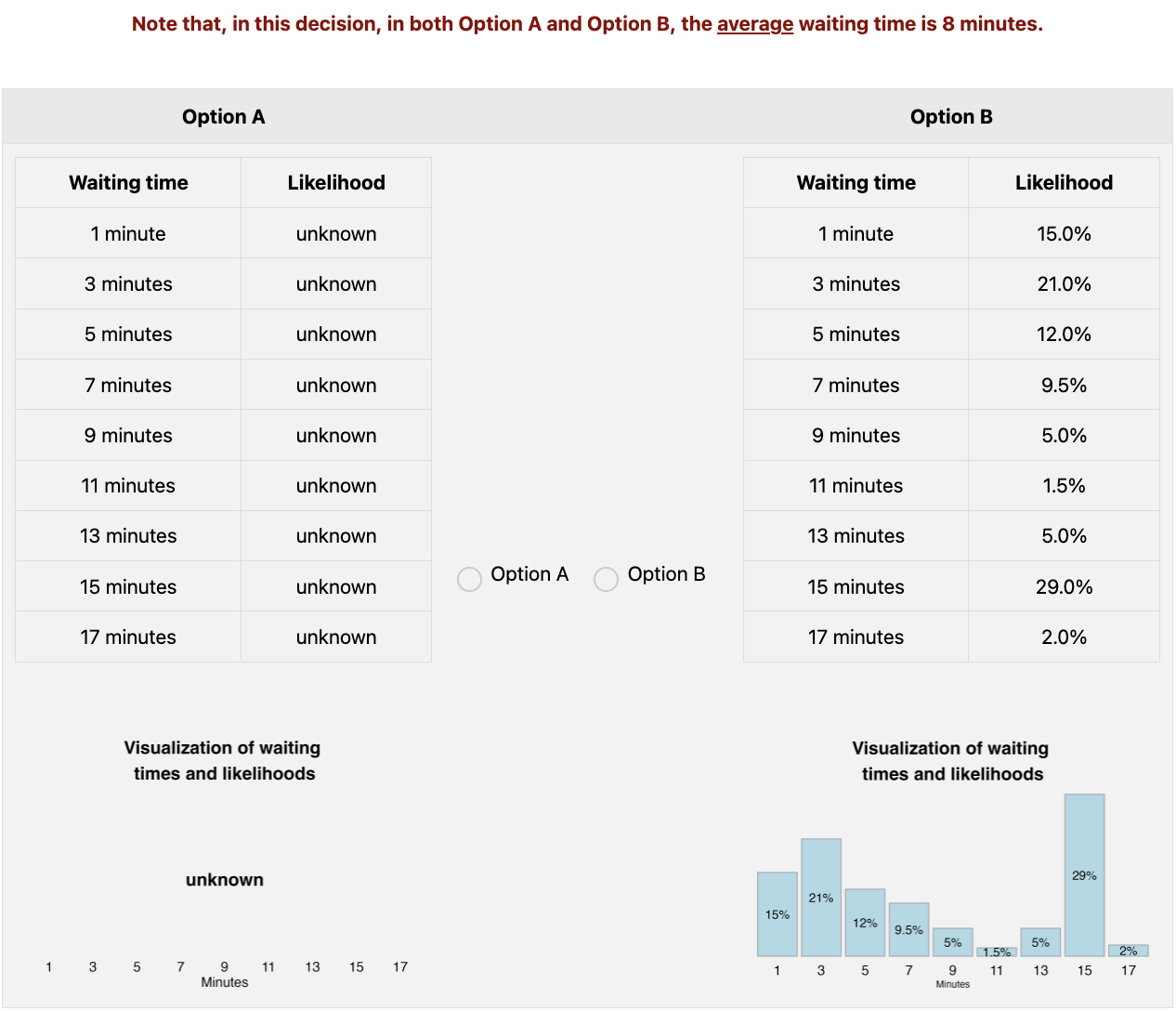}
        \label{fig:decision_type_2}
    \end{subfigure}
    
    \vspace{0.2cm}
    
    \begin{subfigure}[b]{0.48\textwidth}
        \centering
        \caption{Decision Type 3: Unknown vs. Long Tail Partial Info}
        \includegraphics[width=\textwidth]{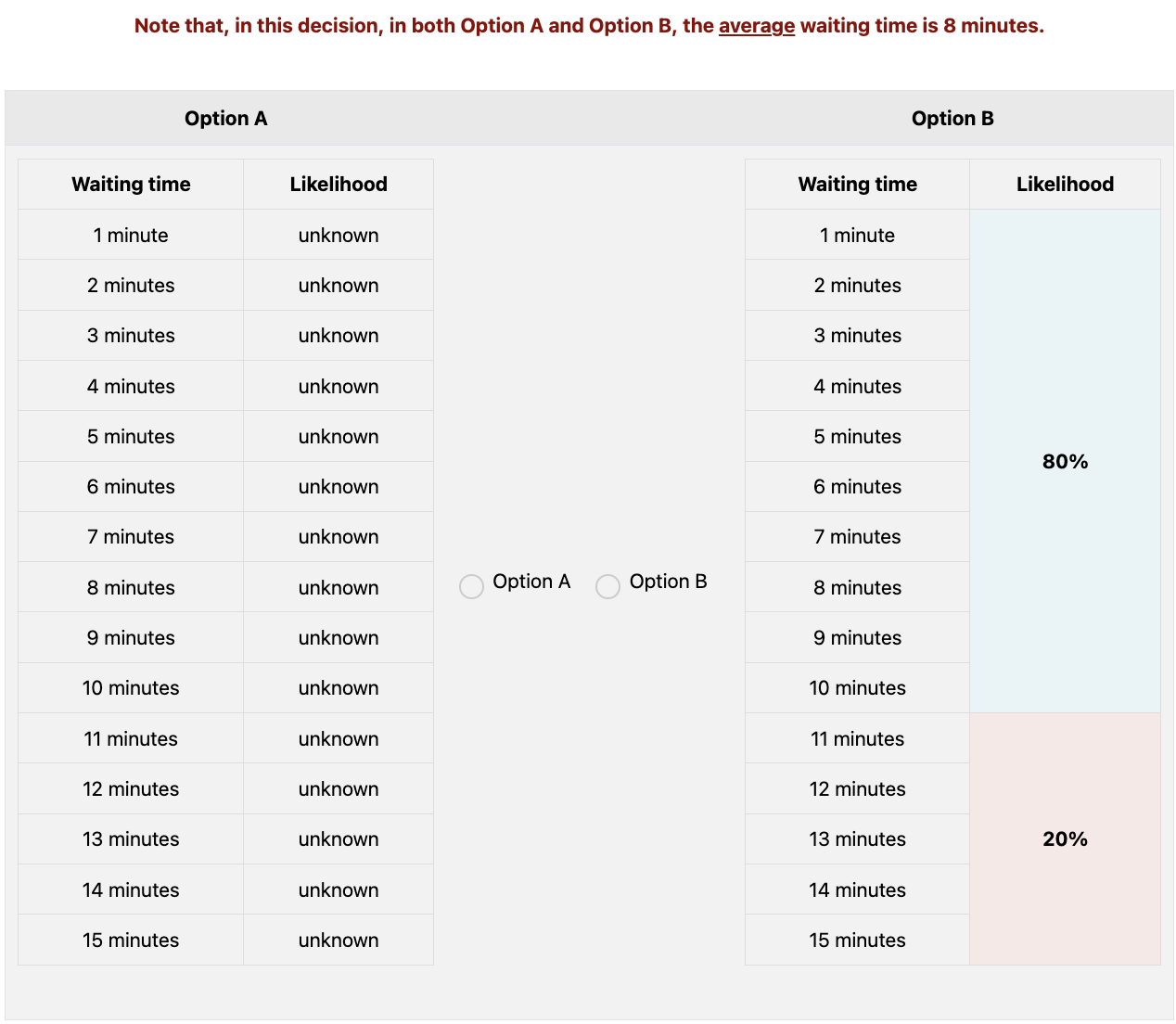}
        \label{fig:decision_type_3}
    \end{subfigure}
    \hfill
    \begin{subfigure}[b]{0.48\textwidth}
        \centering
        \caption{Decision Type 4: Unknown vs. Thick Tail Partial Info}
        \includegraphics[width=\textwidth]{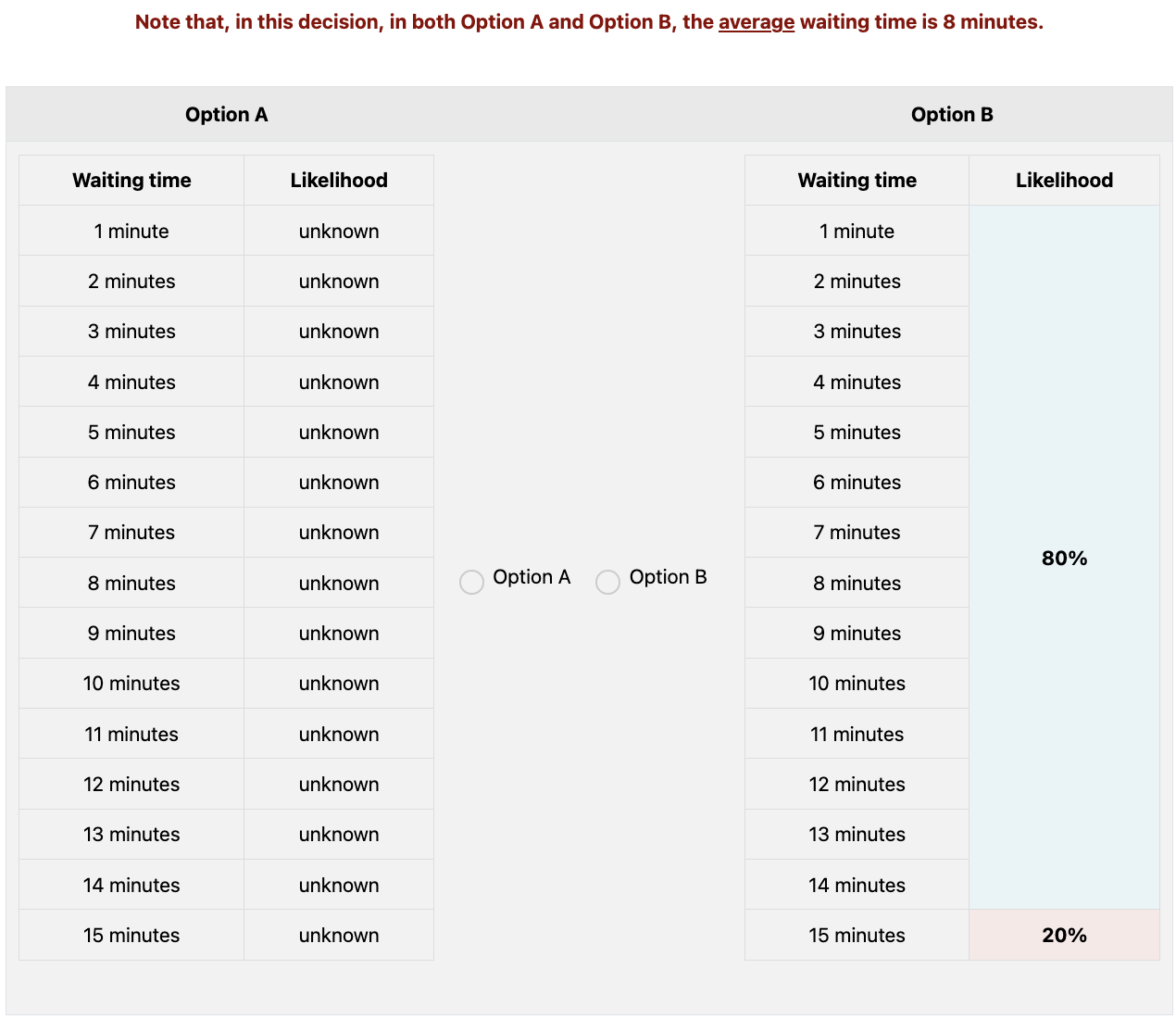}
        \label{fig:decision_type_4}
    \end{subfigure}
\end{figure}

Similar to Studies 2A-B, we used a direct binary comparison design where participants chose between two options. Sample screenshots are in Figure~\ref{fig:study2c_screenshots}. Subjects made a total of eight binary decisions, subdivided into four decision types. In Type 1 decisions, subjects chose between a long-tailed distribution (similar to ones used in Study 2B) and an alternative described as having an ``unknown'' distribution on the same range and with the same mean. Type 2 was similar to Type 1, but used thick right-tailed distributions (again, similar to ones used in Study 2B), comparing full distributional information against an unknown alternative. Type 3 presented both options as unknown distributions with only average waiting times disclosed, but the underlying distributions had 20\% of the probability mass spread across the five highest values for one of the choices (long-right tail). Type 4 similarly had one of the choices with 20\% probability mass concentrated in the highest value (thick-right tail).\footnote{To implement ``unknown'' distributions in the experiment, we adapted the method introduced by \citet{SSD:Ambiguity}. This method generates samples from a distribution that lacks finite quantiles or moments, thus creating genuine ambiguity about the distributional shape for experimental subjects. \citet{SSD:Ambiguity} show that this method produces a more authentic experience of ambiguity compared to alternative methods in the literature.}

In each pairwise comparison, the averages and standard deviations were held constant. Averages were displayed prominently at the top of each decision screen (see Figure \ref{fig:study2c_screenshots}).\footnote{In particular, both the long-tailed and the thick-tailed distributions had $\mu= 8$ and $\sigma=5$ for the first decision, $\mu=9$ and $\sigma=5.5$ for the second decision and $\mu= 9$ and $\sigma=7$ for the third decision within Decision Type 1 and 2.} For decisions involving full distributional information, participants saw both tabular presentations listing waiting times and their corresponding probabilities, as well as histogram visualizations of the probability distributions. For unknown distributions, participants saw only the average waiting time with all specific probabilities listed as ``unknown.'' A total of 158 participants were recruited from Prolific with the same eligibility criteria as in previous studies.
 
\subsubsection{Hypotheses}
Recall that Study 2B revealed that people strongly prefer long-thin over thick-right tails when full distributional information is available. 
Further, we can reasonably expect decision-makers to treat distributional uncertainty (not knowing the distribution) as preferable to unfavorable distributions but inferior to favorable distributions. Therefore, when faced with a known long right-tailed distribution, people should prefer this certainty over an unknown alternative that could potentially be thick-tailed. Conversely, when faced with a known thick right-tailed distribution, the uncertainty of an unknown distribution becomes attractive as it preserves the possibility of a more favorable tail shape. Formally:

\smallskip
\noindent \textit{H2C: Holding mean and bounds constant, there is a preference for distributions with a long-right tail over unknown distributions, and a preference for unknown distributions over distributions with a thick-right tail.}

\subsubsection{Results}
The results of Study 2C are in Table \ref{tab:Study:3A:information:test}. These results suggest a hierarchy of preferences that is largely consistent with H2C. For Type 1 decisions comparing long-tailed distributions with full information against unknown distributions, participants showed an overwhelming preference for the long-tailed option across all three decisions (88.0\%, 81.0\%, and 93.7\%, all $p < 0.001$), with 91.1\% of participants making the majority of their choices consistent with this preference ($p < 0.001$). This suggests that long tails are not only preferred to thick tails when both are fully known, but are also preferred to unknown distributions. In contrast, Type 2 decisions comparing unknown distributions against thick-tailed distributions with full information showed more modest effects. While participants did prefer the unknown option (57.0\%, 57.6\%, and 64.6\%, with $p = 0.040$, $p = 0.028$, and $p < 0.001$ respectively), the preference was somewhat weaker, with 62.0\% making the majority of choices consistent with avoiding thick tails ($p = 0.001$). This suggests that people also dislike ambiguity, and their aversion to thick-tailed distributions must be balanced against their aversion to uncertainty about the alternative.\footnote{Direct comparisons between Type 1 and Type 2 decisions confirm this asymmetry. When comparing the same distributions across information conditions, participants chose the long-tailed option over unknown significantly more often than they chose unknown over the thick-tailed option (e.g., 88.0\% vs. 57.0\% for the first decision pair, difference $= 31.0\%$, $p < 0.001$). Overall, participants showed significantly stronger consistency in preferring long tails over unknown (91.1\%) compared to preferring unknown over thick tails (62.0\%), with a paired $t$-test confirming this difference ($t = -7.84$, $p < 0.001$).} The results for Type 3 and Type 4 decisions, which presented partial information scenarios, reinforce this pattern, with Type 3 decisions showing strong preference for long tails even under partial information (88.0\%, $p < 0.001$) while Type 4 decisions showed no significant preference (43.0\%, $p = 0.960$). 

Taken together, the results are largely consistent with a preference ordering of long-tailed $\succ$ unknown $\succ$ thick-tailed distributions, supporting H2C. However, thick-right-tail aversion does not always dominate the desire for distributional information: when choosing between a thick-right tail distribution with partial information and complete uncertainty (Type 4 decision), participants' preferences were more evenly split. 

\begin{table}[bth]\centering \scriptsize    \renewcommand{\arraystretch}{1.01}
\caption{Study 2C: Hypothesis Tests}  
\label{tab:Study:3A:information:test}
\begin{tabular}{lcccccccccccc} \toprule
       & \multicolumn{2}{C{3.2cm}}{Type 1 Decisions: \newline Long Tail $\succ$ Unknown} && \multicolumn{2}{C{3.3cm}}{Type 2 Decisions:\newline Unknown $\succ$ Thick Tail} && \multicolumn{2}{C{3.2cm}}{Type 3 Decision:\newline Long Tail $\succ$ Unknown} && \multicolumn{2}{C{3.3cm}}{Type 4 Decision:\newline Unknown $\succ$ Thick Tail} \\ \cmidrule{2-3}\cmidrule{5-6}\cmidrule{8-9}\cmidrule{11-12}
  & Freq.        & Test $=50\%$     && Freq.        & Test $=50\%$     && Freq.        & Test $=50\%$     && Freq.        & Test $=50\%$     \\ \cmidrule{1-3}\cmidrule{5-6}\cmidrule{8-9}\cmidrule{11-12}
Decision 1  & 88.0\%      & $<0.001$      && 57.0\%      & 0.040            && 88.0\%      & $<0.001$     && 43.0\%      & 0.960     \\
Decision 2   & 81.0\%      & $<0.001$      && 57.6\%      & 0.028            && --      & --     && --      & --     \\
Decision 3 & 93.7\%      & $<0.001$      && 64.6\%      & $<0.001$            && --      & --     && --      & --     \\
\midrule
Majority & 91.1\%      & $<0.001$      && 62.0\%      & 0.001            && --      & --     && --      & --     \\
All Three & 73.4\%      & $<0.001$      && 33.5\%      & 1.000            && --      & --     && --      & --     
 \\
\bottomrule
\end{tabular}
\begin{minipage}{\textwidth}
        \scriptsize
        \vspace{0.1cm}
        \textit{Note:} ``Majority'' row tests whether a subject makes at least two of the three decisions consistent with the respective hypothesis. ``All Three'' row tests whether a subject makes all three decisions consistent with the respective hypothesis. $p$-values are based on one-sided proportion tests.
    \end{minipage}
\end{table}

\subsection{Discussion\label{s3:disc}}
The results of Studies 2A-2C show that moment-based models often fail to capture how people evaluate waiting time distributions. Instead, tail characteristics appear to drive preferences. In Study 2A, participants showed preferences that contradicted standard risk aversion: they preferred higher variance and higher kurtosis distributions when other moments were held constant. This suggests that statistical moments alone may not always provide an adequate framework for understanding waiting preferences, and that other, psychologically relevant characteristics of probability distributions must be driving behavior. Study 2B helps shed light on what these features might be. This study shows that tail thickness is a primary driver of waiting preferences. While participants showed no consistent preferences for concentrated probability mass or specific mode locations, they showed a strong and robust preference for long tails over thick tails across all comparisons, conditional on mean and variance. This helps explain the results from Study 2A: participants were not responding to variance or kurtosis, but rather to how probability mass was distributed in the tails of the distribution. The preference for long-right tails suggests that people distinguish between different types of tail risk, with a preference for distributions where extreme delays are possible but improbable, rather than distributions where moderately long delays are relatively likely. 

Study 2C confirms that thick-right-tail aversion persists in settings with incomplete information. The preference ordering (long-tailed $\succ$ unknown $\succ$ thick-tailed) shows that the response to the availability of distributional information depends on tail characteristics. When facing long-tailed distributions with full information, 91.1\% preferred the known distribution over uncertainty, suggesting that transparency about favorable tail shapes is beneficial. However, when facing thick-tailed distributions, 62.0\% preferred the unknown alternative. At the same time, when participants were presented with a choice between a partially known thick right-tailed distribution and a completely unknown distribution, there was no statistically significant preference. This suggests that limited information about an unfavorable distribution may not be worse than complete ambiguity.

What psychological mechanisms could plausibly explain thick-right-tail aversion? One potential explanation is the fear of being in ``last place'' -- similar to last-place aversion in queues \citep{buell2021}. In thick right-tailed distributions, probability mass concentrates near the worst-case outcome, so decision-makers face a relatively high chance of experiencing outcomes close to the worst case.  Additionally, thick-tailed distributions increase the \textit{salience} of experiencing poor outcomes relative to other available options. With a peak being in the right portion of the probability distribution, the surrounding outcomes appear relatively unlikely, making the negative outcomes appear more salient, and thus more ``cognitively available'' \citep{TverskyKahneman1973}. These psychological mechanisms offer plausible explanations, though we acknowledge that they are post-hoc interpretations that warrant further research.

\section{Study 3: Information Preferences\label{sec:study:3}}
Studies 2A-C showed that people are averse to thick right-tailed distributions and that this preference persists even with incomplete information. However, these studies did not ask what specific distributional information decision-makers find most valuable when evaluating uncertain waits. If decision-makers are sensitive to right tail characteristics, as we have shown in Study 2, then they may be more interested in learning about the shape of that right tail. This question also has practical relevance given the practical constraints of communication channels (such as limited screen space or customer attention). Study 3 addresses this by examining what distributional information participants are most interested in.

\subsection{Experiment Design}
To elicit information preferences, we return to the multiple-price-list format from Study 1, where Option A is held constant and deterministic for all decisions (wait 1 minute and receive \$1), while Option B is generally stochastic, with the monetary amount, waiting time distribution, and information available about that distribution, varying across decisions. A total of 319  participants were recruited from Prolific with the same eligibility criteria as in Studies 1 and 2. The experiment design is summarized in Table \ref{tab:study:3:scenarios}.\footnote{We use a slightly larger sample size (relative to Studies 1, 2A, 2B and 2C) because of the higher number of different scenarios that participants could experience (12 distributions $\times$ five possible information scenarios).} 

\begin{table}[b!]
    \centering \scriptsize
    \caption{Study 3: Experiment Design}
    \label{tab:study:3:scenarios}
    \renewcommand{\arraystretch}{1.2}
    \begin{tabular}{cc}
    \toprule
     \multicolumn{2}{c}{\textbf{Possible Waiting Times in Option B} (Varied between Treatments)} \\
         \midrule
    \textbf{Low Range Treatment:} $\{1,2,...,6\}$ minutes & \textbf{High Range Treatment:} $\{1,2,...,15\}$ minutes \\
    \midrule
         \multicolumn{1}{C{8cm}}{One scenario with deterministic wait + \newline One scenario with no info (min/max only)} & \multicolumn{1}{C{8cm}}{One scenario with deterministic wait + \newline One scenario with no info (min/max only)} \\
     $\downarrow$ & $\downarrow$ \\
    Info preference elicitation (left tail, midrange or right tail) & Info  preference elicitation (left tail, midrange or right tail)  
     \\$\downarrow$ & $\downarrow$ \\
          \multicolumn{1}{C{8cm}}{Four scenarios randomly sampled from 12 distributions  \newline(Info revealed based on participant preference and chance)} & \multicolumn{1}{C{8cm}}{Four scenarios randomly sampled from 12 distributions \newline(Info revealed based on participant preference and chance)} \\
    \bottomrule
    \end{tabular}
    \begin{minipage}{\textwidth} \scriptsize \vspace{0.1cm}
    Note: The 12 distributions varied in their means, variances, and other moments. See Figure \ref{fig:dist:s3} for the complete set of distributions and Appx.\ A.4 for information elicitation screens. 
    \end{minipage}
\end{table}

Each participant completed six scenarios (multiple price lists), with each scenario presenting them with a different distribution for Option B. All participants first faced one scenario with a deterministic wait and one scenario with no distributional information where only the minimum and maximum wait times were shown. After that, participants faced four scenarios randomly sampled from a set of 12 distributions with varying shapes, means, and other moments (see Figure \ref{fig:dist:s3} in the Appendix). Participants did not know ex ante the set of distributions being considered, but they did know the range of possible waiting times. There were two between-subject conditions: a ``low range'' condition where the possible wait times were $\{1,2,...,6\}$ minutes, and a ``high-range'' condition where it was $\{1,2,...,15\}$ minutes.

The key innovation of Study 3 was in the elicitation of information preferences. After the second scenario, participants were asked what type of distributional information they would most like to receive. As shown in Figure \ref{fig:screenshot:info} (Appendix), participants could choose to see probability mass in different ranges: for the low range treatment, they could select the $\{1,2\}$, $\{3,4\}$, or $\{5,6\}$ minute ranges; for the high range treatment, they could select the $\{1,2,3,4,5\}$, $\{6,7,8,9,10\}$, or $\{11,12,13,14,15\}$ minute ranges. These ranges correspond to the left tail, midrange, and right tail of the distributions, respectively. Participants indicated both their first and second preferences. Subsequently, their top choice was revealed with 75\% probability and their second choice with 50\% probability.\footnote{By creating exogenous variation in the information received, we can identify the causal effects of different types of information on choices while controlling for selection effects. Please see \textsection 4.4 and Appendix C.3 for related analysis.} Notably, different from Studies 1 and 2, the 12 distributions differed in their means. This was done to make information elicitation more consequential (had we fixed the means of the distributions, the participants would be able to infer much of the distributional shape from a single piece of information).  

\subsection{Hypotheses}
The only study we are aware of that examines information acquisition about a probability distribution is \cite{Frechette-et-al-2017}. They examined this question in the money (as opposed to time) domain and found that people are most interested in the left tail of a distribution (representing worst-case outcomes), though they also found substantial heterogeneity in preferences.  Given that preferences in the money domain do not necessarily extrapolate to decisions concerning time \citep{leclerc1995,soman2001}, we take a more cautious approach and hypothesize that heterogeneity in information acquisition preferences will prevail, and people will, on average, be indifferent across the three pieces of information.

\smallskip
\noindent \textit{H3: Decision-makers are indifferent between obtaining probabilistic information about the left tail, midrange, or right tail.}

\noindent Arguments against the null can be derived from our previous results. In particular, the results of Study 2 suggest that people have a strong aversion to thick-tailed distributions, with 86.7\% of participants consistently preferring long-right tails over thick-right tails. This would imply that people may particularly value information about the right tail, as this would help them identify and avoid thick right-tailed distributions.

\subsection{Results}
As before, we focus on presenting the results of hypothesis tests (detailed summary statistics are in Fig.\ \ref{fig:bars:s3_pref} in the Appendix). To test H3 we conduct a series of Binomial tests examining potential differences in the preference ordering. All pairwise statistics on the preference orderings among the three pieces of information, as well as the corresponding $p-$values are in Table \ref{tab:Study:3:H3.2:test}. The test results strongly reject H3: across the two treatments, 78\% of participants prefer to receive right tail information to midrange ($p\ll0.001$), and 68\% of participants prefer right tail to left tail ($p\ll0.001$). Further, there is a slight preference for left tail information over midrange, with the comparison being statistically significant for the pooled data ($p=0.038$). Together, these tests suggest that people have a consistent preference for right-tail information relative to both left-tail and midrange information.

\begin{table}[tb]\centering
\caption{Study 3: Information Preferences (Tests of H3)}
\label{tab:Study:3:H3.2:test} \small
\begin{tabular}{lcccccccc} \toprule
       & \multicolumn{2}{c}{Right Tail $\succ$ Midrange} && \multicolumn{2}{c}{Left Tail  $\succ$ Midrange} && \multicolumn{2}{c}{Right Tail $\succ$ Left Tail} \\ \cmidrule{2-3}\cmidrule{5-6}\cmidrule{8-9}
Treatment  & Frequency        & Test $=50\%$     && Frequency        & Test $=50\%$     && Frequency        & Test $=50\%$     \\ \cmidrule{1-3}\cmidrule{5-6}\cmidrule{8-9}
Low Range Treatment   & 82.78\%      & $\ll 0.001$      && 55.63\%      & 0.096            && 73.51\%      & $\ll 0.001$      \\
High Range Treatment  & 73.20\%      & $\ll 0.001$      && 54.90\%      & 0.129            && 63.40\%      & 0.001            \\
Pooled & 77.96\%      & $\ll 0.001$      && 55.26\%      & 0.038            && 68.42\%      & $\ll 0.001$     \\ \bottomrule
\end{tabular}
\end{table}

\subsection{Follow-up Study: The Role of Information Disclosure}
Practical constraints often prevent service providers from disclosing detailed distributional information. To better understand how our results may be affected by such constraints, we conducted a follow-up study.  The goals of this study were to a) validate the result from Study 2C that, from the customer's perspective, receiving tail  information (in addition to average, min and max information) can be valuable even if that information is not favorable, b) check whether simplified (i.e., coarsened) information about tail probability mass is interpreted similarly to complete probability information, and c) provide additional decision data on a wider set of probability distributions for our structural estimation in \textsection 5.1. To conserve space, we present the experiment design and analysis in Appendix \ref{sec:validation} and report only key results in the main text.

The follow-up study followed exactly the same design as Study 3 with one key difference: there was no information elicitation stage, and complete distributional information was shared with all participants. A total of 204 participants were recruited. To examine the role of different forms of information sharing, we pooled the data from Study 3 and from the follow-up study and created three distinct information regimes: complete information regime (complete distributional information), coarsened information regime (probability mass, rather than each support value, revealed for left tail, midrange, and right tail), and no information regime (only minimum and maximum wait times shown). This allows us to quantify how different levels of information disclosure affect the monetary value participants put on uncertain waits.

The results (Table \ref{tab:s3:inforegs} in the Appendix) confirm that information disclosure significantly affects willingness to accept uncertain waits. However, the direction and the size of the response depend on distributional characteristics. For favorable (heavy left-tail) distributions, sharing information reduces switching amounts by \$0.16 (coarse information) to \$0.18 (complete information), both $p<0.001$. That is, disclosing distributional information leads to a significantly more favorable response to uncertainty. For unfavorable (heavy right-tail) distributions, disclosing tail shape \textit{increases} switching amounts by \$0.04-\$0.08, i.e., makes them less attractive ($p<0.001$ and $p=0.098$). Furthermore, pooling across all distributions, information disclosure reduces switching amounts on average ($p=0.029$ and $p=0.034$), suggesting that the positive response to favorable distributions is stronger than the negative response to unfavorable ones. Notably, participants respond similarly to complete and coarsened information, with the differences between these two regimes not being statistically significant for any distribution type ($p\geq0.164$). In sum, the follow-up study shows that the net effect of disclosure is positive despite the negative response to unfavorable distributions, and that there is little value to information precision once the general distributional shape is revealed.

\subsection{Discussion}
Studies 1 and 2 showed that people are averse to uncertain waits, with preferences driven not only by mean and variance but also by tail shape, with a strong aversion to thick right-tailed distributions. However, these studies left two important gaps. First, if people could choose what information to receive about an uncertain wait, what would they prioritize? Second, how are choices affected by the availability (and precision) of distributional information? Both these questions are relevant for the practical implications of our results. In Study 3 we addressed these questions through a design that allows participants to select which portion of the distribution they want revealed (left tail, midrange, or right tail). We then conducted a follow-up study to measure how different levels of information disclosure affect the cost of an uncertain wait.

Our first result is that people overwhelmingly seek right-tail information. This aligns with the thick-right-tail aversion documented in Study 2. Learning about the probability mass assigned to the worst-case scenario(s) allows people to update their beliefs and form preferences about the distribution. Second, we examined how distributional information affects the valuation of uncertain waits. For distributions with thick-left (i.e., thin-right) tails, revealing probability information reduced switching amounts. Conversely, for distributions with thick-right (i.e., thin-left) tails, information disclosure increased switching amounts but the effect sizes were somewhat smaller (and not always significant). Pooling across all distributions, disclosing distributional information made the wait more attractive. Further, the result that coarse information produced effects that were statistically indistinguishable from complete information suggests that simplified representations of uncertainty are as effective as more precise ones.

These results have practical implications for service operations. The strong preference for right-tail information suggests that when communication constraints exist, and informativeness is the goal, priority should be given to sharing probabilistic information about the upper range of waiting times. Services with long-right tails will benefit from transparency, as disclosure reduces the cost of waiting. However, when right tails are thick, the decision to disclose (or not) distributional information is more complex. In this case,  service providers may need to weigh trust and customer satisfaction (from receiving information), against potential demand losses (from responding to unfavorable information). Finally, the finding that simplified probability display (probability mass in the left tail, in the midrange, in the right tail) are as effective as complete distributions suggests that service providers can achieve the benefits of transparency without sharing complex probabilistic information.

\section{Utility Estimation, Alternative Explanations and Simulation Study \label{sec:utility}}
In this section we propose a plausible mathematical model that rationalizes thick-right-tail aversion, discuss why several alternative explanations cannot explain our results and present a simulation study that illustrates the service design implications of accounting for tail preferences in a utility model.    

\subsection{Tail-Based Utility Models}
We begin by examining several candidate utility models that could explain our results. Given that many different models are potentially plausible, we first tested a wide range of specifications, including pure Value-at-Risk (or percentile-based) models, pure Conditional Value at Risk (CVaR, sometimes called Expected Shortfall) models, moment-based models and models based on standard convex and concave transformations of outcomes (power and exponential utility). After extensive exploration, we narrowed down our search to utility models that aim at characterizing how a decision-maker responds to the outcomes in the right tail of the probability distribution (i.e., worst-case outcomes). Specifically, we focus on percentile-based and CVaR models, both of which are commonly used to model investor attitudes towards risk in the portfolio optimization literature \citep{rockafellar2002conditional, alexander2009minimizing, sarykalin2008value, filippi2018conditional}.  We explored various combinations and parameterizations of both types of models and settled on a combined percentile and CVaR specification, which provides a more complete characterization of tail risk than either measure in isolation, offers an intuitive explanation of our key result (thick-right-tail aversion), and achieves the best empirical fit to our data.  

Table \ref{tab:structural} presents the estimates from tail-based utility models (columns 1-3), as well as more standard moment-based models (columns 4-7). For estimation, we pool the data from Studies 2A, 2B (\textsection 3.1--3.2) and the follow-up Study (\textsection 4.4 and Appendix D).\footnote{We do not include Study 1 (\textsection 2) data because the distributions used there do not have enough variation in tail shape. We also do not include Study 2C (\textsection 3.3) and Study 3 (\textsection 4.1-4.3) data because these studies do not present participants with complete distributional information; therefore, we cannot measure the response to different distributional features.} We first examine model coefficients. All seven models control for mean duration. Unsurprisingly, the coefficient on mean is negative and significant across all specifications (-0.321 to -0.612), indicating aversion to distributions with higher expected waits. Next, consider tail-based models. The coefficient on percentile is also negative and significant (-0.114 to -0.214), suggesting that people are averse towards distributions where the 70th, 80th, or 90th percentile waiting times are higher. In contrast, the coefficients on CVaR are positive (0.116 to 0.331).  That is, conditional on probability mass at high support values, decision-makers prefer when the value above the cutoff is shifted more towards higher values. For example, if we consider the 80th percentile model, distributions with many bad outcomes that lie well above that percentile are more attractive than the thick-tailed distributions with a relatively high 80th percentile and an accumulation of values close to that cutoff. 

\begin{table}[t!]
\centering
\caption{Conditional Logit Regression Results}
\label{tab:structural}
\scriptsize
\begin{tabular}{lcccccccc}
\toprule 
            &\multicolumn{3}{c}{Tail-based Models} & &\multicolumn{4}{c}{Moment-based Models}  \\
\cmidrule(lr){2-4} \cmidrule(lr){6-9}
            &\multicolumn{1}{c}{70th}&\multicolumn{1}{c}{80th}&\multicolumn{1}{c}{90th}& &\multicolumn{1}{c}{Mean}&\multicolumn{1}{c}{Mean+Var}&\multicolumn{1}{c}{Mean+Var+Skew}&\multicolumn{1}{c}{All}\\
\cmidrule(lr){2-4} \cmidrule(lr){6-9}
Mean & $-$0.410*** & $-$0.540*** & $-$0.612*** & & $-$0.409*** & $-$0.403*** & $-$0.321*** & $-$0.393*** \\
& (0.052) & (0.054) & (0.055) & & (0.044) & (0.044) & (0.047) & (0.048) \\
\addlinespace
CVaR & 0.116*** & 0.331*** & 0.328*** & & & & & \\
& (0.025) & (0.024) & (0.025) & & & & & \\
\addlinespace
Percentile & $-$0.114*** & $-$0.214*** & $-$0.149*** & & & & & \\
& (0.010) & (0.017) & (0.033) & & & & & \\
\addlinespace
Variance & & & & & & $-$0.009*** & $-$0.009** & 0.007* \\
& & & & & & (0.003) & (0.003) & (0.004) \\
\addlinespace
Skewness & & & & & & & 0.711*** & 0.392*** \\
& & & & & & & (0.079) & (0.084) \\
\addlinespace
Kurtosis & & & & & & & & 0.545*** \\
& & & & & & & & (0.051) \\
\addlinespace
Constant & $-$0.123** & $-$0.134*** & $-$0.142*** & & $-$0.118** & $-$0.117** & $-$0.136*** & $-$0.127** \\
& (0.049) & (0.050) & (0.051) & & (0.048) & (0.048) & (0.050) & (0.050) \\
\midrule
AIC & 6241.469 & 6112.327 & 6132.453 & & 6372.984 & 6367.318 & 6283.247 & 6163.464 \\
BIC & 6290.728 & 6161.586 & 6181.711 & & 6405.824 & 6408.367 & 6332.506 & 6220.933 \\
\midrule
Observations & 27,168 & 27,168 & 27,168 & & 27,168 & 27,168 & 27,168 & 27,168 \\
Subjects & 540 & 540 & 540 & & 540 & 540 & 540 & 540 \\
\bottomrule
\end{tabular}
\begin{minipage}{0.99\textwidth}
\scriptsize
        \vspace{0.1cm}
Note: Pooled data from Study 2A-B (\textsection 3.1--3.2) and follow-up study (Appendix D)  are used. Standard errors are in parentheses. Estimates are based on weighted random effects logit with standard errors clustered at subject level, with weights to account for the number of decisions. CVaR = Conditional Value at Risk (expected value above specified percentile). Response to monetary amount and round are controlled for. $^*$, $^{**}$ and $^{***}$ denotes significance at the 10\%, 5\% and 1\% level, respectively.
\end{minipage}
\end{table}

The model fit statistics in Table \ref{tab:structural} further suggest that the tail-based approach produces a better fit than the moment-based approach. Indeed, the 80th percentile model has the lowest AIC (6112.327) compared to the best moment-based model (6163.464). This supports the idea that people evaluate waiting based on tail characteristics rather than based on traditional moment-based measures. In \textsection 5.3 we will use the estimates from Table \ref{tab:structural} to characterize customer preferences in a discrete event simulation study of service design (pooled vs. dedicated queue system). 

\subsection{Alternative Explanations}
Several alternative behavioral models could, in principle, explain our results, although each faces difficulties when applied to our setting. First, standard risk-averse utility functions (e.g., power utility, exponential utility etc.) are inconsistent with our findings. This is because risk aversion implies convex disutility in waiting time, whereas Study 1 indicated diminishing sensitivity \citep[consistent with][]{leclerc1995}. Second, reference-dependent models, such as disappointment aversion \citep{loomes1986}, are unlikely to explain our results. Disappointment models tend to predict a preference for distributions with a greater concentration of probability mass in the lower region of the support.\footnote{Disappointment is triggered for outcomes that are worse than the expected utility of the distribution \citep{loomes1986}. Holding means equal, disappointment is thus greater for distributions with a greater concentration of probability mass in the lower region.} However, our Type 2 decisions in Study 2B did not reveal a systematic preference for low-mode distributions. Third, we considered probability weighting functions from prospect theory \citep{tversky1992} as a potential explanation. These models usually posit that small probabilities are overweighted, which would predict that rare long waits in long-tailed distributions should loom larger. However, our results in Study 2B-2C suggest the opposite: distributions with a long-right tail were actually preferred to thick-tailed ones. Taken together, these considerations suggest that the standard mechanisms emphasized in canonical models of choice under risk in the money domain do not translate seamlessly to waiting time contexts. Instead, tail thickness appears to be a salient feature of the waiting time distribution that drives decisions.

\subsection{Simulation Study: Pooled vs.\ Dedicated Queue Design}
To better understand the implications of our behavioral results for service design, we conduct a discrete-event simulation study. We follow the setup of \cite{sunar2021}, who study pooled vs.\ dedicated queues in overloaded systems where delay-sensitive customers observe queue lengths before deciding to join. Following their framework, we simulate two systems: (i) a \textbf{Pooled System} consisting of a single $M/M/N$ queue with $N$ identical servers and (ii) a \textbf{Dedicated System} consisting of $N$ parallel $M/M/1$ queues, where the customer arrives at one of the $N$ queues at random. In both configurations, arriving customers observe the current queue length and join only if their expected utility (given the queue length) is greater than zero. We compare three utility models -- a mean-only utility model \citep[used in][]{sunar2021}, a mean-variance utility model (using parameters estimates from  Table \ref{tab:structural}, col.\ 5) and a tail-based utility model (using parameter estimates from Table \ref{tab:structural}, col.\ 2). For further details, please see Appendix E.

A summary of simulation results is in Table \ref{tab:simulation:summary} and detailed 
results are in Table \ref{tab:simulation_results}. The classic advantage of pooling is that it helps balance the loads across servers, which leads to less idle time and more efficient processing.  Indeed, under mean-only utility, a pooled design dominates at moderate traffic levels ($\rho \leq 1.0$), achieving a higher welfare in 100\% of scenarios. At these congestion levels, pooling reduces balking rates substantially (7.4\% vs.\ 20.0\%). However, in over-capacitated systems ($\rho > 1.0$), dedicated systems are preferred in 61\% of scenarios. This is because pooling creates over-joining: too many customers join pooled queues despite congestion, reducing welfare. This is consistent with \citet{sunar2021}. The results under a mean-variance utility are similar, although pooled systems perform somewhat better when $\rho > 1.0$. 

\begin{table}[t!]
\centering
\caption{Summary of Simulation Results: Pooled vs.\ Dedicated Performance}
\label{tab:simulation:summary}
\scriptsize
\renewcommand{\arraystretch}{1.2}
\begin{tabular}{lcccccc}
\toprule
& \multicolumn{2}{c}{\textbf{Mean-Only}} & \multicolumn{2}{c}{\textbf{Mean-Variance Utility}} & \multicolumn{2}{c}{\textbf{Tail-Based Utility}}\\
& \multicolumn{2}{c}{(Table \ref{tab:structural}, col.\ 4)} & \multicolumn{2}{c}{(Table \ref{tab:structural}, col.\ 5)} & \multicolumn{2}{c}{(Table \ref{tab:structural}, col.\ 2)}\\
\cmidrule(lr){2-3} \cmidrule(lr){4-5} \cmidrule(lr){6-7}
 & \textbf{Pooled} & \textbf{Balk Rate (\%)} & \textbf{Pooled} & \textbf{Balk Rate (\%)} & \textbf{Pooled} & \textbf{Balk Rate (\%)} \\
\textbf{Traffic} & \textbf{Wins (\%)} & \textbf{Pool / Ded} & \textbf{Wins (\%)} & \textbf{Pool / Ded} & \textbf{Wins (\%)} & \textbf{Pool / Ded} \\
\midrule
Moderate: $\rho \leq 1.0$ & 100 & 7.4 / 20.0 & 100 & 7.4 / 20.0 & 70 & 18.0 / 14.4 \\
Overcapacitated: $\rho > 1.0$ & 39 & 13.2 / 24.9 & 56 & 13.1 / 24.9 & 83 & 23.2 / 19.6 \\
\bottomrule
\end{tabular}
\vspace{0.05cm}
\begin{minipage}{0.99\linewidth} \vspace{0.2cm}
\scriptsize \textit{Notes:} Simulation conducted with numbers of servers ($N \in \{2,3,5\}$), utilization levels ($\rho \in \{0.90, 0.95, 1.00, 1.05, 1.10\}$), and service rewards ($R \in \{1,2,3\}$), resulting in a total of 45 scenarios. Balking rates are averaged (in \%) across all scenarios within each traffic condition. ``Pooled Wins'' indicates the percentage of scenarios where the pooled system achieves higher social welfare than the dedicated system. See Table \ref{tab:simulation_results} for details.
\end{minipage}
\end{table}

The results are quite different under tail-based utility. At moderate traffic levels ($\rho \leq 1.0$), dedicated systems perform better in 30\% of scenarios. Dedicated performs especially well when $N=2$. In this scenario, dedicated systems produce long-right tails that customers do not mind (recall the positive coefficient on CVaR in Table \ref{tab:structural}). Note that pooled systems now have \textit{higher} balking rates (18.0\% vs.\ 14.4\%), i.e., lower throughput, suggesting that the tail shape they produce is more aversive than in dedicated systems. Furthermore, pooled systems become more attractive at higher congestion levels, winning 83\% of scenarios at $\rho > 1.0$. In dedicated systems, over-joining at high congestion levels leads to the probability mass becoming more concentrated near high wait times. In contrast, pooling under high congestion reduces tail thickness (through load balancing), and that leads to better system performance.\footnote{For example, with two servers and the reward of 2, as $\rho$ increases from 0.9 to 1.1, a pooled system produces a 9.9\% increase in the 80th percentile but a 5.5\% decrease in CVaR (thinning tail), while a dedicated system produces a 17.1\% increase in 80th percentile and a 14.7\% increase in CVaR (thickening tail).} More generally, this simulation exercise shows that incorporating distributional preferences into a service design model can have important welfare implications and can change preferred service design.

\section{Concluding Remarks}
Service operations are often optimized for average throughput or average wait time, yet service designs (priority rules, load balancing, routing etc.) can reshape the full distribution of waits, including variance and right-tail mass \cite[e.g.,][]{kleinrock1976,whitt2002,stanford2014,do2020}. Empirical studies of service systems such as emergency departments, call centers and transportation reveal a variety of waiting time distributions, often featuring heavy and/or long tails \cite[e.g.,][]{brown2005,oredsson2011,CarrionLevinson2012}. While these distributional patterns are well-documented, their behavioral implications for customers are understudied. We address this gap with a series of pre-registered, incentivized online experiments examining how people respond to different waiting-time distributions. 

\subsection{Summary of Results}
Our first study confirms that, within a given distribution type, people become more averse to waiting times as the mean and variance increase. Although this result has been shown in prior work \citep{leclerc1995, kroll2008, flicker2022}, extending it to a wider range of probability distributions and doing it in an incentivized manner is still valuable. 

Perhaps more importantly, our main contribution is in showing that mean and variance do not fully characterize behavior. The shape of the waiting time distribution, particularly the shape of its right tail drives preferences in ways that traditional moment-based models fail to capture. This is especially notable given that many real-life services involve skewed distributions with thick or long right tails, as documented in call centers where both waiting and service times exhibit these characteristics \citep{brown2005} as well as in transportation settings \citep[e.g.,][]{FosgerauFukuda2012,SusilawatiEtAl2013}.

To be specific, three main findings emerge from our work. First, \textbf{people exhibit strong aversion to thick-right tailed distributions}. Study 2A showed that when we held constant three of four distributional moments, participants sometimes preferred higher variance and higher kurtosis distributions, which contradicts predictions from standard risk preferences. Study 2B revealed the underlying driver: thicker tails need not correspond to greater variance or kurtosis. Indeed, participants overwhelmingly preferred distributions with long, smooth right tails (where extreme delays are possible but unlikely) over thick-right tails (where substantial probability mass concentrates at moderately long delays). This preference was remarkably robust, with 86.7\% of participants consistently choosing long-tailed over thick-tailed distributions when mean and variance were held constant. 

Second, consistent with our first result, \textbf{people actively seek out information about right-tail outcomes}. Study 3 showed that when people can choose what to learn about an uncertain wait, they overwhelmingly choose right-tail information. Across our treatments, 68\% preferred learning about the right tail over the left tail, and 78\% chose it over midrange information. In other words, people want to know whether probability mass accumulates in the upper range of waiting times. 

Third, to ensure that our results hold in more realistic waiting scenarios, we showed that \textbf{preferences persist under incomplete information}. Study 2C showed that even with incomplete information, long-right tailed distributions are preferred to unknown distributions (given the same mean). However, completely unknown distributions were preferred to thick-tailed distributions in some of the comparisons, suggesting that people balance their dislike of thick-right tails against their aversion to complete ambiguity.

\subsection{Implications for Service Design}
Our findings have implications for service providers designing and communicating their offerings. The strong aversion to thick-right tails suggests that service designs that create concentrated probability mass at longer wait times may be problematic. Priority queue systems illustrate this well. By shortening waits for high-priority customers, they often concentrate probability mass at longer delays for low-priority customers, creating thick-tailed distributions. The problem is especially acute in hidden-priority regimes where customers cannot observe their priority level in advance, for example, in ride-sharing where certain customer segments receive priority, or in call centers with skill-based routing.  

A related service design question is the choice between  pooled and dedicated queues. Pooling reduces idleness and leads to greater system throughput; however, when waiting is costly and customers choose to join (or not)  based on expected waiting costs, this may lead to customers ``overjoining'' relative to dedicated systems \citep{sunar2021}. We show in \textsection 5.3 that, as congestion increases, dedicated queues can lead to relatively thick right tails: an unlucky draw (e.g., a complicated transaction) disproportionately affects customers in a dedicated queue, whereas pooling spreads this type of variability across a larger customer base. Our results thus provide an additional, behavioral rationale for pooling in high-congestion settings.

For information disclosure decisions, our results provide some further guidance. When communication constraints  (limited screen space or customer attention) exist, providers should focus on conveying upper-range probabilities rather than just means or full distributions. This result also aligns with recent empirical and analytical work in the design of healthcare services, where right-tail wait-time information is increasingly used as a key metric for patient satisfaction \citep{ansari2022,gurlek2024}. Further, because simplified probability ranges are as effective as complete distributions (\textsection 4.4), simple communication about tail probabilities, e.g., in the form of percentiles, can achieve transparency benefits without overwhelming customers with complex probabilistic information. Service providers with long right-tailed waiting distributions are likely to benefit from transparency, as revealing distributional information substantially increases service attractiveness. However, providers with thick right-tailed distributions face a more complex tradeoff. While customers dislike complete ambiguity, they may penalize thick-tailed distributions once revealed. This suggests that providers need to think carefully about disclosure policies and may need to condition how much to disclose on the type of waiting time distribution produced by their service system.

\subsection{Limitations and Future Directions}
Our studies were conducted in a controlled, context-free experimental setting and with a sample of participants who are both more educated and somewhat younger relative to the general population (See Table \ref{tab:demographics}). Our results may thus not fully generalize to settings where the demographics are different from our sample. Additionally, we list three potential avenues for future refinements. First, we have focused on virtual queue settings where the (remaining) waiting time is known but the queue advancements are not displayed. Thus, our results may not extrapolate to physical queues where the advancement of the queue may be more salient than the passage of time. Second, modeling and experimentally examining socially driven behaviors \citep[as in][]{shunko2018,rosokha2024} may be a useful next step. Third, we have focused solely on decisions made \textit{prior to} joining a service system and did not examine retrospective or in-process evaluations of a wait \citep{luo2022} or potential deviations from customer expectations or reference points  \citep{yu2022,debo2023}. Real-world service experiences often involve dynamic expectations that evolve over time \citep{guda2023,deshmane2023}, which can shape the overall experience in ways not captured by our experiment. Future research could explore how decisions are influenced by the discrepancy between expected and experienced waiting times, and how service providers can manage these evolving expectations through effective communication and service design.

\bibliographystyle{plainnat}
\bibliography{./support_files/mybib}

\newpage 
\setlength\intextsep{5pt}

\appendix
\renewcommand{\thefigure}{\thesection\arabic{figure}}
\renewcommand{\thetable}{\thesection\arabic{table}}

\section*{Appendix to ``Chasing Tails: How Do People Respond to Wait Time Distributions?''}

\noindent This Appendix includes the following sections:
\smallskip

\begin{itemize}
    \item [\ref{appx:instructions}] - Instructions, screenshots and parameters.
    \item [\ref{sec:pilot}] - Details of the pilot study.
    \item [\ref{sec:additional:analysis}] - Additional Results for Studies 1-3.
    \item [\ref{sec:validation}] - Details of the follow-up study.
    \item [\ref{sec:numerical:demo}] - Simulation study illustrating the effects of different customer utility models on service design.
\end{itemize}
\bigskip

 \section{Instructions\label{appx:instructions}}
 \setcounter{figure}{0}
 \setcounter{table}{0}
 \subsection{Study 1 Instructions}

This section includes the stimuli (Instructions, screenshots and protocols) for Study 1. In this set of instructions, a subject first completed the non-deterministic block, and then the deterministic block. The sequence (deterministic$\longrightarrow$ non-deterministic or vice versa) was randomized for each subject. Differences between the two treatments (Uncertain and Ambiguous treatments) are described in square brackets. Instructions for Study 2 and 3 are analogous and are described later in the Appendix. 

\bigskip

\begin{mdframed}[backgroundcolor=white, innertopmargin=5pt, innerbottommargin=5pt, innerleftmargin=10pt, innerrightmargin=10pt]
\setlength{\parskip}{0.5em}
\setlength{\parindent}{0pt}

Welcome to this decision-making study. The study takes approximately 15 minutes to complete. Please pay close attention to the instructions. To ensure that you understand the instructions we will ask you several questions as we go along. You will only be allowed to proceed to the task if you can answer those questions. If you complete this experiment you will receive a participation payment of \$3. In addition, you will receive a bonus payment. Bonus payments range between \$1 and \$5 and depend on your decisions and on chance.

In the following, you'll face 6 decision problems. Each decision problem consists of 21 choices between two options: ``Option A'' and ``Option B''. We now describe how each decision problem works.

As noted, with each decision problem, there will be a list of decisions that you will need to make. Each decision is a paired choice between "Option A" and "Option B". Both options consist of amounts of money and lengths of time you must wait to receive the given amount of money. "Option A" is a relatively small amount of money with a short wait. "Option B" is a relatively larger amount of money, typically with a longer wait.

An example of the decisions you will make is in the table below. This example will not be played for real money or time - it is just for you to get used to the interface.
 \end{mdframed}
\smallskip

[Abbreviated Screenshot:]
\smallskip

\begin{figure}[H]
    \centering
    \includegraphics[width=0.9\textwidth]{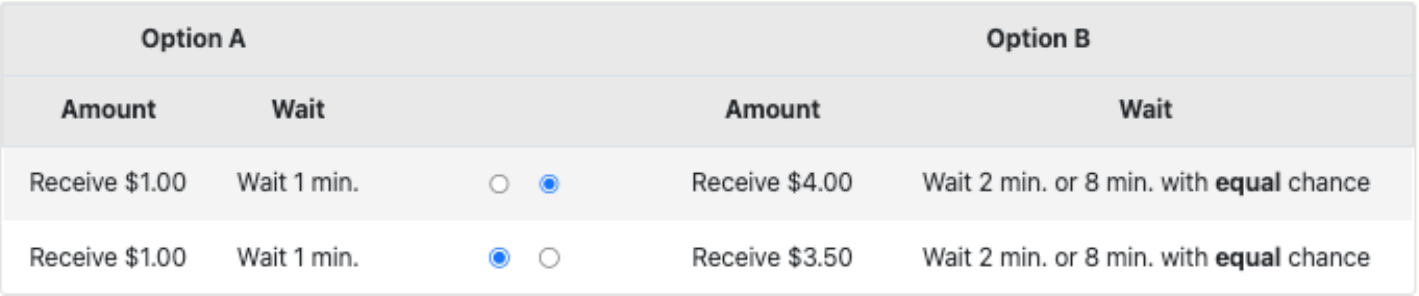}
\end{figure}

\begin{mdframed}[backgroundcolor=white, innertopmargin=5pt, innerbottommargin=5pt, innerleftmargin=10pt, innerrightmargin=10pt]
\setlength{\parskip}{0.5em}
\setlength{\parindent}{0pt}

Your task is to choose which option, A or B, that you prefer for each decision presented to you. Once you have completed all of the 6 decision problems, you will be taken to a new screen in order to determine your payoff. Please click "Next" now to continue with the instructions.

As noted earlier, you will make several choices between option A and option B. After you have made all choices for all decision problems, we will first randomly select (with equal chance) one of the decision problems. For that decision problem, we will use the option that you have chosen, A or B, to determine your payoff. For example, suppose that the row in the table below is selected for payment. Suppose also that you selected Option B for this decision. In this case, after clicking "Next", you will be taken to a waiting screen, were you will be asked to wait for either 2 minutes or 8 minutes (with equal chance). If you successfully wait for the required length of time, you will be paid \$4.00. If you do not successfully wait for the required length of time, then you will not receive any payment.
\end{mdframed}

\smallskip
[Subject sees an example where Option B is chosen]
\smallskip

\begin{mdframed}[backgroundcolor=white, innertopmargin=5pt, innerbottommargin=5pt, innerleftmargin=10pt, innerrightmargin=10pt]
\setlength{\parskip}{0.5em}
\setlength{\parindent}{0pt}

On the other hand, suppose that the row in the table below is selected for payment. Suppose also that you selected Option A for this decision. In this case, you will be asked to wait 1 minute, after which you will receive \$1.00.
\end{mdframed}

\smallskip
[Subject sees an example where Option A is chosen]
\smallskip

\begin{mdframed}[backgroundcolor=white, innertopmargin=5pt, innerbottommargin=5pt, innerleftmargin=10pt, innerrightmargin=10pt]
\setlength{\parskip}{0.5em}
\setlength{\parindent}{0pt}

Note: Because each decision problem and each question within a decision problem is equally likely to be chosen for payment, you have no incentive to lie on any question. If you lied on a question, and if that question is chosen for payment, then you would end up with the option you like less.
\end{mdframed}

\smallskip
[Subject completes a comprehension quiz]
\smallskip

\begin{mdframed}[backgroundcolor=white, innertopmargin=5pt, innerbottommargin=5pt, innerleftmargin=10pt, innerrightmargin=10pt]
\setlength{\parskip}{0.5em}
\setlength{\parindent}{0pt}
As noted earlier, you will make several choices between Option A and Option B. After you have made all choices for all decision problems, we will randomly select (with equal chance) one of the decision problems and implement your choice for that problem. If the choice that has been selected involves waiting, you will be presented with a wait. During the wait we will use attention checks. The following page demonstrates what you would experience in the waiting stage.
\end{mdframed}

\smallskip
[The following paragraph was only shown in the Ambiguous treatment.]
\smallskip

\begin{mdframed}[backgroundcolor=white, innertopmargin=5pt, innerbottommargin=5pt, innerleftmargin=10pt, innerrightmargin=10pt]
\setlength{\parskip}{0.5em}
\setlength{\parindent}{0pt}
As you can see below, waiting times may be uncertain. Suppose you choose option B. To determine the duration we will use a computerized coin whose likelihood of landing on heads or tails is somewhere between 0 and 100\%. You can think about it just as a regular coin, but instead of having 50\% probability of falling on heads or tails, that probability is $X$. In fact, even we (the experimenters) do not know $X$ -- it will be determined by a random algorithm after you make all your choices.
\end{mdframed}

\smallskip
[Screenshot of Waiting Demo]
\smallskip

\begin{figure}[h!]
    \centering
    \includegraphics[width=0.75\textwidth]{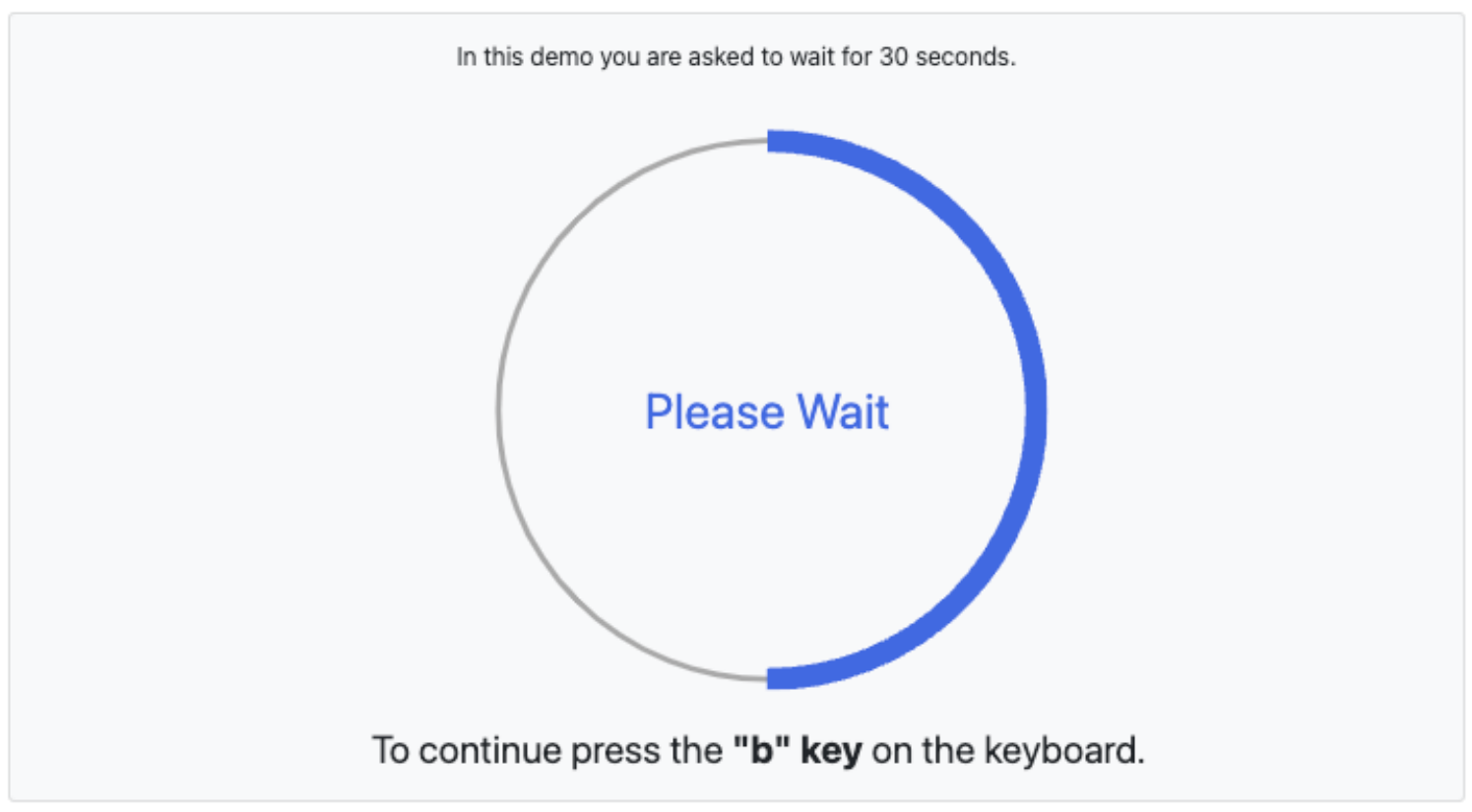}
\end{figure}

\begin{mdframed}[backgroundcolor=white, innertopmargin=5pt, innerbottommargin=5pt, innerleftmargin=10pt, innerrightmargin=10pt]
\setlength{\parskip}{0.5em}
\setlength{\parindent}{0pt}
You are now ready to begin making choices. Remember that one of your choices will be selected for real payment/wait at the end of the study.
\end{mdframed}

\smallskip
[Subject completes the first block of questions (Non-deterministic for this subject).]
\smallskip

\begin{mdframed}[backgroundcolor=white, innertopmargin=5pt, innerbottommargin=5pt, innerleftmargin=10pt, innerrightmargin=10pt]
\setlength{\parskip}{0.5em}
\setlength{\parindent}{0pt}
In the following decision problems you will be asked to make similar choices. However, rather than having a shorter and a longer waiting times, Option B will now involve a no uncertainty. That is, the waiting time for both options will be known ahead of time. Remember that one of your choices may be selected for real payment/wait at the end of the study.
\end{mdframed}
\begin{itemize}

\smallskip
\item[][Subject completes the second block of questions (Deterministic for this subject).]
 
\item[][One of the decisions made by the subject is drawn at random. The subject experiences the time outcome of the selected option (Option A or Option B, depending on the subject's prior choice) for that decision. After completing the wait, the subject proceeds to payment and exit survey.]
\end{itemize}
\smallskip

Subjects see a visual representation of the respective distribution (Binary, Uniform, or Exponential, depending on the treatment) and are trained to develop an intuitive understanding of the risks involved. Quartiles are added for continuous distribution to explain each distribution. After the training subjects answer several additional quiz questions. Additionally, subjects see the graph of the distribution during each decision (parametrized for the respective decision set). The visual representations for the $\mu=5,\sigma=2$ scenario are reproduced below (sizes are scaled to text width). \begin{figure}[htbp]
    \centering  
    \caption{Study 1: Presentation of Probability Distributions. Wait Time in Option B ($\mu=5,\sigma=2$ scenario) }
    \smallskip
     \label{fig:study2_screenshot}
      \begin{subfigure}[b]{0.23\textwidth}
        \caption{Binary Treatment}
        \label{fig:screenshot:bin_treatment}
        \includegraphics[width=\textwidth]{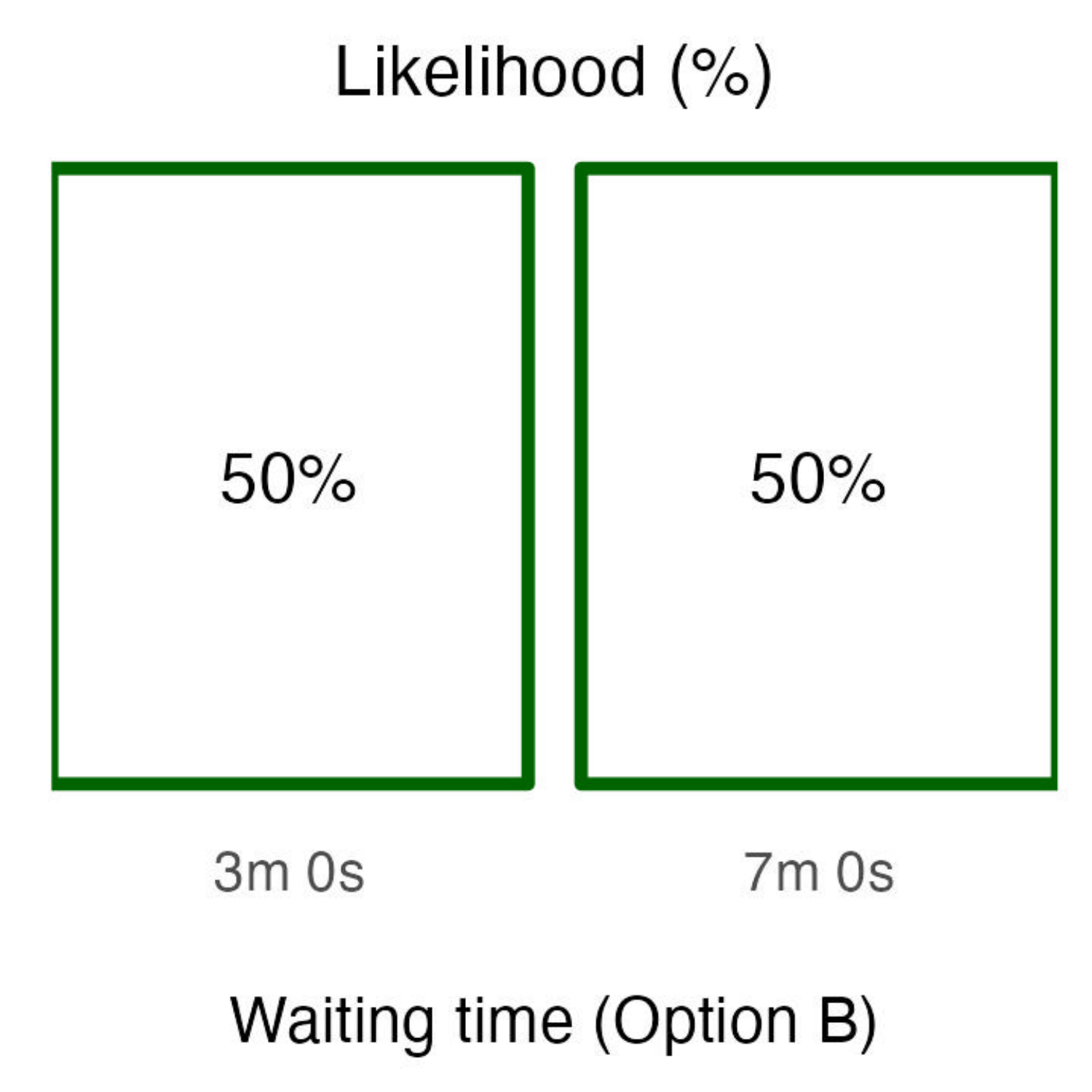}
    \end{subfigure}
    \hfill
    \begin{subfigure}[b]{0.32\textwidth}
        \caption{Uniform Treatment}
                            \vspace{0.93cm}
        \label{fig:screenshot:unif_treatment}
        \includegraphics[width=\textwidth]{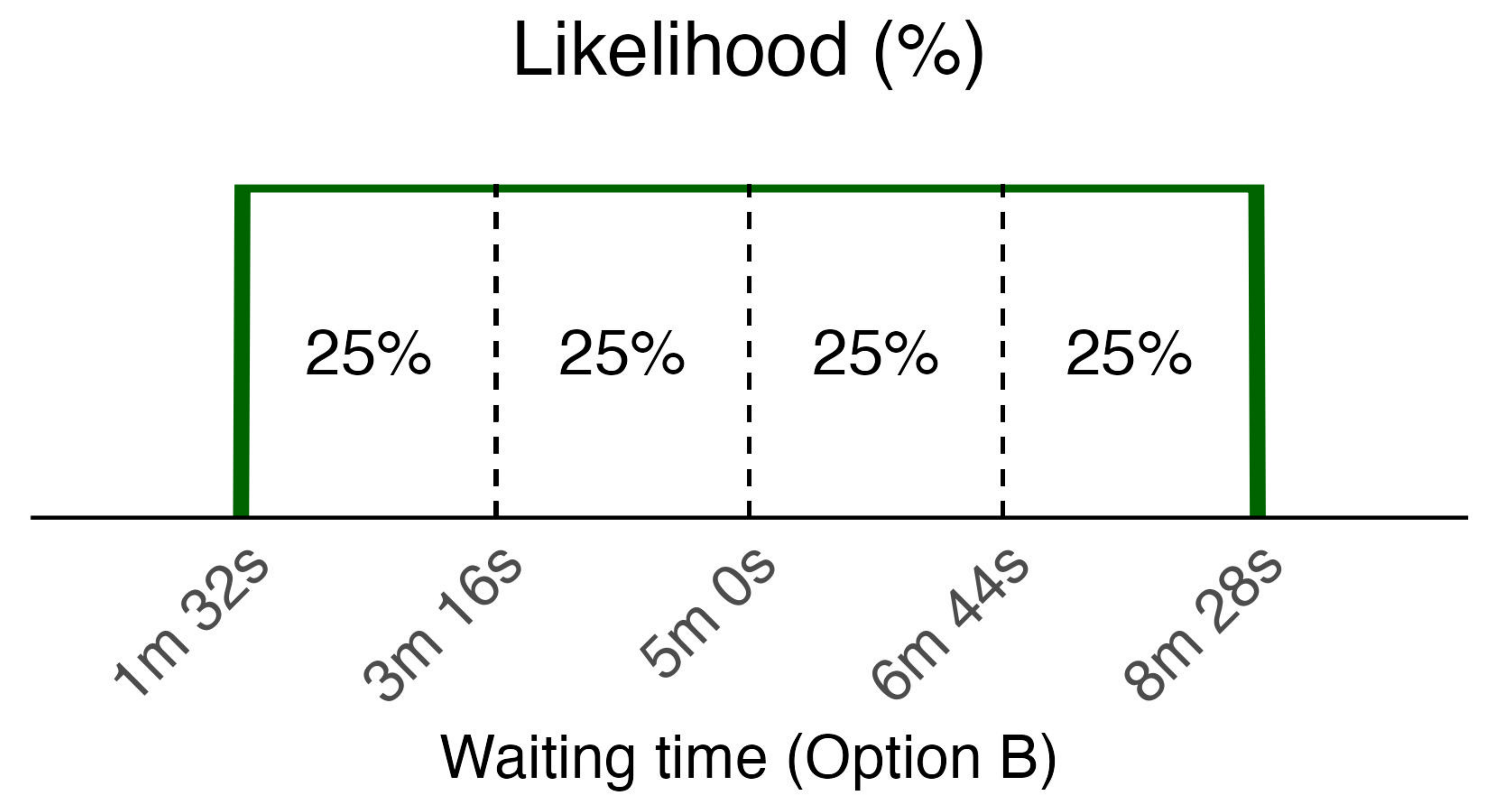}
    \end{subfigure}
    \hfill
    \begin{subfigure}[b]{0.43\textwidth}      
        \caption{Exponential Treatment}
                                    \vspace{0.2cm}
        \label{fig:screenshot:expo_treatment}
        \includegraphics[width=\textwidth]{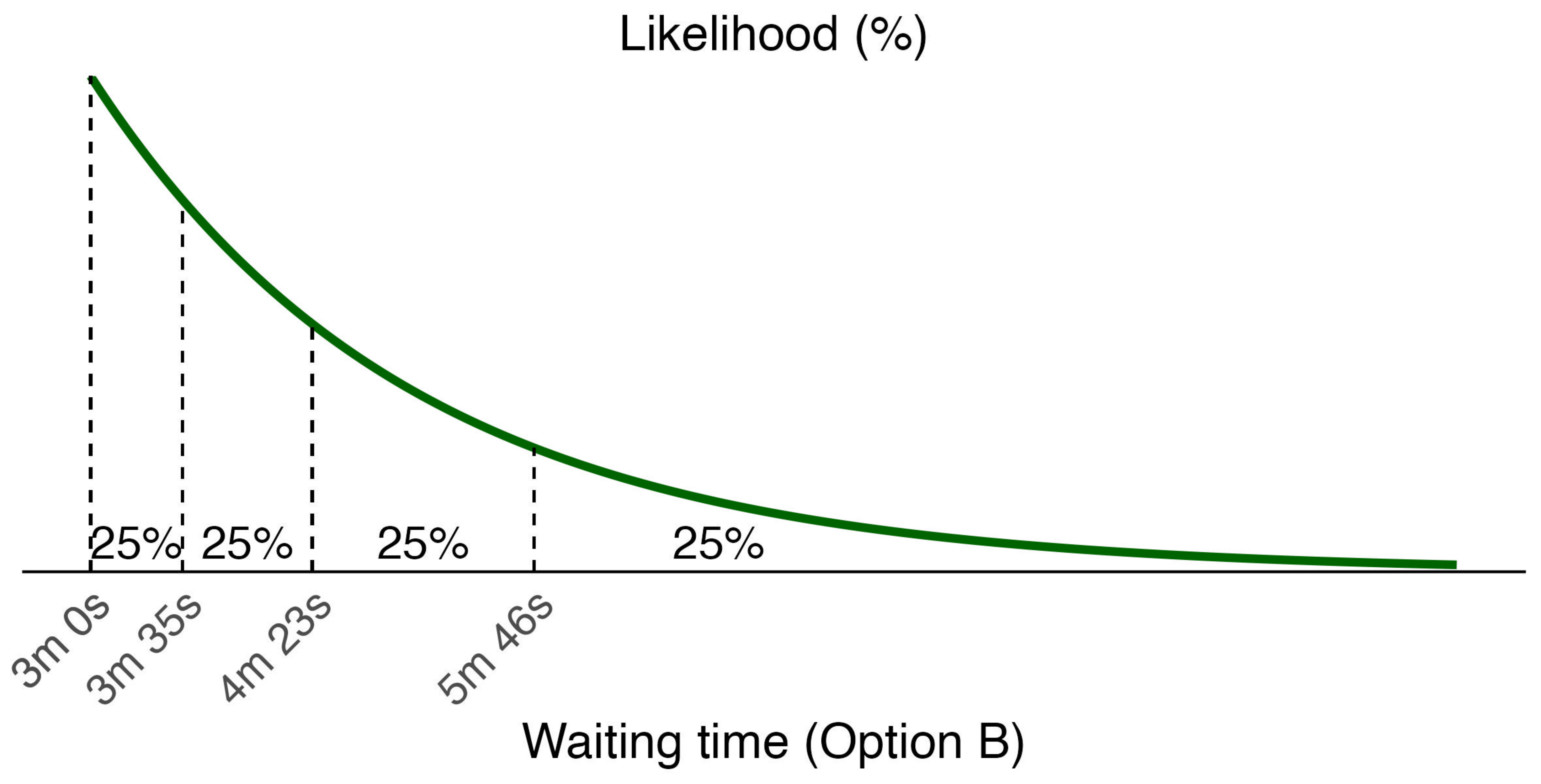}
    \end{subfigure}
\end{figure}

    \FloatBarrier
\newpage  \clearpage

\subsection{Study 2: Distributions}
This section presents the details of the distributions used in Study 2. Note that all decisions involved binary choices between Distribution A and Distribution B. Table \ref{tab:study:2A:distributions} shows the distributions and the decisions in Study 2A. Table \ref{tab:study:2B:distributions} shows the distributions and the decisions in Study 2B. Table \ref{tab:study:2C:distributions} shows the distributions and the decisions in Study 2C. In addition to seeing tables with the distributions subjects saw histograms for each decision they made. The visual representations (screenshots) are in Figures \ref{fig:study2a:screenshots}-\ref{fig:study2c_screenshots} in the main text. 

\begin{table}[h]
\centering
\footnotesize  
\caption{Study 2A: Distribution Characteristics and Decisions}
\label{tab:study:2A:distributions}
\renewcommand{\arraystretch}{1.4}
\begin{tabular}{ccc}
\toprule
\multicolumn{3}{c}{\textbf{Distributions (Support and Probabilities)}}\\
\midrule
\textbf{Name} & \textbf{Support} & \textbf{Probabilities} \\
\midrule
D1  & $\{1,3,5,7,9,11,13,15,17\}$ & $\{0.065, 0.065, 0.120, 0.145, 0.210, 0.145, 0.120, 0.065, 0.065$\} \\
D2  & $\{1,3,5,7,9,11,13,15,17\}$ & $\{0.035, 0.080, 0.210, 0.130, 0.090, 0.130, 0.210, 0.080, 0.035$\} \\
D3  & $\{1,3,5,7,9,11,13,15,17\}$ & $\{0.120, 0.080, 0.120, 0.120, 0.120, 0.120, 0.120, 0.080, 0.120$\} \\
D4  & $\{1,3,5,7,9,11,13,15,17\}$ & $\{0.020, 0.260, 0.090, 0.090, 0.080, 0.090, 0.090, 0.260, 0.020$\} \\
D5  & $\{1,3,5,7,9,11,13,15,17\}$ & $\{0.180, 0.140, 0.080, 0.080, 0.040, 0.080, 0.080, 0.140, 0.180$\} \\
D6  & $\{1,3,5,7,9,11,13,15,17\}$ & $\{0.055, 0.390, 0.035, 0.015, 0.010, 0.015, 0.035, 0.390, 0.055$\} \\
D7  & $\{1,3,5,7,9,11,13,15,17\}$ & $\{0.240, 0.235, 0.010, 0.010, 0.010, 0.010, 0.010, 0.235, 0.240$\} \\
D8  & $\{1,3,5,7,9,11,13,15,17\}$ & $\{0.090, 0.155, 0.210, 0.140, 0.010, 0.025, 0.040, 0.040, 0.290$\} \\
D9  & $\{1,3,5,7,9,11,13,15,17\}$ & $\{0.010, 0.135, 0.285, 0.155, 0.010, 0.010, 0.090, 0.240, 0.065$\} \\
D10 & $\{1,3,5,7,9,11,13,15,17\}$ & $\{0.200, 0.260, 0.050, 0.020, 0.010, 0.010, 0.010, 0.090, 0.350$\} \\
\bottomrule
\end{tabular}

\vspace{0.75em}
\begin{tabular}{cccccccccccccc}
\toprule
\multicolumn{13}{c}{\textbf{Decisions}}\\
\midrule
& & & \multicolumn{5}{c}{\textbf{Option A}} & & \multicolumn{5}{c}{\textbf{Option B}} \\
\textbf{Decision Type} & \textbf{Decision} & & \textbf{Dist} & $\mathbb{E}[t]$ & $\sigma[t]$ & $\mathbb{S}[t]$ & $\mathbb{K}[t]$ & & \textbf{Dist} & $\mathbb{E}[t]$ & $\sigma[t]$ & $\mathbb{S}[t]$ & $\mathbb{K}[t]$ \\
\cmidrule{1-2}\cmidrule{4-8}\cmidrule{10-14}

\multirow{3}{*}{\parbox{3.8cm}{\centering \textbf{Type 1:} $\mu_A=\mu_B$, $\sigma_A \neq \sigma_B$, $\mathbb{S}_A=\mathbb{S}_B$, $\mathbb{K}_A\approx\mathbb{K}_B$}}
 & 1 & & D2  & 9.00 & \cellcolor{gray!30}4.24 & 0.00 & 1.87 && D3  & 9.00 & \cellcolor{gray!30}5.09 & 0.00 & 1.87 \\
 & 2 & & D4  & 9.00 & \cellcolor{gray!30}4.99 & 0.00 & 1.43 && D5  & 9.00 & \cellcolor{gray!30}6.03 & 0.00 & 1.43 \\
 & 3 & & D6  & 9.00 & \cellcolor{gray!30}6.03 & 0.00 & 1.12 && D7  & 9.00 & \cellcolor{gray!30}6.93 & 0.00 & 1.12 \\
\cmidrule{1-2}\cmidrule{4-8}\cmidrule{10-14}

\multirow{3}{*}{\parbox{3.8cm}{\centering \textbf{Type 2:} $\mu_A=\mu_B$, $\sigma_A\approx\sigma_B$, $\mathbb{S}_A \neq \mathbb{S}_B$, $\mathbb{K}_A\approx\mathbb{K}_B$}}
 & 4 & & D4  & 9.00 & 4.99 & \cellcolor{gray!30}0.00 & 1.43 && D9  & 9.00 & 5.00 & \cellcolor{gray!30}0.30 & 1.43 \\
 & 5 & & D5  & 9.00 & 6.03 & \cellcolor{gray!30}0.00 & 1.43 && D8  & 9.00 & 6.00 & \cellcolor{gray!30}0.30 & 1.45 \\
 & 6 & & D7  & 9.00 & 6.93 & \cellcolor{gray!30}0.00 & 1.12 && D10 & 9.00 & 6.99 & \cellcolor{gray!30}0.11 & 1.14 \\
\cmidrule{1-2}\cmidrule{4-8}\cmidrule{10-14}

\multirow{3}{*}{\parbox{3.8cm}{\centering \textbf{Type 3:} $\mu_A=\mu_B$, $\sigma_A\approx\sigma_B$, $\mathbb{S}_A\approx\mathbb{S}_B$, $\mathbb{K}_A \neq \mathbb{K}_B$}}
 & 7 & & D1  & 9.00 & 4.24 & 0.00 & \cellcolor{gray!30}2.37 && D2  & 9.00 & 4.24 & 0.00 & \cellcolor{gray!30}1.87 \\
 & 8 & & D3  & 9.00 & 5.09 & 0.00 & \cellcolor{gray!30}1.87 && D4  & 9.00 & 4.99 & 0.00 & \cellcolor{gray!30}1.43 \\
 & 9 & & D5  & 9.00 & 6.03 & 0.00 & \cellcolor{gray!30}1.43 && D6  & 9.00 & 6.00 & 0.00 & \cellcolor{gray!30}1.12 \\
\cmidrule{1-2}\cmidrule{4-8}\cmidrule{10-14}

 &  10--13  && \multicolumn{11}{c}{Randomly selected pairs drawn from the base distributions D1--D10} \\
\bottomrule
\end{tabular}

\vspace{0.05cm}

\begin{minipage}{0.87\textwidth} 
\scriptsize 
\textbf{Note:}  Decision sequence was randomized within-subject. Notation: $\mathbb{E}[t]$ = expected value, $\sigma[t]$ = standard deviation, $\mathbb{S}[t]$ = skewness, $\mathbb{K}[t]$ = kurtosis. Gray cells indicate parameters that differ between the two distributions in each comparison.
\end{minipage}
\end{table}

\begin{table}[bt]
\centering
\footnotesize  
\caption{Study 2B: Distribution Characteristics and Decisions}
\label{tab:study:2B:distributions}
\renewcommand{\arraystretch}{1.4}

\begin{tabular}{ccc}
\toprule
\multicolumn{3}{c}{\textbf{Distributions (Support and Probabilities)}}\\
\cmidrule{1-3}
\textbf{Name} & \textbf{Support} & \textbf{Probabilities} \\
\cmidrule{1-3}
D1  & $\{1,3,...,15,17\}$ & $\{0.090, 0.155, 0.210, 0.140, 0.010, 0.025, 0.040, 0.040, 0.290$\} \\ 
D2  & $\{1,3,...,15,17\}$ & $\{0.010, 0.135, 0.285, 0.155, 0.010, 0.010, 0.090, 0.240, 0.065$\} \\ 
D3  & $\{1,3,...,15,17\}$ & $\{0.200, 0.260, 0.050, 0.020, 0.010, 0.010, 0.010, 0.090, 0.350$\} \\ 

D4  & $\{1,7,17\}$ & $\{0.210, 0.465, 0.325$\} \\ 
D5  & $\{1,7,17\}$ & $\{0.110,  0.625, 0.265$\} \\ 
D6  & $\{1,7,17\}$ & $\{0.350,  0.240,  0.410$\} \\ 

D7  & $\{1,11,17\}$ & $\{0.325,  0.465, 0.210$\} \\ 
D8  & $\{1,11,17\}$ & $\{0.265,  0.625, 0.110$\} \\ 
D9  & $\{1,11,17\}$ & $\{0.410,  0.240,  0.350$\} \\ 

D10 & $\{1,3,...,23,25\}$ & $\{0.130, 0.120, 0.120, 0.120, 0.110, 0.110, 0.085, 0.060, 0.050, 0.035, 0.030, 0.020, 0.010$\} \\ 
D11 & $\{1,3,...,23,25\}$ & $\{0.050, 0.110, 0.110, 0.160, 0.290, 0.060, 0.060, 0.050, 0.040, 0.030, 0.020, 0.010, 0.010$\} \\ 
D12 & $\{1,3,...,23,25\}$ & $\{0.230, 0.100, 0.100, 0.090, 0.070, 0.070, 0.070, 0.070, 0.060, 0.050, 0.050, 0.020, 0.020$\} \\ 

\bottomrule
\end{tabular}

\vspace{0.75em}

\begin{tabular}{c c c l c c c c c l c c c c}
\toprule
\multicolumn{14}{c}{\textbf{Decisions}}\\
\midrule
 & & & \multicolumn{5}{c}{\textbf{Option A}} & & \multicolumn{5}{c}{\textbf{Option B}} \\
\textbf{Decision Type} & \textbf{Decision} & & \textbf{Dist} & \textbf{Range} & $\mathbb{E}[t^A]$ & $\sigma[t^A]$ & $\mathbb{S}[t^A]$ & & \textbf{Dist} & \textbf{Range} & $\mathbb{E}[t^B]$ & $\sigma[t^B]$ & $\mathbb{S}[t^B]$ \\
\cmidrule{1-2}\cmidrule{4-8}\cmidrule{10-14}

\multirow{3}{*}{\parbox{3.3cm}{\centering \textbf{Type 1:} $\mu_A=\mu_B$, $\sigma_A\approx\sigma_B$, $|\mathrm{supp}(A)|<|\mathrm{supp}(B)|$}}
 & 1 & & D4 & \cellcolor{gray!30}$\{1,7,17\}$ & 8.99 & 6.01 & 0.26 & & D1 & \cellcolor{gray!30}$\{1,3,...,17\}$ & 9.00 & 6.00 & 0.30 \\
 & 2 & & D5 & \cellcolor{gray!30}$\{1,7,17\}$ & 8.99 & 5.15 & 0.55 & & D2 & \cellcolor{gray!30}$\{1,3,...,17\}$ & 9.00 & 5.00 & 0.30 \\
 & 3 & & D6 & \cellcolor{gray!30}$\{1,7,17\}$ & 9.00 & 7.04 & 0.08 & & D3 & \cellcolor{gray!30}$\{1,3,...,17\}$ & 9.00 & 6.99 & 0.11 \\
\cmidrule{1-2}\cmidrule{4-8}\cmidrule{10-14}

\multirow{3}{*}{\parbox{3.3cm}{\centering \textbf{Type 2:} $\mu_A=\mu_B$, $\sigma_A\approx\sigma_B$, $\mathbb{S}_A=-\mathbb{S}_B$}}
 & 4 & & D4 & $\{1,7,17\}$ & 8.99 & 6.01 & \cellcolor{gray!30}0.26 & & D7 & $\{1,11,17\}$ & 9.01 & 6.01 & \cellcolor{gray!30}-0.26 \\
 & 5 & & D5 & $\{1,7,17\}$ & 8.99 & 5.15 & \cellcolor{gray!30}0.55 & & D8 & $\{1,11,17\}$ & 9.01 & 5.15 & \cellcolor{gray!30}-0.55 \\
 & 6 & & D6 & $\{1,7,17\}$ & 9.00 & 7.04 & \cellcolor{gray!30}0.08 & & D9 & $\{1,11,17\}$ & 9.00 & 7.04 & \cellcolor{gray!30}-0.08 \\
\cmidrule{1-2}\cmidrule{4-8}\cmidrule{10-14}

\multirow{3}{*}{\parbox{3.3cm}{\centering \textbf{Type 3:} $\mu_A=\mu_B$, $\sigma_A\approx\sigma_B$, $\max_A \neq \max_B$}}
 & 7 & & D1 & \cellcolor{gray!30}$\{1,3,...,17\}$ & 9.00 & 6.00 & 0.30 & & D10 & \cellcolor{gray!30}$\{1,3,...,25\}$ & 8.99 & 6.04 & 0.58 \\
 & 8 & & D2 & \cellcolor{gray!30}$\{1,3,...,17\}$ & 9.00 & 5.00 & 0.30 & & D11 & \cellcolor{gray!30}$\{1,3,...,25\}$ & 9.00 & 5.05 & 0.86 \\
 & 9 & & D3 & \cellcolor{gray!30}$\{1,3,...,17\}$ & 9.00 & 6.99 & 0.11 & & D12 & \cellcolor{gray!30}$\{1,3,...,25\}$ & 9.00 & 7.02 & 0.51 \\

 & 10--13 & & \multicolumn{11}{c}{Randomly selected pairs drawn from the base distributions D1--D12.} \\
\bottomrule
\end{tabular}

\vspace{0.1cm}
\begin{minipage}{\textwidth} 
\scriptsize 
\textbf{Note:} The sequence of all decisions was randomized within-subject. Notation: $\mathbb{E}[t]$ = expected value, $\sigma[t]$ = standard deviation, $\mathbb{S}[t]$ = skewness, $|\mathrm{supp}(\cdot)|$ = support size, $\max$ = maximum value. Grey cells indicate features that differ between the two distributions in each comparison.
\end{minipage}
\end{table}

 \begin{table}[bt]
\centering
\scriptsize
\caption{Study 2C: Distribution Characteristics and Decisions}
\label{tab:study:2C:distributions}
\renewcommand{\arraystretch}{1.4}

\begin{tabular}{ccc}
\toprule
\multicolumn{3}{c}{\textbf{Distributions (Support and Probabilities/Definition)}}\\
\midrule
\textbf{Name} & \textbf{Support} & \textbf{Probabilities / Definition} \\
\midrule
D1  & $\{1,3,\dots,25\}$ & $\{0.065, 0.150, 0.240, 0.185, 0.095, 0.075, 0.050, 0.040, 0.025, 0.020, 0.020, 0.020, 0.015$\} \\
D2  & $\{1,3,\dots,25\}$ & $\{0.090, 0.115, 0.130, 0.140, 0.130, 0.125, 0.105, 0.050, 0.040, 0.025, 0.025, 0.015, 0.010$\} \\
D3  & $\{1,3,\dots,25\}$ & $\{0.230, 0.100, 0.100, 0.090, 0.070, 0.070, 0.070, 0.070, 0.060, 0.050, 0.050, 0.020, 0.020$\} \\
D4  & $\{1,3,\dots,15\}$ & $\{0.150, 0.210, 0.120, 0.095, 0.050, 0.015, 0.050, 0.290, 0.020$\} \\
D5  & $\{1,3,\dots,17\}$ & $\{0.020, 0.360, 0.050, 0.050, 0.040, 0.050, 0.050, 0.360, 0.020$\} \\
D6  & $\{1,3,\dots,17\}$ & $\{0.200, 0.260, 0.050, 0.020, 0.010, 0.010, 0.010, 0.090, 0.350$\} \\
D7  & $\{1,2,\dots,15\}$ & Unknown distribution \\
D8  & $\{1,2,\dots,17\}$ & Unknown distribution \\
D9  & $\{1,2,\dots,25\}$ & Unknown distribution  \\
D10  & $\{1,2,\dots,15\}$ & Tail info: probability mass $0.20$ on $\{11,12,13,14,15\}$; remainder unspecified \\
D11 & $\{1,2,\dots,15\}$ & Tail info: probability mass $0.20$ at $15$; remainder unspecified \\
\bottomrule
\end{tabular}

\vspace{0.75em}

\begin{tabular}{c c c c c c}
\toprule
\multicolumn{6}{c}{\textbf{Decisions}}\\
\midrule & & & \multicolumn{1}{c}{\textbf{Option A}} & & \multicolumn{1}{c}{\textbf{Option B}} \\
\textbf{Decision Type} & \textbf{Decision} & & \textbf{Dist} & & \textbf{Dist} \\
\cmidrule(lr){1-2}\cmidrule(lr){4-4}\cmidrule(lr){6-6}

\multirow{3}{*}{\parbox{4.9cm}{\centering \textbf{Type 1:} $\mu_A=\mu_B,~\max_A=\max_B$ \\ Long-tailed (full info) vs Unknown}}
 & 1 & & D1 & & D9 \\
 & 2 & & D2 & & D9 \\
 & 3 & & D3 & & D9 \\
\cmidrule(lr){1-2}\cmidrule(lr){4-4}\cmidrule(lr){6-6}

\multirow{3}{*}{\parbox{4.9cm}{\centering \textbf{Type 2:} $\mu_A=\mu_B,~\max_A=\max_B$ \\ Thick-tailed (full info) vs Unknown}}
 & 4 & & D4 & & D7 \\
 & 5 & & D5 & & D8 \\
 & 6 & & D6 & & D8 \\
\cmidrule(lr){1-2}\cmidrule(lr){4-4}\cmidrule(lr){6-6}

\parbox{6cm}{\centering \textbf{Type 3:} $\mu_A=\mu_B,~\max_A=\max_B$ \\ Long-tailed (incomplete info) vs Unknown}
 & 7 & & D7 & & D10 \\  
\cmidrule(lr){1-2}\cmidrule(lr){4-4}\cmidrule(lr){6-6} 

\parbox{6cm}{\centering \textbf{Type 4:} $\mu_A=\mu_B,~\max_A=\max_B$ \\ Thick-tailed (incomplete info) vs Unknown}
 & 8 & & D7 & & D11 \\
\bottomrule
\end{tabular}

\vspace{0.1cm}
\begin{minipage}{0.66\textwidth} 
\scriptsize 
\textbf{Note:} The sequence of all decisions was randomized within-subject. 
\end{minipage}
\end{table}

    \FloatBarrier
\newpage  \clearpage
\subsection{Study 3 Distributions}
\begin{figure}[h!]
    \centering
    \caption{Study 3: Distributions}
    \vspace{0.1cm}
    \includegraphics[width=0.96\textwidth]{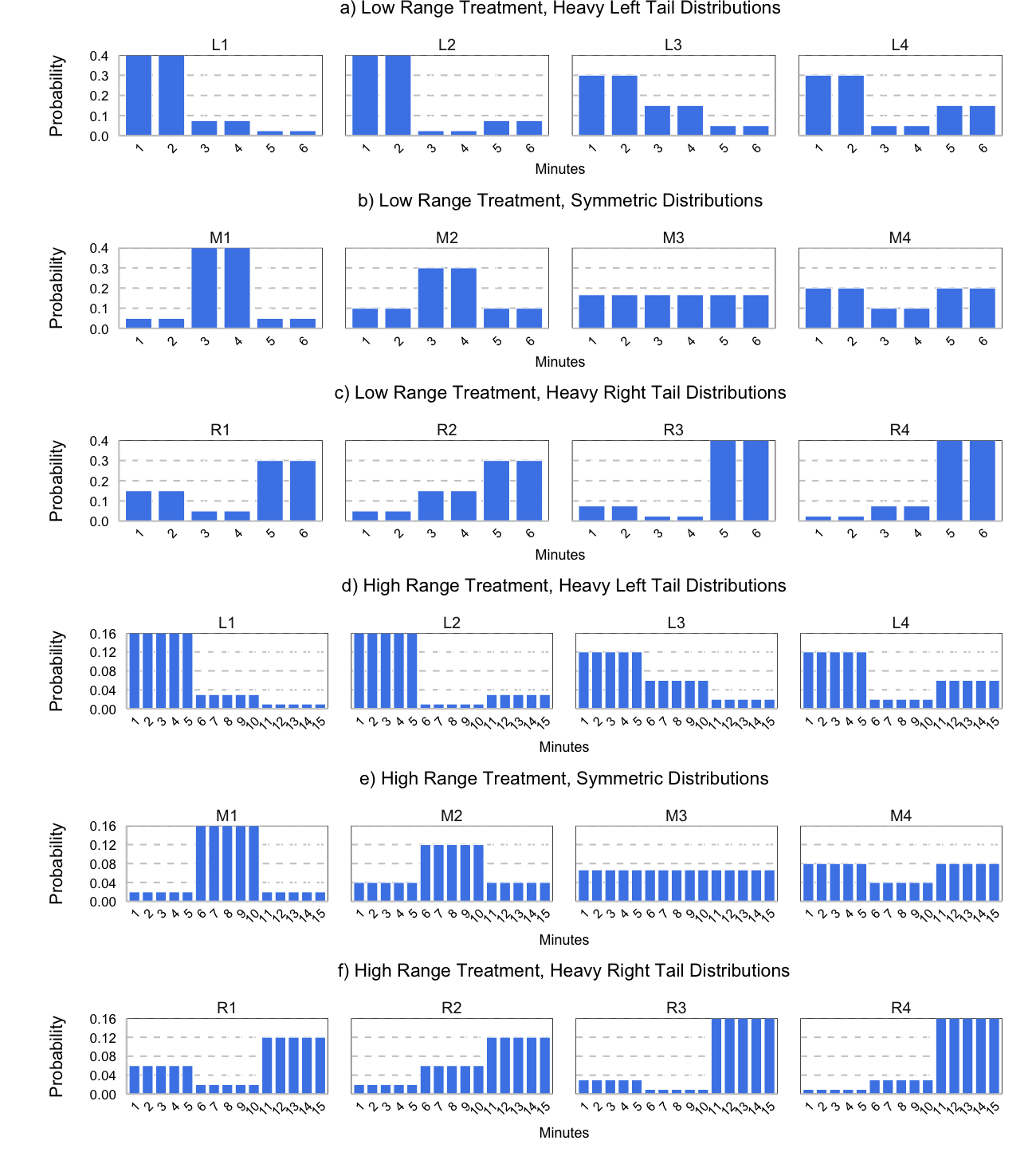}
        \label{fig:dist:s3}
\end{figure}

Figure \ref{fig:dist:s3} shows the probability distributions used in Study 3. For each decision, the distribution was described to participants via a table with all possible waiting time values and their probabilities. 

    \FloatBarrier
\newpage  \clearpage
\subsection{Study 3: Instructions}
Below we present the instructions for Study 3. Instructions for the follow-up study (Appx. D) are the same except that we no longer ask subjects for their information preferences, but simply display the entire probability distribution. General instructions for Study 3 are analogous to Study 1, with the following differences. Before completing the first non-deterministic block (decision set 2) subjects saw the following instructions. 

\bigskip

\begin{mdframed}[backgroundcolor=white, innertopmargin=5pt, innerbottommargin=5pt, innerleftmargin=10pt, innerrightmargin=10pt] 
\setlength{\parskip}{0.5em} 
\setlength{\parindent}{0pt} 

In the remaining decision problems you will be asked to make similar choices. However, the waiting time in Option B will now be uncertain. This means, if you choose Option B, and that choice is selected for real payment/wait, you will not know exactly how much time you will need to wait. As before, the amount of money you will be paid is larger if you choose Option B. The Tables below show the waiting times for Option A and Option B will look like. In particular, in Option B you could wait anywhere between 1 and 15 minutes. However, you do not know how likely it is that you have to wait each amount of time. For example, the chance of waiting exactly 5 minutes could be 5\%, or it could be 75\%. All you know is that the time cannot be less than 1 minute or greater than 15 minutes.
\end{mdframed}
\bigskip

\begin{figure}[h!]
    \centering
    \includegraphics[width=0.75\textwidth]{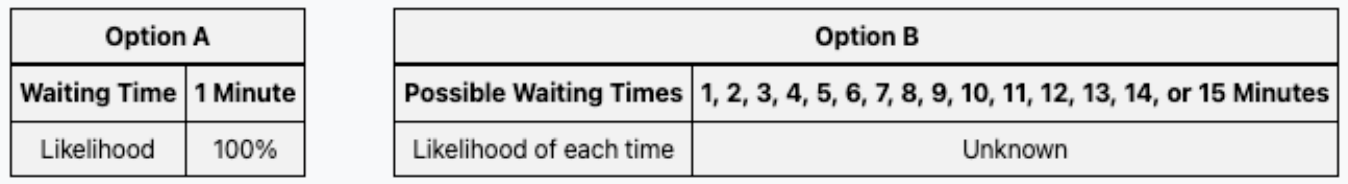}
\end{figure}

\begin{itemize}

\medskip

\item[][Subject completes a short comprehension quiz.]
 
\item[][Subject completes the decision set.]

\end{itemize}
\medskip

\begin{mdframed}[backgroundcolor=white, innertopmargin=5pt, innerbottommargin=5pt, innerleftmargin=10pt, innerrightmargin=10pt] 
\setlength{\parskip}{0.5em} 
\setlength{\parindent}{0pt} 
In the remainder of the experiment, you will receive partial information about Option B. The table below shows three regions: 1-5 minutes (short waiting time), 6-10 minutes (moderate waiting time) and 11-15 minutes (long waiting time).
\end{mdframed}

\medskip

\begin{itemize}
\item[][Subject sees the following:]
\end{itemize}

\begin{figure}[h!]
    \centering
    \includegraphics[width=0.75\textwidth]{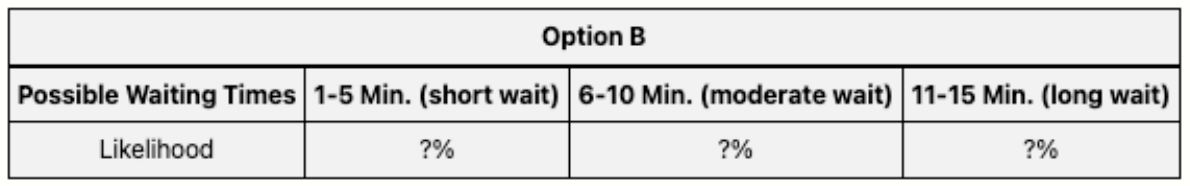}
\end{figure}

\bigskip

\begin{mdframed}[backgroundcolor=white, innertopmargin=5pt, innerbottommargin=5pt, innerleftmargin=10pt, innerrightmargin=10pt] 
\setlength{\parskip}{0.5em} 
\setlength{\parindent}{0pt} 
Below you are asked to indicate the regions whose likelihood (\%) you are most interested in learning. After you do this, we will reveal some of your choices. In particular, in each of the remaining rounds, we will reveal your top choice with a 75\% chance and your second choice with a 50\% chance. In other words, it is not guaranteed that you will receive all the pieces of information that you request, but it is more likely that you receive the information that you rank higher. What information would you like us to reveal?
\end{mdframed}

\begin{figure}[h!]
    \centering
    \caption{Study 3:  Screenshot of Elicitation Procedure}
    \includegraphics[width=0.95\textwidth]{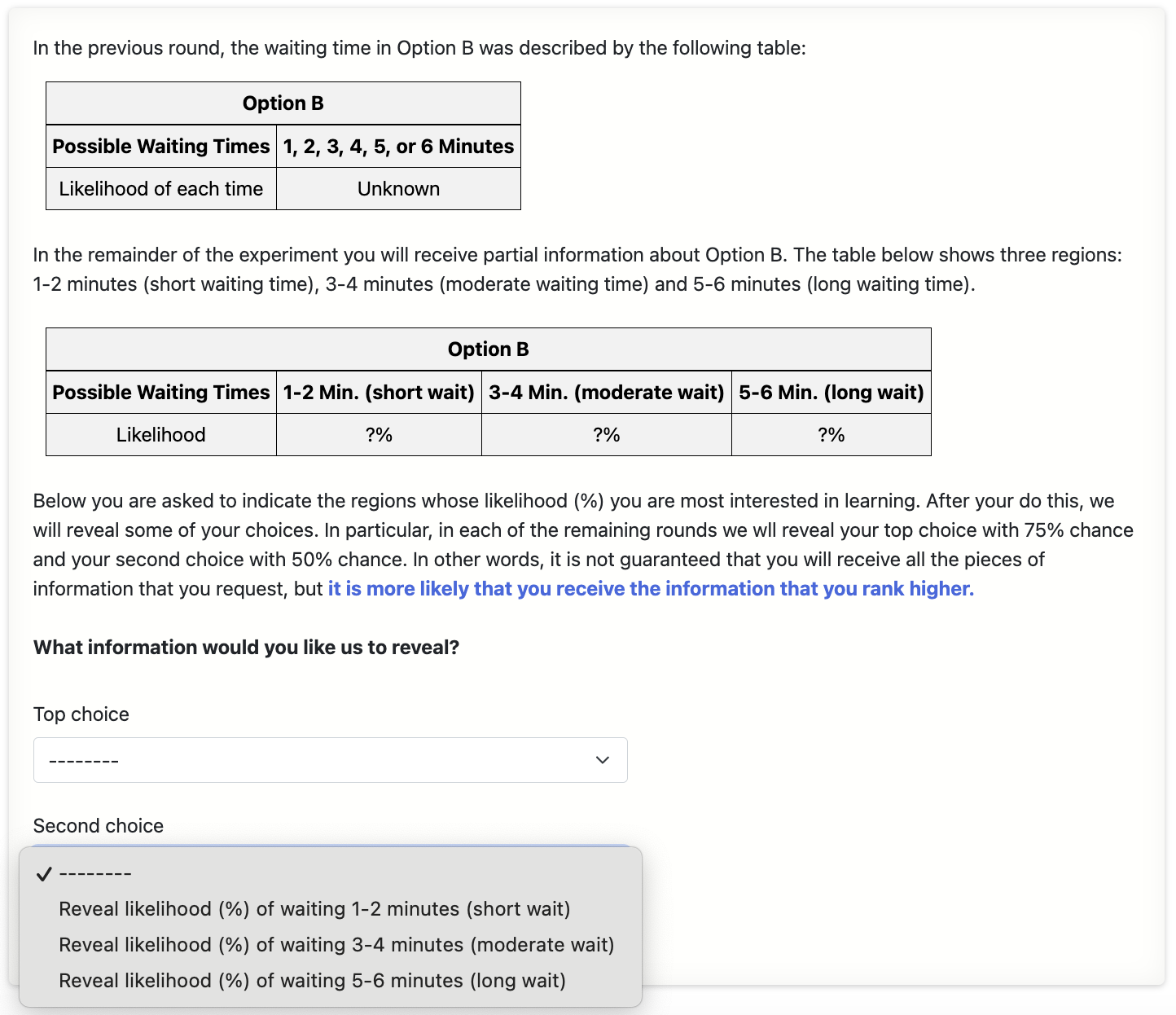}
    \label{fig:screenshot:info}
\end{figure}

\newpage

\begin{itemize}
\item[][Subject indicates their top choice and second choice and continues with the instructions for the partial information scenarios (Scenario 3-6). A screenshot of a scenario with partial information being revealed is reproduced below.]
\end{itemize}

\medskip

\begin{mdframed}[backgroundcolor=white, innertopmargin=5pt, innerbottommargin=5pt, innerleftmargin=10pt, innerrightmargin=10pt] 
\setlength{\parskip}{0.5em} 
\setlength{\parindent}{0pt} 
In this round, the following information (highlighted in blue) is revealed.
\end{mdframed}
\smallskip

\begin{figure}[h!]
    \centering
    \includegraphics[width=0.75\textwidth]{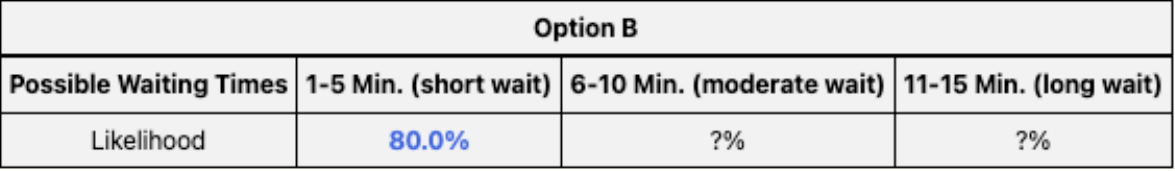}
\end{figure}
 
\begin{itemize}
\item[][Subject continues to making their decisions for the given distribution.]
\end{itemize}

\newpage \clearpage

\section{Pilot Study\label{sec:pilot}}

\subsection{Pilot Study}
As noted in \textsection 2, prior to running Study 1, we conducted a pilot study. We ran this study to examine potential parametrizations and explore an alternative, ambiguous environment where probabilities were unknown. The experimental design, including sample size, treatment conditions, exclusion criteria, hypotheses, and analysis was pre-registered (\url{https://aspredicted.org/d7g2-gfz8.pdf}). All experiments were programmed in oTree \citep{chen2016otree} and conducted online on Prolific. 

\subsubsection{Protocol}\label{sec:protocol}
\begin{figure}[htbp]
    \centering
    \caption{Pilot Study: Screenshot for Uncertain Treatment}\label{fig:study1_screenshot}
        \includegraphics[width=0.9\textwidth]{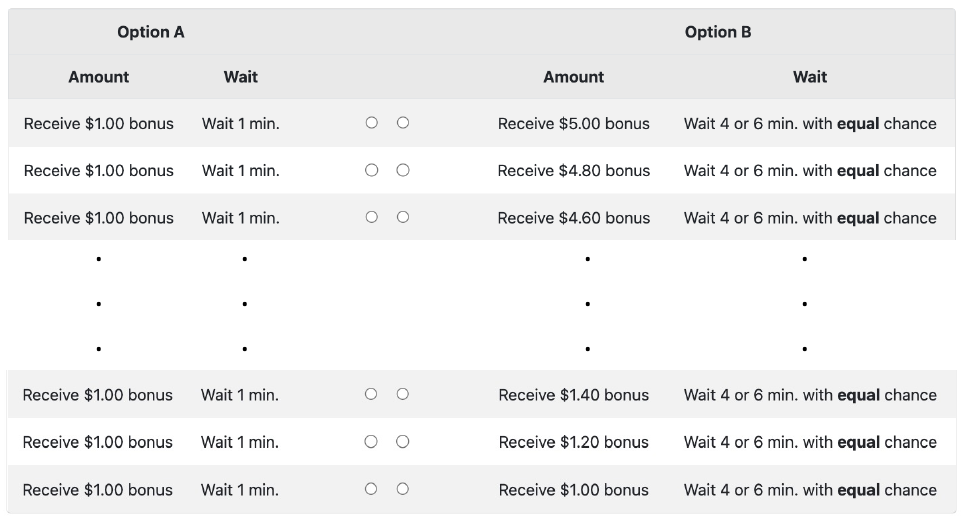}
    

    \begin{minipage}{0.9\textwidth}\scriptsize
    Note: The screenshot shows the case of the Uncertain Treatment. In the Ambiguous Treatment, everything was identical except that the word ``equal'' was replaced with ``unknown'' for the Option B wait scenarios.
    \end{minipage}
\end{figure}

The pilot experiment consisted of six multiple price lists \citep[][]{Holt02}. We will refer to each list as a ``decision set.'' Screenshots of the decision sets are shown in Figure \ref{fig:study1_screenshot}. Each decision set consisted of 21 binary choice scenarios. The waiting scenarios are presented in Table \ref{tab:study:1:scenarios}. In each scenario, Option A was: ``\textit{Receive \$1.00 and wait 1 minute}", while Option B was ``\textit{Receive \$$X$ and wait $Y$ minutes}", where $X$ and $Y$ depended on decision set and the scenario within a decision set. In particular, $X$ was varied within a decision set, while $Y$ was varied across decision sets. In each of the six decision sets, $X$ was varied between \$5.00 and \$1.00 in \$0.20 increments.\footnote{Within a decision set, participants were only allowed at most one switch point from Option B to Option A. This is done to impose strict monotonicity in revealed preferences and is common practice in experiments using multiple price list designs \citep{gonzalez1999,andersen2006}.} Depending on the decision set and treatment, $Y$ could either be deterministic, uncertain (with known probabilities), or ambiguous (with unknown probabilities). In each non-deterministic scenario, we examined a shorter and a longer spread (4/6 and 1/9 minutes for the 5-minute average and 9/11 and 4/16 minutes for the 10-minute average). Decision problems were blocked according to whether they were deterministic or not, and the sequence in which decision sets were presented to participants was randomized both by and within blocks.

\begin{table}[tbh!]
    \centering \scriptsize
    \caption{Pilot Study: Treatments and Waiting Scenarios}
    \label{tab:study:pilot:scenarios}
    \begin{tabular}{ccC{3.6cm}C{3.6cm}} \toprule
    & & \multicolumn{2}{c}{Treatment (varied between-subject)} \\ \cmidrule{3-4}
    Scenario (varied within-subject)   & Wait Times in Option B & Uncertain & Ambiguous\\ \midrule
     1 & 5 min.       &  with prob. 1   &  with prob. 1 \\
     2 & 10  min.     &  with prob. 1   &  with prob. 1 \\
     3 & 4 min. or 6 min.  &  with prob. 0.5 & with unknown prob. \\
     4 & 1 min. or 9 min.  &  with prob. 0.5 & with unknown prob. \\ 
     5 & 9 min. or 11 min.  &  with prob. 0.5 & with unknown prob. \\
     6 & 4 min. or 16 min.  &  with prob. 0.5 & with unknown prob. \\ \midrule 
    \end{tabular} 
    \begin{minipage}{0.94\textwidth} \scriptsize
    Note: The sequence of scenarios was determined at random for each participant. First, the sequence of blocks (deterministic, i.e., scenarios 1-2, or non-deterministic, i.e., scenarios 3-6) was determined at random. Second,  the sequence of scenarios within each block was determined at random. 
    \end{minipage}
\end{table}

The progression of the experiment was as follows. First, participants read through the instructions for the first randomized block (either deterministic or non-deterministic), and completed several comprehension checks. After that, they submitted decisions for each decision problem within the block. Upon completion of the first block, they were informed of a change in the Option B waiting scenarios and made decisions for the remaining problems in this block. After completing all decision problems, one decision set was selected at random, and one decision within that set was selected at random to count for payment, with the wait from that decision being implemented in real time. Depending on their choice, participants either had to wait for one minute (Option A) or $Y$ minutes, where $Y$ depended on the waiting scenario (Option B) for the selected problem.\footnote{A total of three participants dropped out during the wait. They did not receive payment and are not included in our data.}

\subsubsection{Treatments}
We conducted two between-subjects treatments: an Uncertain treatment, in which participants were given the probabilities for the short and long waiting time in Option B, as well as an Ambiguous treatment, in which they were not. We used 50\% probabilities in the Uncertain treatment. To implement ambiguity in the experiment, we used the \citet{SSD:Ambiguity} method.

\subsubsection{Sample, Exclusions and Incentives\label{sec:sample}}
All experiments were conducted on weekdays between 10 am and 6 pm Eastern Time.  A total of 246 Prolific workers were recruited (average age: 36.5, 54\% female). Only US-based workers with an approval rating of at least 99\% were eligible. A total of 40 participants were excluded based on pre-registered comprehension and attention checks, resulting in a sample size of 206.\footnote{The pre-registered exclusion criteria were as follows. First, we removed participants from the sample who made three or more errors on the quiz. Second, we removed participants who selected a dominated option (i.e., choose Option B when the monetary payment was the same, but the waiting time was longer than for Option A, see the bottom decision in Fig. \ref{fig:study1_screenshot}).} Recruitment was stopped once the target sample of 100 participants per treatment was reached. As noted earlier, participants were incentivized to report their preferences truthfully by experiencing one of their choices at the end of the experiment. During the wait every 30 seconds the time stopped and they were prompted to enter a (randomly chosen) key on the keyboard. This was done to ensure that participants were actively waiting (rather than leaving their computers during the wait). After the wait participants were redirected to the exit survey and payment.  In addition to their earnings from the experiment, all participants received a \$3.00 show-up fee.

\subsection{Results}
We use ``switching amounts'' as our dependent variable. Table \ref{tab:regression:study:1} reports the regression coefficients. First, unsurprisingly, there is a significant effect of increasing average waiting time in Option B (denoted by \texttt{averageWait} in Table \ref{tab:regression:study:1}): participants are willing to trade each additional minute for approximately 11.5 cents ($p\ll0.001$). Second, the effects of increasing uncertainty (\texttt{sdWait}) do not appear to be linear. Behavior is quite similar between the deterministic (\texttt{sdWait=0}) and the low-variance (\texttt{sdWait=1}) settings. However, a larger variance (conditions with $\texttt{sdWait}=4$ and $\texttt{sdWait}=6$) prompts participants to demand a greater monetary compensation (both coefficients significant at $p\ll0.001$). Further, we see that the coefficient on \texttt{Ambiguous Wait} is both quantitatively small and not significantly different from zero, which indicates that, on average, subjects do not view uncertain waits differently from ambiguous waits (in both specifications, $p > 0.2$). 
\begin{table}[h!]
\centering 
\renewcommand{\arraystretch}{1.1}
\small
\caption{Pilot Study: Regression Results}
\begin{tabular}{ld{3.6}@{}ld{3.6}@{}l} 
\toprule
 \multicolumn{5}{r}{Dependent variable: \texttt{Switching Amount} (USD)} \\ 
\midrule
Omitted Category: \texttt{Uncertain Wait,} & \multicolumn{2}{c}{\multirow{2}{*}{(1)}} & \multicolumn{2}{c}{\multirow{2}{*}{(2)}} \\ 
\texttt{averageWait=5, sdWait=1.} & & \\ 
\midrule
\texttt{averageWait} & 0.115\sym{***} & (0.007) & 0.115\sym{***} & (0.007) \\
\texttt{sdWait} &  &  &  &  \\
$~~~~$ \texttt{0 (Deterministic)} & -0.044 & (0.033) & -0.044 & (0.033) \\
$~~~~$ \texttt{4} & 0.171\sym{***} & (0.035) & 0.171\sym{***} & (0.035) \\
$~~~~$ \texttt{6} & 0.205\sym{***} & (0.041) & 0.205\sym{***} & (0.041) \\
\texttt{Ambiguous Wait} & -0.049 & (0.133) & -0.013 & (0.140) \\
\texttt{Round} &-0.020  &  (0.015) & -0.020 & (0.015) \\
\texttt{Age} &  &  & 0.005 & (0.006) \\
\texttt{Male} &  &  & -0.243\sym{*} & (0.140) \\
\texttt{Education} &  &  & -0.005 & (0.077) \\
\texttt{Income} &  &  & -0.047 & (0.049) \\
\texttt{Constant} & 1.382\sym{***} & (0.097) & 1.390\sym{***} & (0.248) \\
\midrule
$R^2$ & 0.082 &  & 0.102 &  \\
No. Observations & \dor{1236} &  & \dor{1236} &  \\
No. Participants & \dor{206} &  & \dor{206} &  \\
\bottomrule
\end{tabular}
\label{tab:regression:study:1}
\vspace{0.1cm}
\begin{minipage}{.75\textwidth}\scriptsize
Notes: Random effects regression coefficients are reported (Standard errors in parentheses). $^*$, $^{**}$ and $^{***}$ denotes significance at the 10\%, 5\% and 1\% level, respectively.
\end{minipage}
\end{table}

\newpage \clearpage
\section{Additional Analysis for Studies 1-3 \label{sec:additional:analysis}}
In this section we present demographic survey data for the participants of all studies (Appx. C.1) as well as the detailed analysis for Study 1 (Appx. C.2) and for Study 3 (Appx. C.3-C.4).

\subsection{Demographic Details}

\begin{table}[htbp]
\centering
\small
\caption{Demographic Characteristics Across Studies}
\label{tab:demographics}  
\begin{tabular}{lcccccccc}
\toprule
& Pilot & 1 & 2A & 2B & 2C & 3 & Follow-up & Total \\
\midrule
\textbf{Age Range} & & & & & & & & \\
18--25 & 15.53 & 18.35 & 3.21 & 6.06 & 8.86 & 20.39 & 11.43 & 13.61 \\
26--40 & 54.37 & 43.67 & 44.87 & 47.27 & 43.67 & 42.76 & 51.14 & 46.44 \\
41--55 & 21.36 & 24.37 & 29.49 & 34.55 & 35.44 & 23.36 & 26.03 & 26.75 \\
56+ & 8.74 & 13.61 & 22.44 & 12.12 & 12.03 & 13.49 & 11.42 & 13.21 \\
\midrule
\addlinespace
\textbf{Income} & & & & & & & & \\
\$0--\$19,999 & 26.70 & 24.68 & 15.38 & 15.15 & 15.82 & 25.99 & 20.55 & 21.77 \\
\$20,000--\$39,999 & 25.24 & 21.20 & 19.87 & 24.24 & 20.89 & 20.07 & 17.81 & 21.22 \\
\$40,000--\$59,999 & 18.93 & 17.41 & 22.44 & 20.61 & 25.95 & 18.75 & 18.72 & 19.82 \\
\$60,000--\$79,999 & 16.50 & 12.97 & 14.74 & 16.36 & 10.13 & 13.82 & 18.26 & 14.59 \\
\$80,000--\$99,999 & 4.37 & 10.13 & 16.67 & 3.64 & 10.13 & 9.54 & 9.13 & 9.06 \\
Above \$100,000 & 8.25 & 13.61 & 10.90 & 20.00 & 17.09 & 11.84 & 15.53 & 13.55 \\
\midrule
\addlinespace
\textbf{Education} & & & & & & & & \\
None & 2.43 & 1.27 & 0.64 & 0.61 & 0.63 & 1.64 & 0.00 & 1.13 \\
High School & 37.86 & 33.86 & 30.77 & 39.39 & 29.11 & 25.99 & 28.31 & 31.80 \\
Undergraduate & 37.38 & 45.25 & 46.15 & 36.36 & 48.73 & 44.74 & 45.66 & 43.62 \\
Graduate & 15.53 & 12.66 & 15.38 & 12.73 & 15.19 & 17.76 & 13.70 & 14.80 \\
Postgraduate & 6.80 & 6.96 & 7.05 & 10.91 & 6.33 & 9.98 & 12.33 & 8.66 \\
\midrule
\addlinespace
\textbf{Gender} & & & & & & & & \\
Male & 42.72 & 39.24 & 50.00 & 52.12 & 53.16 & 41.78 & 47.03 & 45.22 \\
Female & 54.37 & 59.81 & 48.72 & 46.67 & 44.30 & 55.92 & 52.05 & 53.06 \\
Other & 2.91 & 0.95 & 1.28 & 1.21 & 2.53 & 2.30 & 0.91 & 1.72 \\
\bottomrule
\end{tabular}
\end{table}

\subsection{Study 1: Detailed Results}
\begin{figure}[b!]
    \centering
    \caption{Study 1: Average Switching Amounts}
    \includegraphics[width=\textwidth]{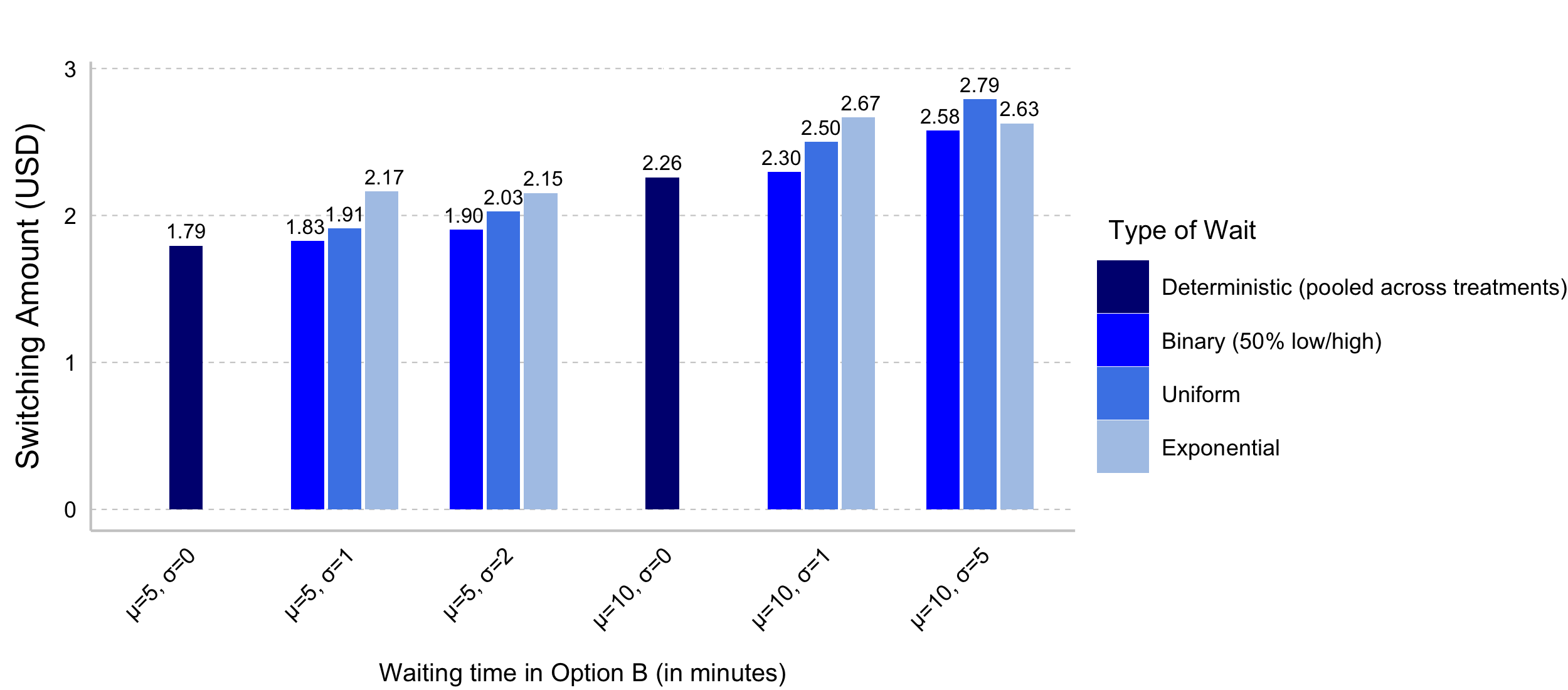}
    \label{fig:bars:s2}
\end{figure}

The average switching amounts are shown in Figure \ref{fig:bars:s2}.  Consistent with the pilot study, we see that switching amounts are higher for longer average waits. Further, holding the average wait constant, switching amounts increase with the standard deviation of the wait. This is statistically significant for most comparisons: for $\mu=5$ conditions, deterministic vs. $\sigma=1$ (rank sum test, $p = 0.04$) and deterministic vs. $\sigma=2$ ($p \ll 0.01$), and for $\mu=10$ conditions, deterministic vs. $\sigma=1$ ($p = 0.06$) and deterministic vs. $\sigma=5$ ($p \ll 0.01$). However, there also appear to be some differences between the distributions. In particular, we observe only minimal differences between the deterministic and the $\sigma=1$ scenarios when the distribution of the wait is binary (\$1.79 vs. \$1.83, $p = 0.58$ and \$2.26 vs. \$2.30, $p = 0.99$). However, this is no longer true for the remaining two distributions. Even a small amount of variability appears to increase switching points relative to the deterministic scenarios for exponential distributions (\$1.79 vs. \$2.17, $p \ll 0.01$; \$1.79 vs \$2.15, $p \ll 0.01$; \$2.26 vs. \$2.50, $p = 0.33$; and \$2.26 vs. \$2.67, $p \ll 0.01$), though the effects for uniform distributions are more mixed. Indeed, the switching amount for exponential waits is higher than for binary waits in most cases ($\mu=5$, $\sigma=1$: $p = 0.02$; $\mu=5$, $\sigma=2$: $p = 0.09$; $\mu=10$, $\sigma=1$: $p = 0.02$, $\mu=10$, $\sigma=5$: $p = 0.95$), while uniform vs. binary differences are directionally similar (in favor of binary) but not statistically significant (all $p > 0.35$). Finally, participants appear to be quite insensitive to standard deviation when dealing with exponential waits (\$2.17 vs. \$2.15, $p = 1.00$ and \$2.67 vs. \$2.63, $p = 0.76$).

Table \ref{tab:regression:study:2} shows the tests of whether switching amounts differ based on the features of the waiting time distribution. To test our main pre-registered hypothesis, H1B, we conduct the joint test that the coefficients on \texttt{Binary Wait}, \texttt{isUniform}, and \texttt{isExponential} are all equal. The results of these tests are reported in the last row of the table. H1B is rejected -- behavior is not the same for all waiting distributions, with the gap being significantly different from zero at $p < 0.05$, both with and without controls.
\begin{table}[tb!]
\centering \scriptsize\renewcommand{\arraystretch}{1.1}
    \caption{Study 1: Regression Results}
    \begin{tabular}{ld{3.6}@{}ld{3.6}@{}l} \toprule
     \multicolumn{5}{r}{Dependent variable: \texttt{Switching Amount} (USD)} \\ 
\midrule
Omitted Category: \texttt{Binary Wait,} & \multicolumn{2}{c}{\multirow{2}{*}{(1)}} & \multicolumn{2}{c}{\multirow{2}{*}{(2)}} \\ 
\texttt{avgB=5, stdev=1} & & \\ 
\midrule
\texttt{avgB}       &    0.098\sym{***}&  (0.005)&    0.098\sym{***}&  (0.005)\\
\texttt{stdev}         &            &       &            &      \\
$~~~$0  (deterministic)       &   -0.117\sym{***}&  (0.038)&   -0.118\sym{***}&  (0.038)\\
$~~~$2         &    0.046\sym{**} &  (0.020)&    0.046\sym{**} &  (0.020)\\
$~~~$5        &    0.193\sym{***}&  (0.032)&    0.193\sym{***}&  (0.032)\\
\texttt{isUniform}      &    0.153\sym{**} &  (0.062)&    0.151\sym{**} &  (0.062)\\
\texttt{isExponential}   &    0.091         &  (0.068)&    0.090         &  (0.068)\\
\texttt{Age}            &                  &         &    0.006         &  (0.004)\\
\texttt{Male}        &                  &         &   -0.080         &  (0.104)\\
\texttt{Education}            &                  &         &    0.067         &  (0.061)\\
\texttt{Income}          &                  &         &    0.006         &  (0.033)\\
\texttt{Round}        &    0.001         &  (0.010)&    0.001         &  (0.010)\\
\texttt{Constant}        &    1.407\sym{***}&  (0.060)&    1.074\sym{***}&  (0.209)\\
\midrule
$R^2$ & 0.088 &  & 0.098 &  \\
No. Observations & \dor{1896} &  & \dor{1896} &  \\
No. Participants & \dor{316} &  & \dor{316} &  \\
\midrule \midrule
\texttt{Binary Wait = isUniform = isExponential} &  \multicolumn{2}{c}{$p=0.044$} &  \multicolumn{2}{c}{$p=0.045$}  \\
\bottomrule
\end{tabular}
\label{tab:regression:study:2}
\vspace{0.1cm}
\begin{minipage}{0.95\textwidth}\scriptsize
Notes: Random effects regression coefficients are reported (Standard errors in parentheses). $^*$, $^{**}$ and $^{***}$ denotes significance at the 10\%, 5\% and 1\% level, respectively. The bottom row reports the $p-$value of the joint test that all three uncertain wait coefficients are equal (pre-registered hypothesis). 
\end{minipage}
\end{table}

\subsection{Study 3: Summary Statistics and Additional Hypothesis Tests}
\subsubsection{Summary Statistics on Information Preferences}
\begin{figure}[bp]
    \centering
    \caption{Study 3: Information Preferences}
    \includegraphics[width=\textwidth]{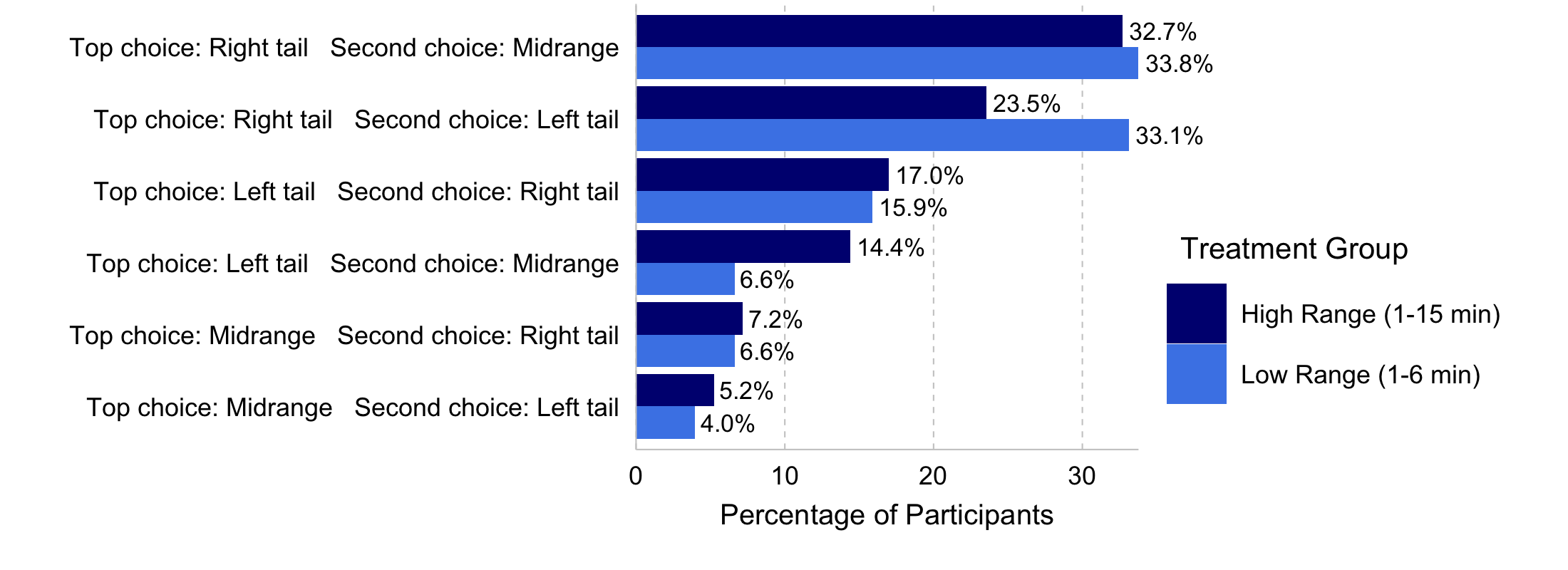}
    \label{fig:bars:s3_pref}
\end{figure}

Figure \ref{fig:bars:s3_pref} presents the relative frequencies of top and second choices by treatment. In both the High Range and Low Range treatments, the top choice for the majority of participants (56.2\% and 66.9\%, respectively) is the right tail of the waiting time distribution, suggesting a tendency to focus on the upper bound of potential wait times, rather than the central tendency or lower tail.  However, the second choice preferences are somewhat different across treatments, with the midrange being the most common second choice (47.1\%) in the High Range treatment, while the Low Range treatment shows a more even split, with midrange (40.4\%) and left tail (37.1\%) being closely tied. This suggests that as the overall range of potential waiting times increases, there is a slight shift away from the tails and towards the midrange, though the interest in the right tail dominates across both treatment conditions.

\begin{figure}[htbp]
    \centering
    \caption{Study 3: Switching Amounts by Treatment and Type of Wait}
    \includegraphics[width=\textwidth]{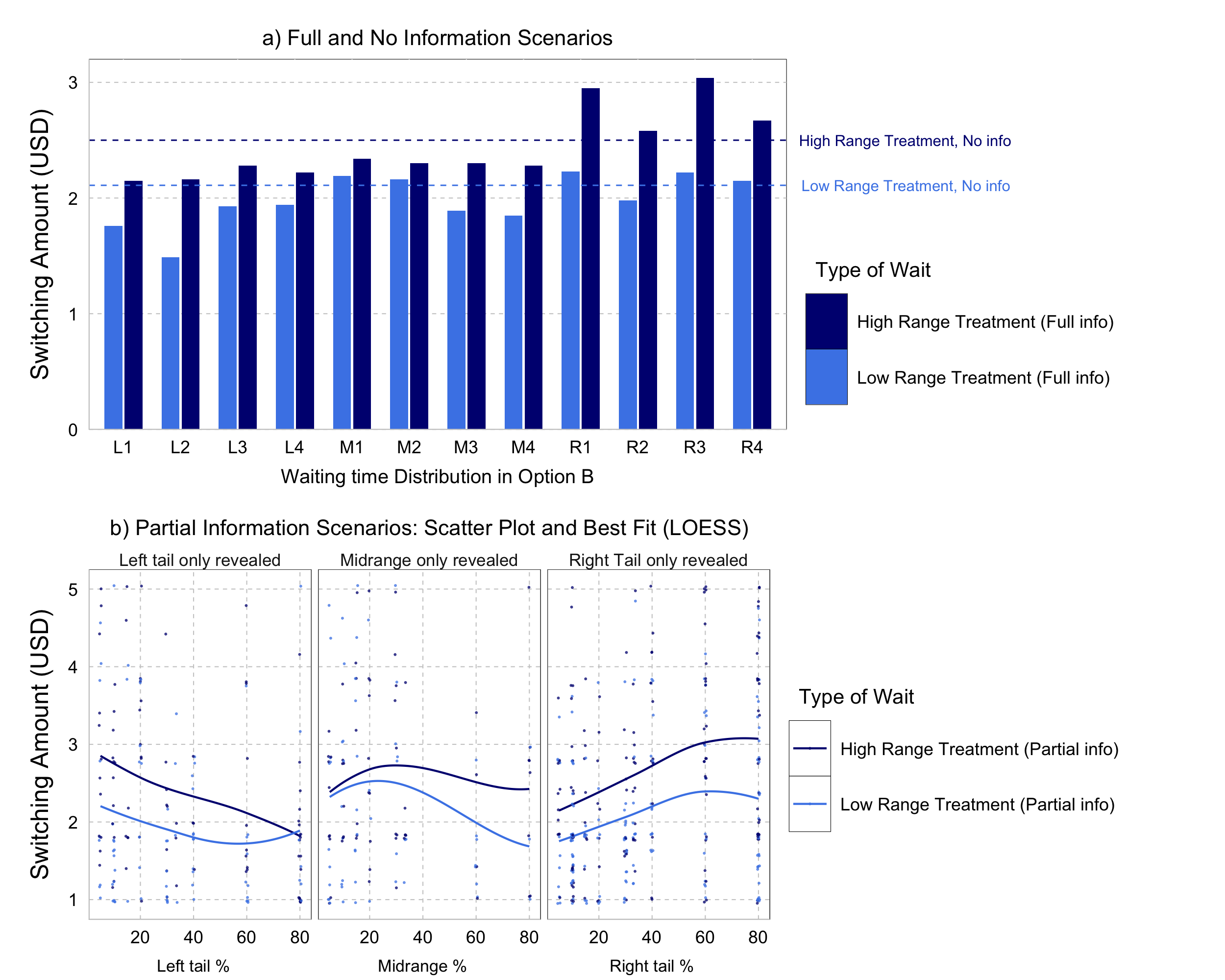}
    \label{fig:bars:s3A}
\end{figure}

Figure \ref{fig:bars:s3A} shows the switching amounts for different time distributions. Recall that a higher switching point indicates a higher cost of waiting in a given scenario. Panel (a) focuses on scenarios in which either no information (dashed line) or all three pieces of information were revealed. The data show that, unsurprisingly, revealing information that makes a wait appear more desirable results in lower switching amounts. Consider first the leftmost four sets of bars. These bars show scenarios in which the waiting time distribution is left-tail heavy; that is, lower waiting times (1-2 minutes in the Low Range treatment and 1-5 minutes in the High Range treatment) are more likely than longer times. Unsurprisingly, revealing this information leads to a lower cost of waiting relative to the no-information scenarios. Next, consider the middle four sets of bars. These are symmetric distribution scenarios (flat, bell-shaped, and U-shaped, respectively). In these scenarios, revealing waiting time information does not necessarily result in a lower switching amount in the Low Range treatment, but does so in the High Range treatment.  Finally, consider the rightmost four sets of bars, which present the response to less favorable (right-tail heavy) distributions. In the  High Range treatment, participants respond quite negatively to the revealed information with all four bars extending above the no-information scenario. In contrast, the response in the Low Range treatment appears to be minimal. In sum, participants appear to be incorporating the information into their decisions, with the response being somewhat stronger upon seeing the right tail, and in the High Range treatment.

Next, consider panel (b) of Figure \ref{fig:bars:s3A}. This panel plots the partial information (probability mass in the left tail, midrange, or right-tail) revealed to a participant against the switching amount, along with a LOESS (LOcally Estimated Sums of Squares) fit. Each scatter point corresponds to one decision. The left part of panel (b) shows that participants in both treatments require a smaller monetary payment as the probability mass in the left tail increases. However, the response in the Low Range treatment appears to be substantially more flat and noisy than in the High Range treatment. The middle part of panel (b) shows the response to midrange information. The response appears quite noisy, though there is a decreasing trend in the Low Range treatment. The right part of panel (b) shows that, as the right tail probability mass goes up, people demand a greater monetary payment. Consistent with our hypotheses, the response to the right tail appears to be somewhat more defined than the response to the left tail or midrange, particularly for the High Range treatment.

\subsection{Study 3: Additional Hypotheses}

\noindent \textit{H3-APPX: }
\begin{itemize}
\item[\textit{a)}]\textit{Response to left tail (right tail) probability mass increase is positive (negative).}
\item[\textit{b)}]\textit{Response to midrange probability mass increase is positive.}
\item[\textit{c)}]\textit{Response to tails is stronger than response to midrange.}
\item[\textit{d)}]\textit{Response to right tail is stronger than response to left tail.}
\item[\textit{e)}]\textit{Response to right tail is stronger for long waits.}
\item[\textit{f)}]\textit{Response is stronger to one's top choice than to one's second or third choice.}
\end{itemize}

\subsubsection{Tests of H3-APPX}
We next test H3-APPX, i.e., the responses to different types of incomplete information about a waiting time distribution. To test H3-APPX-a and H3-APPX-b, we perform random effects regressions with a single explanatory variable (tail/midrange probability mass) and control for treatment. The results fully support H-APPX-a and partially support H-APPX-b. In particular, the response to midrange probability mass increases is not always significantly different from 0, suggesting that people do not respond as strongly to a reduction in variance, as they do to a reduction in means. Further, the data partially support H-APPX-c and reject H-APPX-d. The results partially support H-APPX-e. The response to the right tail is significantly stronger for long waits, but not under all information scenarios. Finally, we do not find significant support for H-APPX-f; that is, information preferences do not predict the strength of one's response to that information.

\begin{figure}[bt]
    \centering
    \caption{Study 3: Tail/Midrange Effects on Switching Amount}
    \includegraphics[width=0.97\textwidth]{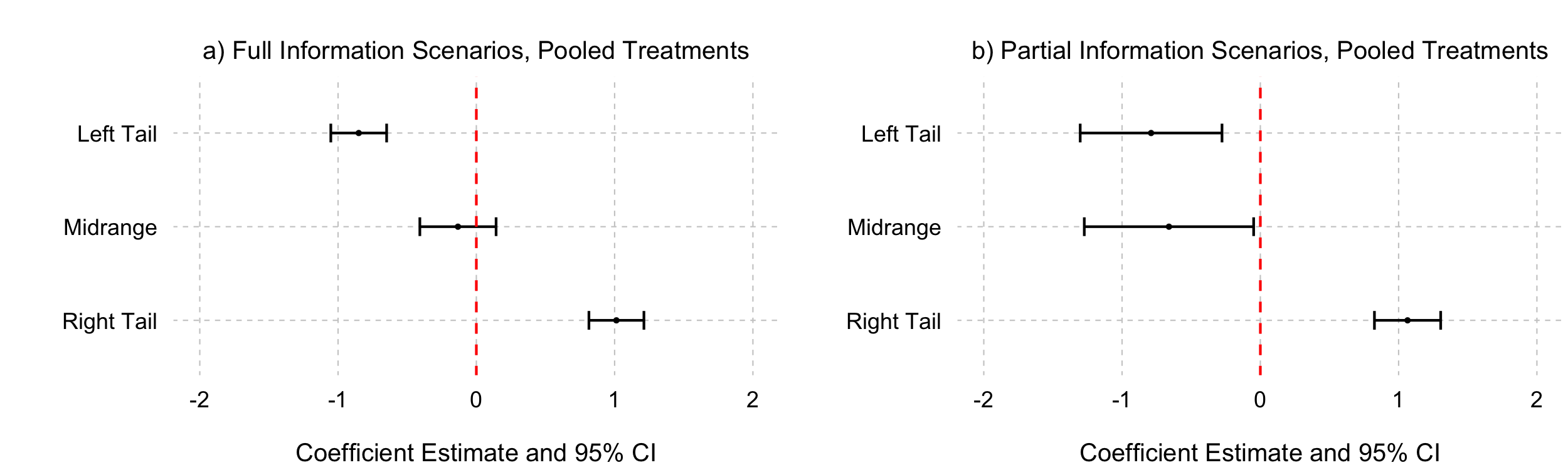}
    \label{fig:marginal:s3:pooled}
\end{figure}

We begin with H-APPX-a. Panel a) of Figure \ref{fig:marginal:s3:pooled} focuses on full information scenarios and shows the marginal effects of changes in the left tail probability, midrange, and right tail probability on switching amounts. Panel b) repeats the analysis for partial information scenarios. In both cases, we perform random effects regressions with a single explanatory variable (tail/midrange) and control for treatment.  
The results can be summarized as follows. First, the results support H-APPX-a. That is, participants react positively (negatively) to increases in the left (right) tail probability mass. All four coefficients on the tails are significantly different from 0 ($p<0.01$).  Second, the results partially support H-APPX-b. In both full information (panel a) and partial information (panel b) scenarios, switching amounts decrease as the midrange probability mass goes up. However, the effect is not statistically significant in the full information scenario and quite close to the 0.05 cutoff in the partial information scenario ($p=0.345$ and $p=0.034$, respectively). Third, the results partially support H-APPX-c. The response to tails is significantly stronger than the response to midrange in the full information case (both $p\ll0.001$) but not in the partial information case ($p=0.751$ and $p=0.229$). Fourth, the results reject H-APPX-d.  The response to right tail is not significantly stronger (in absolute value terms) than the response to left tail ($p=0.342$ and $p=0.262$). Overall, these comparisons suggest that people update their beliefs about the waiting time distribution when provided with tail and midrange information and weigh both tails similarly. However, when they receive detailed information about the waiting time distribution, they tend to focus more on the tails than on the midrange.

\begin{figure}[tb]
    \centering
    \caption{Study 3: Right Tail Effects on Switching Amount by Treatment}
    \includegraphics[width=0.97\textwidth]{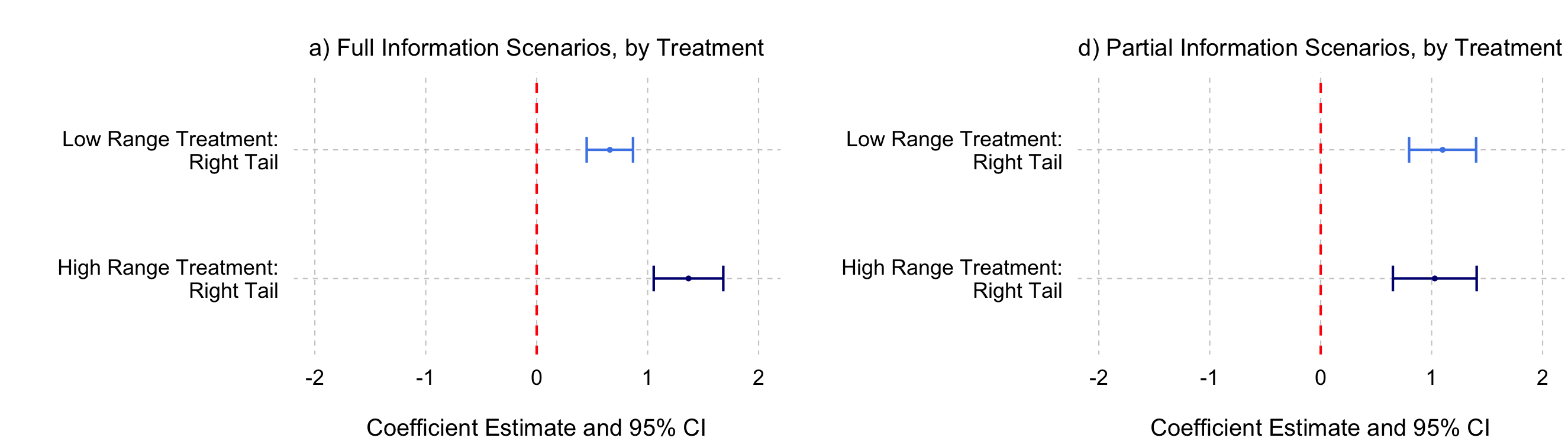}
    \label{fig:marginal:s3:right}
\end{figure}

Next, we turn to H-APPX-e. The marginal effects of the right tail probability mass on switching points in each treatment are shown in Figure \ref{fig:marginal:s3:right}. We have hypothesized that the right tail effect is stronger for the High Range treatment than in the Low Range treatment. Figure \ref{fig:marginal:s3:right} suggests that this is true for the full information scenario, but not for the partial information scenario. Indeed, the right tail effect is approximately twice as strong in the High Range treatment than in the Low Range treatment in the full information scenario (the difference between coefficients is significantly different from zero, $p\ll0.001$). However, the effect is approximately the same across treatments in the partial information scenario ($p=0.714$). Finally, we test H-APPX-f in which we hypothesized that people may respond more strongly to the information they ranked higher. Contrary to H-APPX-f, our tests show that switching points are not significantly related to information preferences (all $p-$values above 0.2). 

\newpage \clearpage \raggedbottom

\section{Follow-up Study\label{sec:validation}}
\subsection{Experiment Design}
This Appendix presents the design and results of our follow-up study, supplementing the summary of results  in \textsection 4.4. The follow-up study followed the same design as Study 3 (\textsection 4) with the difference that there was no information elicitation stage, and complete distributional information was shared with all participants. A total of 204 participants were recruited for this study, following the same recruitment procedures and exclusion criteria as the previous studies. The pre-registration for this study is available at \url{https://aspredicted.org/p4xn-p4kh.pdf}.

\subsection{Summary Statistics}
\begin{figure}[bthp]
    \centering
    \caption{Follow-up Study: Switching Amounts by Treatment and Type of Wait}
    \includegraphics[width=\textwidth]{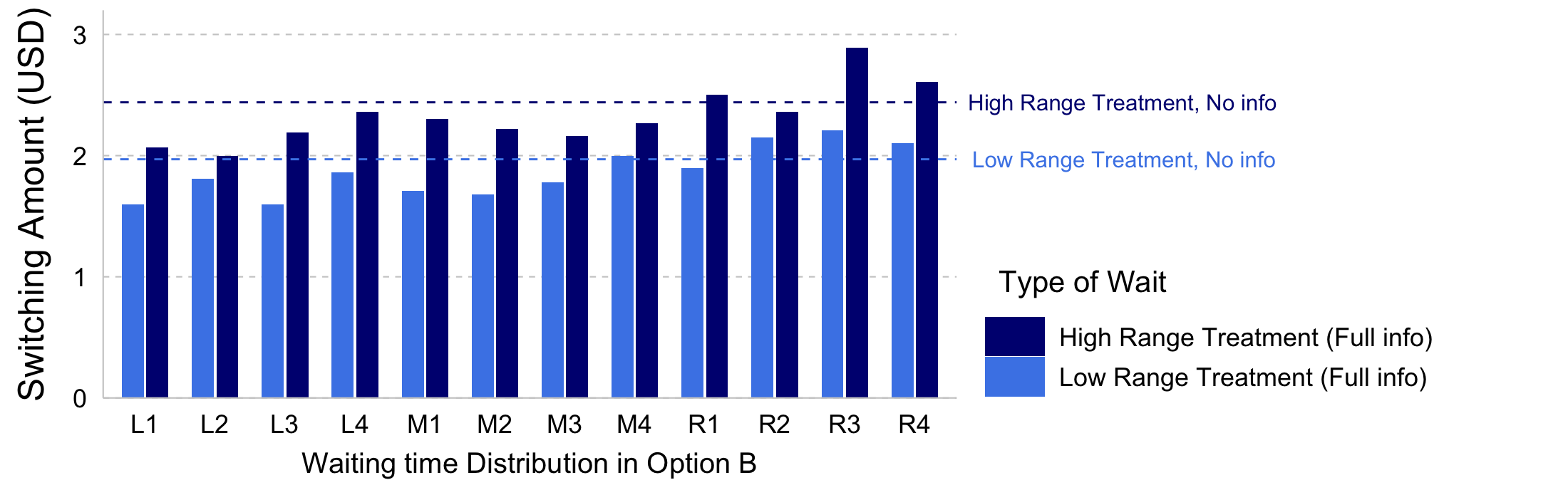}
    \label{fig:bars:s3B}
\end{figure}

Figure \ref{fig:bars:s3B} shows the switching amounts for different time distributions. Recall that a higher switching point indicates a higher cost of waiting in a given scenario. Recall further that one of the six price lists asked participants to evaluate waits in the absence of the distributional information (min/max only). These decisions are shown as horizontal dashed lines. The remaining 24 bars show the switching amounts for each of the 24 distributions presented in Figure \ref{fig:dist:s3}.

Consider first the leftmost four sets of bars. These bars show scenarios L1-L4, i.e., scenarios in which the waiting time distribution is left-tail heavy; that is, lower waiting times (1-2 minutes in the Low Range treatment and 1-5 minutes in the High Range treatment) are more likely than longer times. Unsurprisingly, revealing this information leads to a lower cost of waiting relative to the no-information scenario. Next, consider the middle four sets of bars, M1-M4. These are symmetric distribution scenarios (U-shaped, flat, and bell-shaped, respectively). In these scenarios, revealing waiting time information does not necessarily result in a lower switching amount in the Low Range treatment, but does so in the High Range treatment.  Finally, consider the rightmost four sets of bars, R1-R4, which present the response to less favorable (right-tail heavy) distributions. In the High Range treatment, participants respond quite negatively to the revealed information with all four bars extending above the no-information scenario. In contrast, the response in the Low Range treatment appears to be minimal. Overall the statistics are strikingly similar to the ones in Figure C3a (Incomplete information scenarios in Study 3), which suggests that participants respond to detailed probabilistic information similarly as they do to coarsened information in Study 3. We will next formally test how decision-makers respond to different types of information provided. 

\medskip

\subsection{Empirical Approach}
Study 2C showed that people respond to the availability of distributional information, although the direction and strength of the response varied based on tail shape. The multiple price list format used in Study 3 allows us to quantify the monetary value of receiving such information. In particular, the pooled data of Study 3 and the follow-up study creates several information regimes: \textbf{complete information}, \textbf{incomplete (coarse)  information} (probability mass revealed for all three ranges: left tail, midrange, and right tail) and \textbf{no information} (only minimum and maximum wait times shown).  
Building on Study 2, we expect more complete disclosure to be valued positively (i.e., more information $\longrightarrow$ lower cost of waiting). At the same time, we expect that the effect depends on the shape of the distribution that is being disclosed. In particular, Study 2C showed a preference for long right-tailed distributions over completely unknown ones, as well as (some) preference for unknown distributions over known, thick right-tailed ones. 



We next examine whether the presence and type of distributional information indeed affects decisions. Recall from Study 1 that switching amount is the maximum US Dollar amount  that participants are willing to give up to avoid the uncertain wait (and switch to the certain wait of 1 minute and a smaller monetary payment of \$1), with higher values indicating greater aversion to the uncertain option. To test H3B, we examine how different levels of distributional information disclosure affect participants' willingness to accept uncertain waiting times using the regression specification:
\vspace{-.2cm}
\begin{multline*}
\texttt{SwitchAmount}_{it} = \beta_0 + \boldsymbol{\beta_1}(\texttt{CoarseInfoRevealed} \times \textbf{\texttt{Distribution}})_{it} \\
+ \boldsymbol{\beta_2} (\texttt{CompleteInfoRevealed} \times \textbf{\texttt{Distribution}})_{it} + \gamma \texttt{Controls}_{it} + \alpha_i + \epsilon_{it}
\end{multline*}
\vspace{-.74cm}

\noindent where  $\boldsymbol{\beta_1}$ captures the effect of sharing coarse information and $\boldsymbol{\beta_2}$ captures the effect of sharing complete information, both relative to not sharing any distributional information. The  $\boldsymbol{\beta_1}$ and $\boldsymbol{\beta_2}$ coefficients are vector-valued because the information shared is different for each distribution seen by a subject. By estimating this regression for subsets of distributions (thick left-tailed, thick right-tailed, symmetric, pooled), and then computing the average marginal effects, we can examine the effects of information disclosure on choices. 

\subsection{Results}
The results confirm that information disclosure increases the willingness to accept uncertain waits. The average marginal effects are in Table \ref{tab:s3:inforegs}. For thick left-tail distributions (col. 1), which are favorable because they concentrate probability mass on shorter waits, both information disclosure regimes (coarse and full) significantly reduce switching amounts (by \$0.16--\$0.18). That is, disclosure of favorable distributional characteristics substantially increases acceptance of uncertain waits.  Symmetric distributions (col. 2) show negative but relatively modest effects. For thick right-tail distributions (col. 3), coarse information disclosure actually increases switching amounts by \$0.076. Complete information produces a smaller and statistically insignificant effect in this case (\$0.035, $p=0.098$). The pooled results (panel d) show that information disclosure reduces switching amounts \textit{on average}, if we consider all 12 distributions. Thus, while more (less) favorable information leads to greater (lower) uptake of the uncertain option, the overall effects of information disclosure are significantly positive. Further, while decision-makers respond more favorably to receiving full (vs. coarse) information, the differences between the two are not statistically significant for any of the four specifications (lowest $p=0.164$). That is, decision-makers attach little value to the precision of information, once the overall shape is revealed. 


\smallskip

\begin{table}[h!]
\centering 
\renewcommand{\arraystretch}{1}
\scriptsize
\caption{Study 3: Information Regimes Regression Results}
    \label{tab:H3B:reg}
\begin{tabular}{ld{3.6}@{}ld{3.6}@{}ld{3.6}@{}ld{3.6}@{}l} 
\toprule
 \multicolumn{9}{c}{Dependent variable: \texttt{Switching Amount} (USD)} \\ 
\midrule
\multicolumn{1}{r}{Distributions included:} & \multicolumn{2}{c}{L1-L4} & \multicolumn{2}{c}{M1-M4} & \multicolumn{2}{c}{R1-R4} & \multicolumn{2}{c}{All}\\ 
 & \multicolumn{2}{c}{(1)} & \multicolumn{2}{c}{(2)} & \multicolumn{2}{c}{(3)} & \multicolumn{2}{c}{(4)} \\ 
\midrule
Information Regime & & & & & & & & \\
~~~~\texttt{CoarseInfo}   & -0.163\sym{***} & (0.022) & -0.075\sym{***} & (0.020) & 0.076\sym{***} & (0.021) & -0.071\sym{**} & (0.029) \\
~~~~(avg. marginal effect) & & & & & & & & \\
~~~~\texttt{FullInfo}  & -0.175\sym{***} & (0.023) & -0.082\sym{***} & (0.021) & 0.035\sym{*} & (0.021) & -0.116\sym{***} & (0.034) \\
~~~~(avg. marginal effect) & & & & & & & & \\ [2ex]
Interactions & \multicolumn{2}{c}{} & \multicolumn{2}{c}{} & \multicolumn{2}{c}{} & \multicolumn{2}{c}{} \\
~~~~\texttt{CoarseInfo} $\times$ \texttt{Distr.} & \multicolumn{2}{c}{Yes} & \multicolumn{2}{c}{Yes} & \multicolumn{2}{c}{Yes} & \multicolumn{2}{c}{Yes} \\
~~~~\texttt{FullInfo} $\times$ \texttt{Distr.} & \multicolumn{2}{c}{Yes} & \multicolumn{2}{c}{Yes} & \multicolumn{2}{c}{Yes} & \multicolumn{2}{c}{Yes} \\
\midrule
$R^2$ (Overall) & \multicolumn{2}{c}{0.077}  & \multicolumn{2}{c}{0.049}  &  \multicolumn{2}{c}{0.058}    & \multicolumn{2}{c}{0.078}   \\
No. Observations & \multicolumn{2}{c}{1,118}  &\multicolumn{2}{c}{1,113}  & \multicolumn{2}{c}{1,124}  & \multicolumn{2}{c}{2,023}   \\
No. Groups & \multicolumn{2}{c}{523}  & \multicolumn{2}{c}{523}   & \multicolumn{2}{c}{523} & \multicolumn{2}{c}{523}  \\
\bottomrule
\end{tabular}
\label{tab:s3:inforegs}
\vspace{0.1cm}
\begin{minipage}{0.92\textwidth}\scriptsize
Notes: Average marginal effects are reported (Standard errors in parentheses). All specifications include round controls and full interactions between information regimes and distribution shown. $^*$, $^{**}$ and $^{***}$ denotes significance at the 10\%, 5\% and 1\% level, respectively.
\end{minipage}
\end{table}

\clearpage \newpage

\section{Simulation Study of Queue Joining Behavior\label{sec:numerical:demo}}
This section supplements the discrete-event simulation results reported in \textsection 5.3. In this simulation, we use our utility estimation results in \textsection 5.1 to study how the choice of a utility model of customers joining a queueing system affects the preferred service system design (pooled vs. dedicated system). We next present the methodology and the key results of the simulation. 

\subsection{Simulation Methodology}
We follow the setup of \citet{sunar2021} who study the choice between pooled and dedicated queues in overloaded systems where delay-sensitive customers observe queue lengths before deciding to join. Following their framework, we simulate two service configurations:

\begin{itemize}
    \item \textbf{Pooled System:} A single $M$/$M$/$N$ queue with $N$ identical servers
    \item \textbf{Dedicated System:} $N$ parallel $M$/$M$/$1$ queues, each with a single server
\end{itemize}

\noindent In both configurations, arriving customers observe the current queue length and join only if their expected utility is positive. In the dedicated queue case customers arrive at one of the queues at random and then choose to join (or not). Outside utility of not joining the queue is normalized to zero. We compare three utility models:

\begin{enumerate}
    \item \textbf{Mean-Only Utility:} $U = R - 0.411 \cdot \mathbb{E}[W]$, based on our estimates in \textsection 5.1  (Table 10, col. 1),
    \item \textbf{Mean-Variance Utility:} $U = R - 0.406 \cdot \mathbb{E}[W] - 0.009 \cdot \text{Var}(W)$, based on our estimates in \textsection 5.1  (Table 10, col. 5), and
    \item \textbf{Tail-Based Utility:} $U = R - 0.544 \cdot \mathbb{E}[W]  + 0.331 \cdot \text{CVaR}_{80}(W) - 0.215 \cdot P_{80}(W)$, based on our estimates in \textsection 5.1  (Table 10, col. 2),
\end{enumerate}

\noindent where $R$ is the service benefit, $W$ is the random variable denoting sojourn time, $P_{80}$ is the 80th percentile, $\text{CVaR}_{80}$ is the conditional value-at-risk at the 80\% level, and $\text{Var}(W)$ is the variance. Each customer arriving to the service system observes the queue length and infers the distribution of $W$ based on queue length and service time distribution. 

We vary three key parameters:
\begin{itemize}
    \item Traffic intensity: $\rho \in \{0.9, 0.95, 1.0, 1.05, 1.1\}$
    \item Number of servers: $N \in \{2, 3, 5\}$
    \item Service benefit: $R \in \{1, 2, 3\}$
\end{itemize}

\noindent Each configuration is run for 20 replications $\times$ 5,000 time units each, with a 1,000-unit warm-up period. Social welfare is computed as the sum of expected utilities at joining for all joining customers (Outside option of not joining is normalized to zero). 

\subsection{Simulation Results}

Table \ref{tab:simulation_results} presents results for all 45 parameter combinations. The table shows throughput percentages and social welfare under pooled (Pool) and dedicated (Ded) systems, with the winning configuration highlighted in green.

\begin{table}[htbp]
\centering
\caption{Simulation Study: Comparison of Pooled vs. Dedicated Systems 
\label{tab:simulation_results}}  
\scriptsize \renewcommand{\arraystretch}{1.1}
\begin{tabular}{C{0.6cm}cL{0.6cm}@{\hspace{0.3em}}cccc@{\hspace{1em}}cccc@{\hspace{1em}}cccc}
\toprule
& & & \multicolumn{4}{C{3.2cm}}{\textbf{Mean-Only Utility} \citep{sunar2021}} & \multicolumn{4}{C{3.2cm}}{\textbf{Mean-Variance Utility} (Table 10, col. 5)} & \multicolumn{4}{C{3cm}}{\textbf{Tail-Based Utility} (Table 10, col. 2)} \\
\cmidrule(lr){4-7} \cmidrule(lr){8-11} \cmidrule(lr){12-15}
& & & \multicolumn{2}{c}{\textbf{Thru (\%)}} & \multicolumn{2}{c}{\textbf{Welfare}} & \multicolumn{2}{c}{\textbf{Thru (\%)}} & \multicolumn{2}{c}{\textbf{Welfare}} & \multicolumn{2}{c}{\textbf{Thru (\%)}} & \multicolumn{2}{c}{\textbf{Welfare}} \\
\cmidrule(lr){4-5} \cmidrule(lr){6-7} \cmidrule(lr){8-9} \cmidrule(lr){10-11} \cmidrule(lr){12-13} \cmidrule(lr){14-15}
$\boldsymbol{\rho}$ & $\boldsymbol{N}$ & $\boldsymbol{R}$ & \textbf{Pool} & \textbf{Ded} & \textbf{Pool} & \textbf{Ded} & \textbf{Pool} & \textbf{Ded} & \textbf{Pool} & \textbf{Ded} & \textbf{Pool} & \textbf{Ded} & \textbf{Pool} & \textbf{Ded} \\
\midrule
0.90 & 2 & 1 & 81 & 70 & \cellcolor{green!20}2502 & 1987 & 81 & 70 & \cellcolor{green!20}2481 & 1960 & 75 & 79 & 3737 & \cellcolor{green!20}3871 \\
0.90 & 2 & 2 & 93 & 84 & \cellcolor{green!20}6918 & 6202 & 93 & 84 & \cellcolor{green!20}6877 & 6154 & 82 & 90 & 7740 & \cellcolor{green!20}8098 \\
0.90 & 2 & 3 & 97 & 92 & \cellcolor{green!20}12051 & 10088 & 97 & 92 & \cellcolor{green!20}11951 & 9955 & 86 & 94 & 11966 & \cellcolor{green!20}12642 \\
0.90 & 3 & 1 & 89 & 70 & \cellcolor{green!20}3766 & 2989 & 89 & 70 & \cellcolor{green!20}3716 & 2943 & 78 & 79 & \cellcolor{green!20}6459 & 5833 \\
0.90 & 3 & 2 & 96 & 84 & \cellcolor{green!20}11093 & 9322 & 97 & 84 & \cellcolor{green!20}11162 & 9207 & 83 & 90 & \cellcolor{green!20}13509 & 12091 \\
0.90 & 3 & 3 & 99 & 92 & \cellcolor{green!20}19501 & 15118 & 99 & 92 & \cellcolor{green!20}19589 & 14994 & 90 & 94 & \cellcolor{green!20}18969 & 18895 \\
0.90 & 5 & 1 & 95 & 70 & \cellcolor{green!20}6849 & 4969 & 94 & 70 & \cellcolor{green!20}6775 & 4918 & 82 & 79 & \cellcolor{green!20}12195 & 9706 \\
0.90 & 5 & 2 & 99 & 84 & \cellcolor{green!20}20703 & 15546 & 99 & 84 & \cellcolor{green!20}20816 & 15430 & 89 & 90 & \cellcolor{green!20}23058 & 20286 \\
0.90 & 5 & 3 & 100 & 92 & \cellcolor{green!20}36590 & 25226 & 100 & 92 & \cellcolor{green!20}36581 & 25009 & 92 & 94 & \cellcolor{green!20}33974 & 31551 \\
\midrule
0.95 & 2 & 1 & 79 & 68 & \cellcolor{green!20}2529 & 2023 & 79 & 69 & \cellcolor{green!20}2500 & 1990 & 73 & 77 & 3760 & \cellcolor{green!20}3955 \\
0.95 & 2 & 2 & 91 & 82 & \cellcolor{green!20}6713 & 6208 & 91 & 82 & \cellcolor{green!20}6701 & 6170 & 79 & 88 & 7766 & \cellcolor{green!20}7969 \\
0.95 & 2 & 3 & 95 & 90 & \cellcolor{green!20}11200 & 9839 & 96 & 90 & \cellcolor{green!20}11342 & 9715 & 84 & 92 & 11934 & \cellcolor{green!20}12143 \\
0.95 & 3 & 1 & 88 & 68 & \cellcolor{green!20}3701 & 3020 & 87 & 69 & \cellcolor{green!20}3657 & 2984 & 76 & 77 & \cellcolor{green!20}6521 & 5899 \\
0.95 & 3 & 2 & 95 & 82 & \cellcolor{green!20}10552 & 9308 & 95 & 82 & \cellcolor{green!20}10566 & 9242 & 81 & 88 & \cellcolor{green!20}13524 & 11898 \\
0.95 & 3 & 3 & 97 & 90 & \cellcolor{green!20}17975 & 14721 & 97 & 90 & \cellcolor{green!20}18366 & 14723 & 87 & 92 & \cellcolor{green!20}18371 & 18017 \\
0.95 & 5 & 1 & 93 & 68 & \cellcolor{green!20}6519 & 5042 & 93 & 68 & \cellcolor{green!20}6497 & 4986 & 80 & 77 & \cellcolor{green!20}12387 & 9850 \\
0.95 & 5 & 2 & 97 & 82 & \cellcolor{green!20}18658 & 15558 & 98 & 82 & \cellcolor{green!20}18728 & 15440 & 87 & 88 & \cellcolor{green!20}22808 & 19941 \\
0.95 & 5 & 3 & 99 & 90 & \cellcolor{green!20}33236 & 24551 & 99 & 90 & \cellcolor{green!20}33113 & 24317 & 90 & 92 & \cellcolor{green!20}33242 & 30237 \\
\midrule
1.00 & 2 & 1 & 78 & 67 & \cellcolor{green!20}2538 & 2047 & 77 & 67 & \cellcolor{green!20}2502 & 2009 & 71 & 75 & 3788 & \cellcolor{green!20}3996 \\
1.00 & 2 & 2 & 90 & 80 & \cellcolor{green!20}6491 & 6227 & 90 & 80 & \cellcolor{green!20}6487 & 6169 & 78 & 86 & 7771 & \cellcolor{green!20}7840 \\
1.00 & 2 & 3 & 93 & 88 & \cellcolor{green!20}10301 & 9486 & 93 & 87 & \cellcolor{green!20}10603 & 9316 & 81 & 90 & \cellcolor{green!20}11732 & 11563 \\
1.00 & 3 & 1 & 85 & 67 & \cellcolor{green!20}3590 & 3067 & 85 & 67 & \cellcolor{green!20}3586 & 3015 & 74 & 75 & \cellcolor{green!20}6546 & 5956 \\
1.00 & 3 & 2 & 93 & 80 & \cellcolor{green!20}9745 & 9321 & 93 & 80 & \cellcolor{green!20}9856 & 9251 & 79 & 86 & \cellcolor{green!20}13620 & 11714 \\
1.00 & 3 & 3 & 95 & 88 & \cellcolor{green!20}16238 & 14320 & 95 & 87 & \cellcolor{green!20}16384 & 14134 & 85 & 90 & \cellcolor{green!20}17900 & 17178 \\
1.00 & 5 & 1 & 90 & 67 & \cellcolor{green!20}6206 & 5108 & 91 & 67 & \cellcolor{green!20}6184 & 5034 & 77 & 75 & \cellcolor{green!20}12428 & 9942 \\
1.00 & 5 & 2 & 96 & 80 & \cellcolor{green!20}16399 & 15584 & 96 & 80 & \cellcolor{green!20}16415 & 15402 & 84 & 86 & \cellcolor{green!20}22360 & 19779 \\
1.00 & 5 & 3 & 97 & 88 & \cellcolor{green!20}26422 & 23720 & 97 & 88 & \cellcolor{green!20}26929 & 23541 & 88 & 90 & \cellcolor{green!20}32148 & 28949 \\
\midrule
1.05 & 2 & 1 & 75 & 65 & \cellcolor{green!20}2542 & 2073 & 75 & 65 & \cellcolor{green!20}2525 & 2031 & 69 & 73 & 3808 & \cellcolor{green!20}4019 \\
1.05 & 2 & 2 & 87 & 78 & 6156 & \cellcolor{green!20}6190 & 87 & 78 & \cellcolor{green!20}6165 & 6141 & 75 & 83 & 7671 & \cellcolor{green!20}7672 \\
1.05 & 2 & 3 & 91 & 85 & \cellcolor{green!20}9416 & 9132 & 91 & 86 & \cellcolor{green!20}9529 & 9083 & 80 & 88 & \cellcolor{green!20}11645 & 10904 \\
1.05 & 3 & 1 & 83 & 65 & \cellcolor{green!20}3514 & 3094 & 83 & 65 & \cellcolor{green!20}3505 & 3051 & 72 & 73 & \cellcolor{green!20}6613 & 5990 \\
1.05 & 3 & 2 & 90 & 78 & 9018 & \cellcolor{green!20}9271 & 90 & 78 & 9091 & \cellcolor{green!20}9215 & 77 & 83 & \cellcolor{green!20}13562 & 11550 \\
1.05 & 3 & 3 & 92 & 86 & 13707 & \cellcolor{green!20}13822 & 92 & 85 & \cellcolor{green!20}13736 & 13497 & 83 & 88 &\cellcolor{green!20} 17407 & 16592 \\
1.05 & 5 & 1 & 88 & 65 & \cellcolor{green!20}5862 & 5159 & 88 & 65 & \cellcolor{green!20}5822 & 5093 & 75 & 73 & \cellcolor{green!20}12567 & 10030 \\
1.05 & 5 & 2 & 93 & 78 & 13705 & \cellcolor{green!20}15572 & 93 & 78 & 14192 & \cellcolor{green!20}15361 & 82 & 84 & \cellcolor{green!20}22026 & 19274 \\
1.05 & 5 & 3 & 94 & 85 & 19914 & \cellcolor{green!20}22882 & 94 & 85 & 20300 & \cellcolor{green!20}22638 & 86 & 88 & \cellcolor{green!20}30759 & 27277 \\
\midrule
1.10 & 2 & 1 & 73 & 64 & \cellcolor{green!20}2540 & 2091 & 74 & 64 & \cellcolor{green!20}2521 & 2057 & 68 & 71 & 3825 & \cellcolor{green!20}4023 \\
1.10 & 2 & 2 & 85 & 76 & 5939 & \cellcolor{green!20}6175 & 85 & 76 & 5875 & \cellcolor{green!20}6095 & 74 & 82 & \cellcolor{green!20}7634 & 7579 \\
1.10 & 2 & 3 & 88 & 83 & 8523 & \cellcolor{green!20}8846 & 88 & 83 & \cellcolor{green!20}8760 & 8714 & 78 & 85 & \cellcolor{green!20}11419 & 10435 \\
1.10 & 3 & 1 & 81 & 63 & \cellcolor{green!20}3415 & 3131 & 81 & 63 & \cellcolor{green!20}3411 & 3083 & 70 & 71 & \cellcolor{green!20}6631 & 6034 \\
1.10 & 3 & 2 & 88 & 76 & 8265 & \cellcolor{green!20}9269 & 88 & 76 & 8242 & \cellcolor{green!20}9152 & 75 & 82 & \cellcolor{green!20}13583 & 11328 \\
1.10 & 3 & 3 & 89 & 83 & 11811 & \cellcolor{green!20}13223 & 89 & 83 & 11930 & \cellcolor{green!20}12906 & 81 & 85 & \cellcolor{green!20}16804 & 15474 \\
1.10 & 5 & 1 & 85 & 63 & \cellcolor{green!20}5464 & 5209 & 85 & 63 & \cellcolor{green!20}5435 & 5118 & 74 & 71 & \cellcolor{green!20}12622 & 10061 \\
1.10 & 5 & 2 & 90 & 76 & 11409 & \cellcolor{green!20}15489 & 90 & 76 & 11560 & \cellcolor{green!20}15277 & 80 & 81 & \cellcolor{green!20}21584 & 18789 \\
1.10 & 5 & 3 & 91 & 83 & 14944 & \cellcolor{green!20}21856 & 91 & 83 & 15349 & \cellcolor{green!20}21545 & 84 & 85 & \cellcolor{green!20}29417 & 25764 \\
\bottomrule
\end{tabular}
\end{table}

Under mean-based utility (left panel of Table \ref{tab:simulation_results}), pooled dominates completely at $\rho \leq 1.0$, winning all 27 scenarios in this range. Joining customers observe substantially higher queue lengths in pooled systems (e.g., at $\rho=0.90, N=2, R=1$: average queue length is 1.63 customers in pooled vs. 0.47 in dedicated), but the efficiency gains from pooling outweigh these longer queues. This is a classic result in many queueing systems: pooling reduces server idle time through load balancing and produces greater welfare. 

Next, consider higher congestion levels ($\rho \geq 1.05$). At these levels, dedicated systems win 15 of 18 scenarios. Throughput ranges from 73-99\% for pooled systems and 64-92\% for dedicated systems, with pooled consistently achieving higher throughput. At overcapacity with $R>1$, dedicated systems are better because pooling creates congestion levels that are too high: too many customers join pooled queues despite long waiting times. Observed queue lengths in pooled systems can exceed 20 customers while dedicated queues remain short (e.g., at $\rho=1.10, N=5, R=3$: queue lengths are 26.4 in pooled vs. 3.4 in dedicated). This is consistent with \cite{sunar2021}, who show that pooled systems generate excessive throughput in overloaded systems, due to overjoining.

Mean-variance utility (middle panel of Table \ref{tab:simulation_results}) produces similar results to mean-based utility. Pooled wins all scenarios at $\rho \leq 1.0$, while dedicated wins 13 of 18 scenarios at $\rho \geq 1.05$. Throughput and welfare values are close to mean-based utility. 

Tail-based utility (right panel of Table \ref{tab:simulation_results}) shows a different pattern of results, with pooled systems winning 37 of 45 cases (82\%). This is because different queue designs produce different tail shapes. At low congestion ($\rho=0.90$) with few servers ($N=2$), dedicated wins across all service benefit levels ($R=1,2,3$). In this regime, dedicated queues produce long thin tails: wait times can occasionally extend to extreme values, but these occurrences are spread smoothly across the tail rather than concentrated near the maximum. Indeed, dedicated systems lead to low 80th percentile outcomes but produce quite long tails, while pooled systems have shorter tails with less extreme outcomes. This leads to a 4-12 percentage points higher throughput in dedicated systems. Notably, the advantage of dedicated systems only holds with $N=2$, because the efficiency advantage of pooling is relatively small in this case. 

At higher congestion ($\rho \geq 1.05$) with $N=2$, coordination failure (overjoining) in dedicated systems begins to produce thicker tails: probability mass now often concentrates near the maximum wait time rather than being spread smoothly. For example, at $\rho=1.10$, $N=2$, $R=2$, dedicated systems exhibit severe thick-tail problems with $P_{80}=5.17$ compared to pooled $P_{80}=3.13$. In this case, the thick right tail becomes so aversive that pooled systems win despite 8 percentage points lower throughput (74\% vs 82\%). This is because the utility advantage from avoiding thick moderate tails outweighs the throughput disadvantage. Pooling provides load balancing that smoothens out the tail, preventing the concentration of probability mass near extreme wait times that customers find so aversive. 

As the number of servers increases to $N=5$, pooled dominates across nearly all scenarios. With more servers, pooling distributes arrivals across a larger shared capacity, which prevents the queue from building thick right tails. This is because load balancing in pooled systems helps smoothen the tails of distributions even at higher congestion levels. 

The high-level takeaway from the simulation exercise is that when customers have tail-based preferences (averse to thick right tails but do not mind long-thin tails), the optimal service design can be markedly different from that under the classic mean-based utility. Failing to account for these preferences can lead to substantial welfare losses.
\raggedbottom
\clearpage

\end{document}